\documentclass[draft]{agujournal2019}
\usepackage{url}
\usepackage{amsfonts, latexsym}
\usepackage{amsmath, amssymb}
\usepackage{lineno}
\usepackage{graphicx}
\usepackage{csquotes}
\usepackage{array}
\newcolumntype{?}[1]{!{\vrule width #1}}
\usepackage[export]{adjustbox}
\usepackage{lscape} 
\usepackage{textcomp}
\usepackage{numprint}
\usepackage{xcolor,colortbl}
\usepackage{multirow}
\usepackage{comment}
\usepackage{xr-hyper}
\usepackage{apacite}

\draftfalse

\journalname{JGR: Solid Earth}

\begin{document}

\title{Attention network forecasts time-to-failure in laboratory shear experiments}


\authors{Hope Jasperson\affil{1}, David C. Bolton\affil{2}, Paul Johnson\affil{3}, Robert Guyer\affil{3,4}, Chris Marone\affil{2,5}, Maarten V. de Hoop\affil{6}}

\affiliation{1}{Rice University, EEPS}
\affiliation{2}{Pennsylvania State University, Department of Geosciences}
\affiliation{3}{Los Alamos National Laboratory, Geophysics Group}
\affiliation{4}{University of Nevada, Reno, Department of Physics}
\affiliation{5}{Sapienza Universitá di Roma, Dipartimento Scienze della Terra}
\affiliation{6}{Rice University, CAAM}

\correspondingauthor{Hope Jasperson}{haj2@rice.edu}

\date{}
	
\begin{abstract}

Rocks under stress deform by creep mechanisms that include formation and slip on small-scale internal cracks. Intragranular cracks and slip along grain contacts release energy as elastic waves termed acoustic emissions (AE). AEs are thought to contain predictive information that can be used for fault failure forecasting. Here we present a method using unsupervised classification and an attention network to forecast labquakes using AE waveform features. Our data were generated in a laboratory setting using a biaxial shearing device with granular fault gouge intended to mimic the conditions of tectonic faults. Here we analyzed the temporal evolution of AEs generated throughout several hundred laboratory earthquake cycles. We used a Conscience Self-Organizing Map (CSOM) to perform topologically ordered vector quantization based on waveform properties. The resulting map was used to interactively cluster AEs. We examined the clusters over time to identify those with predictive ability. Finally, we used a variety of LSTM and attention-based networks to test the predictive power of the AE clusters. By tracking cumulative waveform features over the seismic cycle, the network is able to forecast the time-to-failure (TTF) of lab earthquakes. Our results show that analyzing the data to isolate predictive signals and using a more sophisticated network architecture are key to robustly forecasting labquakes. In the future, this method could be applied on tectonic faults monitor earthquakes and augment current early warning systems.

\end{abstract}

\section*{Plain Language Summary}
Earthquake forecasting is a daunting task, but advances in machine learning can help us towards this goal. In this study, we combine machine learning with expert knowledge about earthquake formation to forecast synthetic earthquakes made in a laboratory. Prior to an earthquake, rocks emit small bursts of energy as they are stressed, which are called acoustic emissions (AE). We identified all the AEs in the raw seismic data from our lab earthquakes and then used a machine learning algorithm called a Conscience Self-Organizing Map to divide them into groups based on their characteristics. We examined the groups to determine which ones would be useful for earthquake forecasting. Finally, we used the data from the chosen groups along with another machine learning network (LSTM and attention networks) to forecast the amount of time until the next earthquake. Our results show that you can reduce earthquake forecasting error by processing seismic data to find predictive signals and by using machine learning networks that are specifically made for time series data.

\section{Introduction}

Earthquake forecasting is a very active topic in geoscience research and recent progress in this area has been rapid, thanks in part to developments in machine learning and the availability of laboratory seismic data containing very large number of labquakes. Quite recently, many researchers have used a variety of machine learning methods in an attempt to forecast labquakes and have had moderate success \cite<e.g.>{rouetLeduc2017, corbi2019, hulbert2019, tanyuk2019, johnson2021}. In a recent Kaggle competition hosted by Los Alamos National Lab and the DOE Office of Science, the top scoring team used a combination of gradient-boosted trees, support vector regression, and a convolutional neural network to forecast time-to-failure (TTF) from raw laboratory seismic data for a final mean absolute error (MAE) of 2.26s \cite{johnson2021}. This network and others achieve respectable results, but their success is limited because their training procedures lack domain expert knowledge and generally fail to fully utilize the power of recurrence. In this work, we prepare  laboratory seismic data for a forecasting network by identifying predictive signals via catalog creation and clustering. By doing this, we remove noise and spurious signals that make training a machine learning network more difficult. We use these predictive signals to train a variety of recurrent networks and achieve a superior MAE of 1.04s using an attention network. Attention is meant to mimic cognitive attention by enhancing the important aspects of the input data and muting the remaining aspects. Determining which aspects of the data are important is learned via model training data by gradient descent. Our results show that analyzing the data to isolate predictive signals and using a more sophisticated network architecture are key to robustly forecasting labquakes.

A primary motivation for this study comes from the engineering field of nondestructive testing (NDT). The goal of NDT is to continuously assess the health of a material or implement while it is in operation \cite{Farrar2013}. NDT has many practical applications, including identifying when machine parts need to be repaired and forecasting infrastructure failure. NDT techniques have been successfully applied to a wide range of materials, such as reinforced concrete \cite{calabrese2013}, manufacturing tools \cite{yen2013}, composite materials \cite{li2015}, wood \cite{diakhate2017} and general structural health monitoring including structures \cite{Farrar2013}.

NDT AE monitoring relies on continuously recording acoustic data as the material of interest is stressed. Acoustic data is recorded at least until the material reaches failure. In the lab, the failure process can be sped up by artificially subjecting the material to stress \cite<e.g.>{calabrese2013,deOliveira2008,diakhate2017, huguet2002,li2015}. Once the material has reached failure, discrete acoustic emissions are isolated within the acoustic data. A discrete AE is a finite-duration elastic wave that is produced by the formation of small-scale internal cracks and slip along grain contacts. Thus, AEs provide a record of how the material responds internally to stress. Next, AEs can be clustered according to damage mechanism using an unsupervised clustering algorithm. This step groups AEs according to the type of crack or deformation by which they were produced. In some studies, transmitted light \cite<e.g.>{li2015} or scanning electron microscopy \cite<e.g.>{fallahi2016} is used to determine the true labels for each cluster. Finally, AE production throughout the failure cycle is analyzed in order to identify temporal patterns and make inferences about the remaining life (RUL) of the material. Some studies have gone further and used acoustic data combined with machine learning methods to forecast RUL with varying degrees of success \cite<e.g.>{Farrar2013, elforjani2018, louis2020, zheng2017}.

In this study, our goal is to follow the example of NDT in developing a deep-learning based forecasting procedure trained and tested on laboratory-generated seismic data. Though the earthquake rupture process is generally different from the material failure studied in NDT, the underlying physics of AE production are thought to be the similar or the same \cite<e.g.>{scholz1968, lockner1977}. In bi-axial shear experiments we analyze, grain-to-grain and grain-to-block displacements are the origin of the AE signals \cite{trugman2020}. Previous studies on similar laboratory datasets sought to identify temporal trends and precursors in acoustic data via clustering \cite{bolton2019} and forecast TTF applying decision tree approaches using the continuous AE signal \cite{rouetLeduc2017} and AE catalogs \cite{lubbers2018}, demonstrating that the laboratory seismic signal contains predictive information. A similar approach applied to the phenomenon of 'slow slip' in the laboratory shear experiments \cite{hulbert2019} and the Cascadia subduction zone, demonstrated that seismic signals in Earth are imprinted with information regarding fault slip rate and upcoming failure \cite{rouetLeduc2019}. Here, we apply a very different ML approach and identify components of the seismic signal that carry predictive information through clustering and use machine learning to continuously forecast TTF and fault shear stress.

Our forecasting method consists of four steps: catalog creation, clustering, cluster selection, and forecasting via deep learning. In the first step, we create a catalog of AEs as a way to eliminate noise and focus on potentially useful waveforms. This can be thought of as segmentation and a kind of coarse clustering, such as that of \citeA{seydoux2020}. Next, we perform a fine grained clustering procedure on the catalog to further categorize the AEs. We introduce a cluster selection process to identify useful clusters and eliminate clusters that do not contain predictive information. Finally, we use an agglomorate of recurrent neural networks to forecast TTF and shear stress at the fault using data from the previously selected waveform clusters.

The layout of this paper is as follows. In section~\ref{benchmark}, we first test our proposed method on a synthetic benchmark dataset. This dataset is generated by the brittle-ductile friction model, also known as the 'broom model' \cite{daub2011}. We train an LSTM network on the benchmark data, which successfully forecasts time to failure. With the success of this test, the remainder of the paper is focused on applying this same forecasting procedure to laboratory seismic data, which is described in section~\ref{labData}. In section~\ref{clustering}, we discuss the procedure used to prepare the AEs for the forecasting network. As in the NDT process, we further refine the catalog by clustering AEs according to waveform features. After analyzing the resulting clusters, we find that only two clusters appear to contain predictive information. In section~\ref{forecasting}, we use these two clusters to train a collection of networks to forecast time to failure and shear stress. Finally, in section~\ref{comparison}, we discuss prior work and provide comparisons with our method. Additional comparisons to historical methods are provided in Section 6 
of the Supporting Information (SI).

\section{LSTM and attention networks}

Our forecasting system is based around the use of recurrent networks, particularly LSTM \cite{hochreiter1997} and attention networks \cite{bahdanau2015}. This type of network is well-suited to time series forecasting because of its ability to store past information for future use. In SI Section 1, 
we establish that rate and state friction can be written as a recurrent network from first principles, making this a physics-informed approach. For training, we use a seq2seq approach \cite{sutskever2014} in which the input sequence consists of seismic features and the output sequence is the forecasted values for shear stress and TTF.

The LSTM network consists of several unidirectional LSTM cells stacked so that the hidden state from one becomes the input vector for the next (Figures S16 and S17 
in SI Section 4). 
A final fully connected layer outputs the time-to-failure and shear stress at the current time step. 

Attention networks are similar to LSTM networks, with the addition of a special layer that gives the network greater ability to observe past data trends. These networks consist of an encoder-decoder structure and generally use gated recurrent units (GRUs) rather than LSTM cells (Figures S19 and S20 
in SI Section 4). 
During the training process, the network learns to assign a relative importance to each data point, which is used as a weighting metric when making forecasts. For these networks, we use local self-attention where at time step t, the network only looks at a fixed window of previous time steps, in which importances are calculated for each AE.

The input vectors for each network consist of cumulative waveform features. Part of our goal is to train each network only on waveforms that contain predictive information. We expect that eliminating non-predictive signals and noise from the training data will increase network accuracy. Therefore, we choose not to use the raw trace or features calculated from sliding windows. We use cumulative features for training as opposed to the raw values to make it easier for the networks to learn from temporal patterns.

We used several common network modifications in order to speed up training and enhance generalization. We trained with mini-batches wherein the network sees multiple quakes at once rather than sequentially. Weight updates are performed after each mini-batch, and thus occur multiple times per epoch. This has the benefit of speeding up training and balancing the training loss between hypersensitivity (i.e. updating after each quake) and insensitivity (i.e. updating after a full epoch). Mini-batches are determined randomly at the beginning of each epoch and thus may contain quakes of very different recurrence intervals. To account for the variable recurrence interval, we pad shorter cycles with zeros to match the longest recurrence interval of the batch. The padding is not included in error calculations. We experimented with different mini-batch sizes to find one that was both fast and accurate.

Similarly, we use a pre-determined sequence length to control the number of weight updates. Updating the network weights after each AE would be time-consuming and excessive, but waiting until the end of a cycle risks averaging out the data and preventing learning. Instead, we update after a pre-determined number of AEs have been seen. After each update, this count restarts. As with the other hyper-parameters, we experimented with different values to strike a balance between too few and too many AEs per update.

Finally, we used gradient clipping to reduce the maximum value of the weight gradients \cite{pascanu2013}. Recurrent networks are prone to excessively large gradients and this can especially be a problem with unbalanced datasets such as the lab data, where short and long recurrence intervals are underrepresented. Gradient clipping speeds up convergence by reducing unwanted fluctuations in the network weights. We experimented with several clipping thresholds to find one that was well-suited to the data.

To test the universality of these networks, we train on all stress conditions and recurrence intervals simultaneously. The cycles are randomly shuffled at the beginning of each epoch so that each training batch contains a random selection of cycles. We train each network using the Adam optimizer \cite{kingma2017, loshchilov2019} and an adaptive learning rate. Loss during training is calculated using the smooth L1 loss function.

\section{Forecasting benchmark} \label{benchmark}

We begin by testing our proposed forecasting method on a benchmark dataset, specifically the Broom model of \citeA{daub2011}. Our goals here are twofold: to test the capability of recurrent networks to learn patterns from metastable dynamical systems and to assess the robustness of the forecasting method. We perform these tests using a LSTM network. We do this because the LSTM is simpler and faster to train than the attention network. However, the LSTM network is also less powerful than the attention network \cite{bahdanau2015}. If the LSTM successfully learns the benchmark data, then we assume that the attention network would work as well.

Next we will briefly describe the results of the benchmark tests. The full results are in SI Section 2. 
We find that the LSTM network is clearly capable of forecasting failure from the Broom model data. In general, the network forecasts time to failure very accurately and does so consistently across all stress levels and recurrence intervals. With these tests, we show that LSTM networks have the ability to learn from irregularly sampled metastable seismic systems. The Broom model is a simple example of such a system, but the excellent network performance suggests that a similar method could be used on more complex data.

\section{Laboratory experiments and previous studies}  \label{labData}

With the success of the forecasting tests on the Broom model data, the next task is to test the network's ability to learn a more complex system. For this task, we focus on laboratory generated seismic data. The laboratory data represents fault conditions on a single frictional patch in the real earth with minimal noise, scattering, attenuation, etc. Furthermore, unlike seismogenic fault zones, the characteristics of laboratory earthquakes can be systematically modulated via the boundary conditions of the experiment. Hence, hundreds of laboratory earthquakes can be produced within a single experiment that have approximately the same properties (assuming the loading conditions are held constant). In addition, the laboratory data also provide high-resolution AE data from a network of acoustic sensors. The acoustic sensors are placed 22 mm from the edge of the fault zone, which gives a reasonable approximation of the dynamical system. This type of data is a step up in complexity compared to the Broom model, but is still far from the true complexity in tectonic fault systems where a fault may comprise a very large number of frictional patches.

The laboratory datasets were created using a biaxial shearing system in a double-direct shear (DDS) configuration \cite<e.g.>{anthony2005,riviere2018,bolton2020}. The DDS configuration consists of two granular fault zones positioned between three steel loading platens. The central block is driven at constant loading rate producing a series of stick-slip events (i.e, laboratory earthquakes). Adjacent to the DDS loading platens are two steel blocks containing an array of  piezoelectric transducers (Figure S10 
in SI Section 3). 
AE data are recorded continuously throughout the experiment at about 4 MHz. Experiments were conducted with soda-lime glass beads due to their highly reproducible seismic and frictional properties \cite{mair2002, anthony2005, scuderi2014}. In this work, we analyze data from an experiment where the normal stress was systematically increased stepwise from 2-8 MPa in steps of 1 MPa, and subsequently decreased back to 2 MPa (see Figure S11 
in SI Section 3). 

Many previous studies have worked to analyze the laboratory data and describe the underlying physics. \citeA{riviere2018} documented changes in the Gutenberg-Richter b-value over lab seismic cycles. \citeA{bolton2020} examined how AE energy release is impacted by factors such as shear velocity, slip displacement, stress drop, etc. They found that acoustic variance is tied to the slip rate of the fault, which results in higher variance closer to failure similar to the results described in \citeA{rouetLeduc2017,rouetLeduc2018} by analyzing the continuous signal. It was also observed in these works that this phenomenon is stronger at high stress than at low stress. In simulations applying DEM and FDEM \cite[respectively]{ren2019,gao2018}, it was shown how the movement of individual gouge grains can be used to predict macroscopic friction.

Other studies have examined the role of AEs as precursors to failure. \citeA{bolton2019} clustered AEs in PC space and found a progression of clusters that indicate the slip stage of the fault. \citeA{shreedharan2020} concluded that changes in acoustic transmissivity (a variation on acoustic amplitude) are a precursor to fault failure. \citeA{lubbers2018} found that there is an evolution of micro events that increase in density as failure is approached. \citeA{trugman2020} located AEs within the fault gouge and observed their spatio-temporal evolution over the seismic cycle. They found that each cycle is different and thus there was not a unifying common pattern.

In this study, we use acoustic data from two laboratory experiments (p4581 and p4583) for a total of nearly 400 stick-slip events. These events take place under a variety of stress conditions with shear stresses ranging from roughly 1 to 3.5 MPa. We identified events using the recorded shear stress and discarded events that took place during transitions between normal stress levels. We focus only on data from the inter-seismic period, which is defined as the minimum shear stress of the previous stick-slip event to the maximum shear stress of the current stick-slip event. The recurrence interval of the stick-slip cycles range from 5-25 seconds, with an average of 10 seconds (note this is averaged over multiple stress levels) (Figure S11 
in SI Section 3). 

In the laboratory, it is common for the main stick-slip event to be preceded by smaller events that have measurable stress drops (see Figure S11b 
in SI Section 3). 
We refer to these small events that precede the main-stick slip event as foreshocks. These events present a particular challenge that was not present in the Broom dataset. Generally it is not clear-cut if foreshocks should be included as separate failure cycles. At high stress, foreshocks tend to have very small stress drops compared to the main slip event and the overall shear stress evolution is usually unchanged. In these cases, we include the foreshocks as part of the longer failure cycle. At low stress, foreshocks tend to be more impactful in that they have proportionally larger stress drops relative to the co-seismic events and the shear stress enters a clear recovery phase as a result. In this case we treat the foreshocks as separate failure cycles. We prepared a separate dataset where foreshocks at high stress were also treated as separate cycles and found that it resulted in a clear decline in forecasting accuracy.

AEs were detected by the same process described in \citeA{lubbers2018}, 
though an unsupervised machine learning method like that of \citeA{seydoux2020} could also be used. The detection process uses a thresholding procedure to scan through the continuous data and catalog events according to their peak amplitude \cite{riviere2018}. The creation of an AE catalog is the first step in the process of reducing noise and identifying useful signals. This process is vital to producing accurate forecasts because it decreases the complexity of the data that the forecasting network must learn. The cataloging process is by no means perfect and some AEs are likely missed due to smaller events that may occur in close proximity to larger events. Furthermore, it is also possible that the magnitude of completeness increases closer to failure because of higher event rates and temporal clustering of larger events. 
However, the Broom tests showed that the LSTM network is robust in this situation. Figure S11d 
in SI Section 3 
is a snapshot of the acoustic data with AEs denoted by red circles. The catalog for the two datasets combined contains about 8 million AEs and each stick-slip cycle consists of several thousand AEs. AEs that nucleate during the co-seismic slip phase are not used.

\section{Feature engineering and AE clustering} \label{clustering}

The second step of the noise removal process is to cluster the AEs. Later, we will examine cluster evolution over time, which will be used to separate predictive and nonpredictive signals. This procedure is inspired by NDT studies in which clustering algorithms are used to provide information about AE production and importance. In these studies, clusters are thought to be associated with different micro-mechanical processes (i.e. damage mechanisms). In this sense, the clusters provide insights into the physical processes occurring inside the stressed material and helps to discriminate between important and unimportant signals. Here, we employ a clustering scheme to further remove noisy signals, which should map to one or more noise clusters, and identify AEs that may contain predictive information.

\subsection{Handcrafted features} \label{dataProc}

\begin{figure*}[!htb]
	\includegraphics[width=\textwidth]{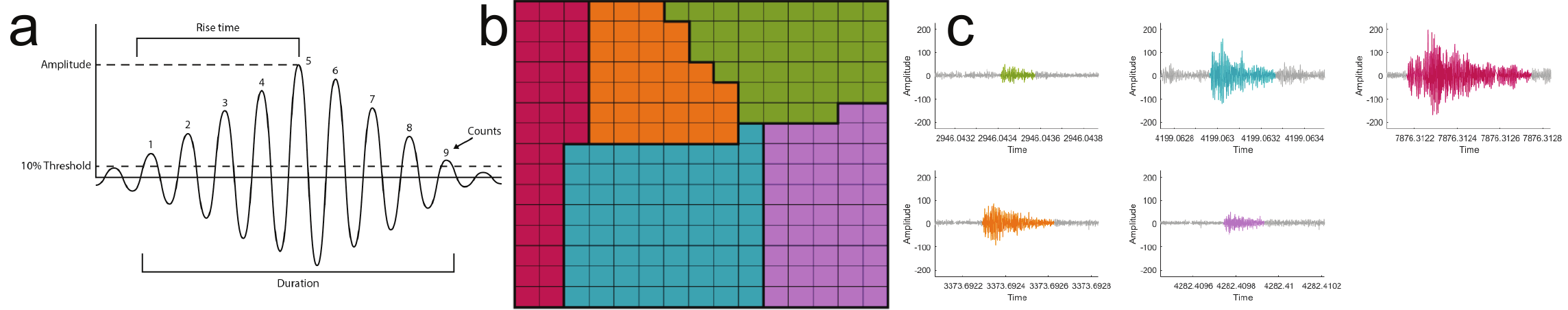}
	\caption{A) Four of the AE features used for clustering. B) Trained SOM lattice divided into five clusters. C) Typical waveforms for each of the five clusters. The colors are the same as in (B).}
	\label{som}
\end{figure*}

For the damage mechanism clustering, we use a set of five features typically used in NDT studies: maximum amplitude, counts, duration, energy, and rise time (Figure~\ref{som}a). The NDT literature shows that these five features are sufficient for AE clustering \cite{grosse2022}. Counts is the number of positive peaks over the duration of the AE and rise time is the length of time until the maximum amplitude is reached. In addition, we use two frequency features, average frequency and peak frequency, which brings the total number of features to seven.

In order to calculate features such as energy, the start and end times of the AEs must be known. Start and end times were determined using an amplitude threshold as described in \citeA{godin2004}. During the processing procedure, we found that a simple 10\% threshold was inadequate due to the large spread of maximum amplitudes. As a result, we used a decreasing threshold scheme whereby the percentage used is decreased as the maximum amplitude increases. For very low amplitude events, the minimum amplitude is fixed to be slightly above the noise level. Despite these adjustments, amplitude thresholding is an imprecise tool, which may warrant further improvements in the future. 

As with the cataloging procedure, this process wrongly eliminates some AEs. For example, amplitude thresholding eliminates AEs that overlap in time or have poorly defined start or end points. In these cases, wrongful eliminations would likely happen even if an expert filtered the AEs by hand. As long as the number of eliminations is not too great, the forecasting network should still be able to learn from the data. In the two lab datasets, less than 10\% of AEs were rejected for these reasons.

It is important to note that we do not expect a large number of clusters due to the nature of the AE catalog. The data processing involved in the catalog creation purposely filters out as much noise as possible in an attempt to leave only signals that have the potential to be seismically important. Without such a catalog, we would expect to encounter many more noise clusters.

Unlike many NDT studies, we are not able to determine the true mechanism labels for the AEs. Though this means we will not be able to provide a full account of changes occurring in the fault gouge, labels are not required for our purposes. Our main goal is to identify elements of the acoustic signal that contain predictive information, which does not necessitate knowledge of the true labels.

\subsection{CSOM and clustering}

For the AE clustering we use a conscience self-organizing map (CSOM) \cite{deSieno1988}, which is a more efficient variant of the original Kohonen self-organizing map (KSOM) \cite{kohonen1982}. In general, SOMs perform a topologically ordered mapping from a multidimensional data space to a 2D lattice space. Data points that map closely together on the lattice are also relatively close together in the data space. SOMs perform a combination of vector quantization and dimensionality reduction similar to PCA and autoencoders. Once the unsupervised SOM training process is complete, clusters are defined on the lattice using either a clustering algorithm, such as k-means, or interactively (by hand) using a variety of visualizations.

We chose to use a CSOM over other clustering methods largely due to the data complexity. The nature of the biax is very different from the NDT laboratory setups, so initially we weren't sure how many clusters to expect or even if we should expect more than one. If we did end up with only a single cluster, the benefit of a CSOM is that the lattice would provide a way of describing the continuum. Even if we did see clusters, we speculated that they would be complicated shapes with fuzzy boundaries in the data space. With an eye towards the future, we also expected that with the addition of more receivers this picture would only become more complex. Thus, we concluded that this problem is not well suited to algorithms like k-means. An alternative, and likely more precise, method would be to use a coarse clustering network and GMMs \cite{seydoux2020} in combination with fine grained clustering using a CSOM.

We trained a CSOM using a 15x15 lattice on over 8 million AEs for 20 million learning steps. The training parameters are listed in Table S8 
in SI Section 4. 
Once training was complete, we used k-means and interactive interpretation to partition the CSOM lattice into clusters. We evaluated many potential cluster configurations using cluster validity indices (CVIs) including CH-VRC~\cite{calinski1974}, Conn\textunderscore Index~\cite{tasdemir2011}, DBI~\cite{davies1979}, the Gap statistic~\cite{tibshirani2001}, GDI~\cite{bezdek1998}, PBM~\cite{pakhira2004}, and Silhouette~\cite{kaufman1990} and visualizations such as the U-matrix~\cite{ultsch1990} (and modified U-matrix~\cite{merenyi2007}), octagonal erosion~\cite{cottrell1996}, and CONNvis~\cite{tasdemir2009}. 

Ultimately we found that an interactive clustering with a k value of 5 produced the best results. The trained and interpreted CSOM lattice is shown in Figure~\ref{som} along with the typical waveforms for each cluster. Note that since we do not have the true labels of these clusters, they will henceforth be referred to by color. We show the typical waveforms to illustrate both the complexity of the signals as well as the differences between the clusters. The red cluster contains long, high amplitude events. The blue cluster is similar, but shorter and with lower amplitude. The purple and green clusters look similar in that they are both short with low amplitude, but the purple cluster has a significantly higher average and peak frequency. The orange cluster is somewhat of a catch-all for whatever signals remain, with a middling length, frequency, and amplitude.

The complexity, and decay seen in the waveforms representative of the clusters can be understood as follows. Each gouge layer is a linear elastic system 0.4 × 10 × 10 cm3. This elastic system, really a mechanical cavity or wave guide, is embedded in a steel environment (the drive block and side blocks) that has sound speed up to an order of magnitude larger than the gouge. The acoustic contrast between mechanical cavity and environment corresponds to a mechanical index of refraction of order 6. An elastic wave in a gouge layer will be reflected many times in the gouge layer before exiting into the steel environment \cite<see>{trugman2020}. Consequently a pulse received at a sensor has a distinctive pattern.

\begin{figure*}[!htb]
    \includegraphics[width=\textwidth]{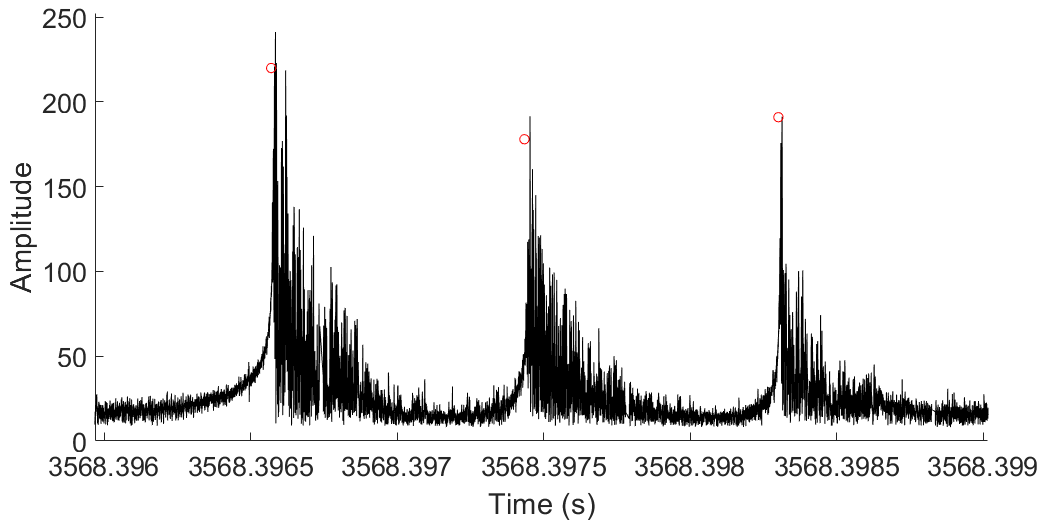}
    \caption{The envelope of the seismic data showing an instance of the same area of the gouge rupturing three times, each with decreased amplitude.}
    \label{envelope}
\end{figure*}

At a simplistic level the nucleation of AEs in a granular material are likely the result of grain fracturing, particle sliding/rolling, and the breaking of force chains. However, because our experiments were conducted at low normal stresses, we assume that grain fracturing is insignificant \cite<e.g.>{mair2002, anthony2005, scuderi2014}.  These micro-mechanical processes, sensibly represented by the sudden appearance of a force dipole in the gouge layer, launches an elastic wave that drives the gouge layer in a low lying shear mode. This motion of the gouge layer rings down slowly, decaying at a rate determined by how rapidly the acoustic contrast allows leakage into the steel environment. Note, the detector itself rings down as well, complicating the net behavior. This leaked signal goes on to the sensors and is an acoustic pulse contributing to the acoustic emission. The envelope (Figure~\ref{envelope}) of these AEs has a sharp onset, followed by exponential decay, and contains energy within the 100+ kHz range. However, the characteristics of the source may complicate this behavior, resulting in AEs with long rise times and long ring down times. Taken together, these factors result in the complex waveforms seen in many of the AEs.

\subsection{Cluster selection} \label{temporal}

Next, we examined the production of AEs by cluster throughout the failure cycle to look for temporal patterns. This is the first step in the process of identifying which clusters are likely to be useful for forecasting and which can be set aside.

\begin{figure*}[!htb]
	\includegraphics[width=\textwidth]{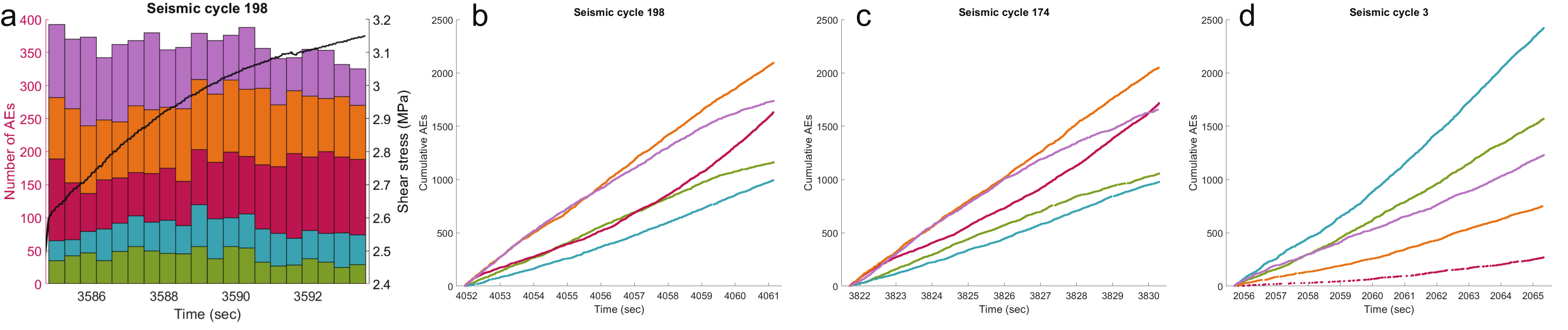}
	\caption{A) (Left axis) Histogram of the number of AEs by cluster over one seismic cycle. Colors correspond to those of Figure~\ref{som}. (Right axis) Shear stress. B-C) Cumulative AEs by cluster for two events of similar high stress. D) Cumulative AEs for a low stress event.}
	\label{cumulAE}
\end{figure*}

As we did with the Broom data, we plotted the cumulative number of AEs over the failure cycles. Three representative cycles are shown in Figure~\ref{cumulAE}. In general, the orange cluster contains the largest number of AEs and the blue and green contain the fewest. Most of the clusters display nonlinear behavior that may be useful for determining the percentage of the recurrence interval that has elapsed. However, not all clusters appear to have predictive capability; the orange cluster is nearly linear and thus will not be useful for failure forecasting. 

Our results suggest that AE production is tied to the normal stress of the system. We observe that stick-slip events with comparable normal stress exhibit very similar cumulative AE patterns, as in Figure~\ref{cumulAE}b and c. For both quakes, the rate of purple and green AE production decreases towards failure. The red AE production rate decreases shortly after stress onset and then increases again in the second half of the cycle. The same is true for the blue cluster to a lesser extent. Even the relative proportions of the clusters are the same between the two quakes; purple and red intersect or nearly intersect just before failure, purple and orange track closely for the first half of the cycle, etc. Though only two quakes are shown here, these similarities appear in all quakes of a similar normal stress. 

In addition, we find that quakes at different normal stresses produce different temporal patterns of AEs (Figure~\ref{cumulAE}a and b vs. Figure~\ref{cumulAE}c, with additional quakes in Figure S13 
in SI Section 3). 
The two quakes in Figures~\ref{cumulAE}a and b occur at relatively high stress where as the quake in Figure~\ref{cumulAE}c occurs at low stress. In the low stress quake, every cluster is nearly linear. We do not see the curved patterns that are evident in the high stress quakes. Additionally, the relative proportions of the clusters are completely different at low stress. Now the largest cluster is the blue cluster, followed by the green, purple, orange, and red clusters. Again, though we only show one quake here, this pattern is repeated among other low stress quakes.

The similarity of patterns at similar stress indicates that AE production is connected to the underlying physics of the system. This notion is supported by the laboratory AE analysis in \citeA{bolton2020}. The difference in patterns between different stress conditions effectively eliminates the possibility of time-to-failure forecasting based exclusively on temporal patterns of AE production. This is because the linear patterns at low stress are unlikely to contain useful information. In Earth the stress state of the fault is generally not known and that presents a challenge.

\begin{figure*}[!htb]
	\includegraphics[width=\textwidth]{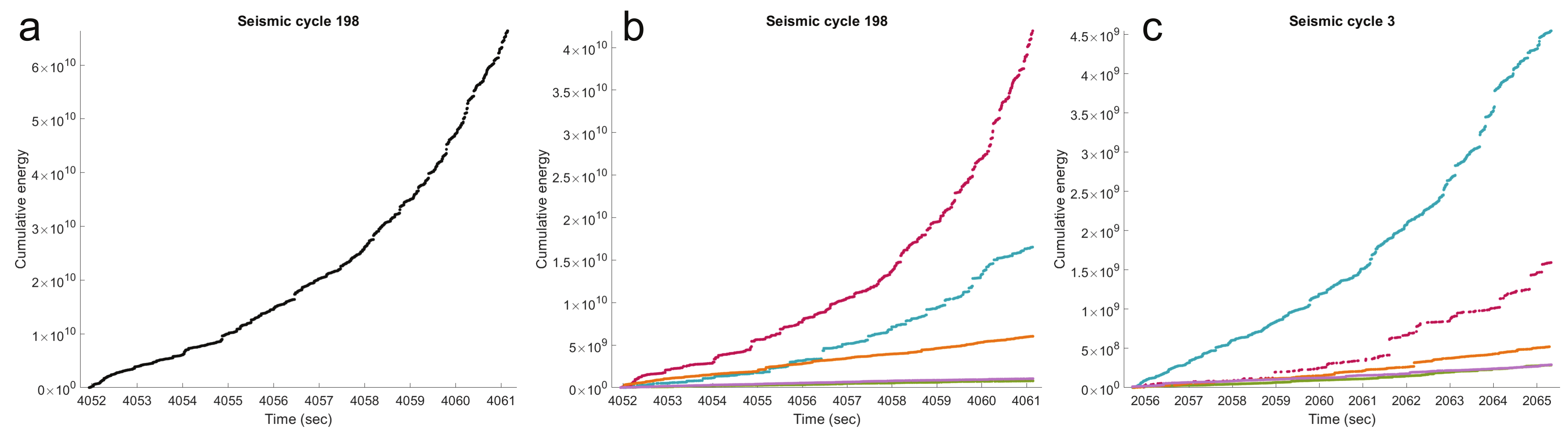}
	\caption{A) Cumulative energy for a single stick-slip event. B) Cumulative energy by cluster for a high stress event. Colors correspond to those of Figure~\ref{som}. C) Cumulative energy by cluster for a low stress event.}
	\label{energy}
\end{figure*}

Next we examined the cumulative AE energy over each failure cycle. Figure~\ref{energy}a is a plot of the cumulative energy from all AEs for a single stick-slip event. As the system progresses towards failure, the amount of energy contained in the AEs increases according to a power law, as was also shown in \citeA{bolton2020}. It is likely that acoustic energy derived exclusively from AEs can be used to estimate failure times as was described by \citeA{lubbers2018}. If true, then the use of the continuous AE signal is not a necessary ingredient to yield successful predictions with the caveat that the earthquake magnitude of completeness is sufficiently small. Furthermore, this highlights the importance of obtaining high-resolution event catalogs that are rich in small magnitude events \cite<e.g.>{ross2019,Mousavi2020}.  This pattern can also be seen in the cumulative energy from events generated by the Broom model (Figure S1 
in SI section 2). 
The power law cumulative energy was vital to the forecasting success using the Broom data, which suggests that the AEs contain the necessary information for TTF forecasting. If this is true, then we have no need for the rest of the acoustic signal and can simply discard it.

Continuing the task of cluster selection, we divide the cumulative energy by AE cluster, as in Figure~\ref{energy}b. At first glance, it appears that the red cluster is all that is needed for failure forecasting because it contains the vast majority of the total AE energy and retains the power law shape. However, as with cumulative AEs, the stress state of the fault plays a role in the energy contained within in each cluster. Figure~\ref{energy}c shows the cumulative energy by cluster for a low stress event. Compared to the high stress event in Figure~\ref{energy}b, the proportion of the total energy contained in the red cluster is reduced. As a result, the blue cluster contains a much larger proportion of the total energy. The power law shape is not as prominent at this stress level, but it is still present in the blue and red clusters. This indicates that the blue cluster is a vital source of predictive information at low normal stress. The three remaining clusters contain very little energy and have linear trends at both stress levels, which indicates that they will not be useful for forecasting. For additional examples, see Figure S14 
in SI Section 3.

We also examined the cumulative trends for the other features used in the clustering scheme (frequency, duration, counts, etc.). These features show mostly linear patterns with some nonlinearity introduced as a byproduct of the energy (e.g. in maximum amplitude). We also observe consistent patterns among quakes of the same stress level and different patterns between different stress levels. Like with the cumulative number of events in the Broom model, we continue to include these features because the raw values may provide the forecasting network clues as to the overall stress level.

Taken together, these findings suggest that only the red and blue clusters are seismically useful for forecasting time-to-failure at all relevant stress levels. Though the orange cluster is the largest at high stress, it consistently displays low, linear cumulative energy. As a result, this cluster is unlikely to contain much, if any, predictive information, so we will set it aside. The purple cluster is nonlinear at high stress and has slightly more energy at higher frequencies, which in turn, produces a strong boundary between it and the other clusters in the CSOM lattice. This cluster is also unlikely to be predictive and we will not use it. The green cluster is perhaps the least remarkable; it is relatively small and has low, linear energy. Green AEs are proportionally larger at low stress than at high stress. This cluster may consist of spurious signals that were mistakenly included during the cataloging process. As such, we will not include it in the forecasting scheme. The blue and red clusters, on the other hand, display the nonlinear energy we saw in the Broom data at all stress levels. The differing proportions of each at various stress levels mean that both clusters are required to capture the full stress spectrum. Therefore, the inputs to the forecasting network will include only the blue and red clusters and the other three will be discarded.

\section{Forecasting} \label{forecasting}

\subsection{Training and testing data}

For time-to-failure and shear stress forecasting, we only use AEs from the red and blue clusters. Network inputs consist of the seven cumulative features used to train the CSOM as well as the cumulative number of AEs. To improve network performance, we experimented with three additional input features: timestamps, shear stress, and shear stress slope (Table S10 
and Figure S22 
in SI Section 4). 
In the real earth, where shear stress measurements are not available, the latter two features could be obtained for TTF forecasting via an additional network. For the timestamps, we subtracted the respective cycle start time from each AE time so that each cycle began at time zero. Shear stress was included as recorded at the time of each AE. We calculated shear stress slope using a best fit line on the shear stress data. For each AE, we calculated the best fit line to the shear stress vs. time data on a window of 50 AEs centered on the AE in question. We found that the best results came from including the stress slope during training. Including timestamps slightly improved the MAE for long cycles at the expense of short cycles. Including shear stress had little impact on network performance.

All input features were normalized to [0,1] according to their maximum and minimum values across the entire dataset. In order to maintain the temporal ordering of AEs, each input vector contains the eight features for each of the two clusters plus stress slope, bringing the input size to 17. In addition, each input vector contains the features for the current AE as well as the features from the previous AE from the other cluster. Over time, the inputs resemble a step function where the features for each cluster are held constant until an AE from that cluster occurs.

\begin{figure*}[!htb]
    \centering
    \includegraphics[width=4in]{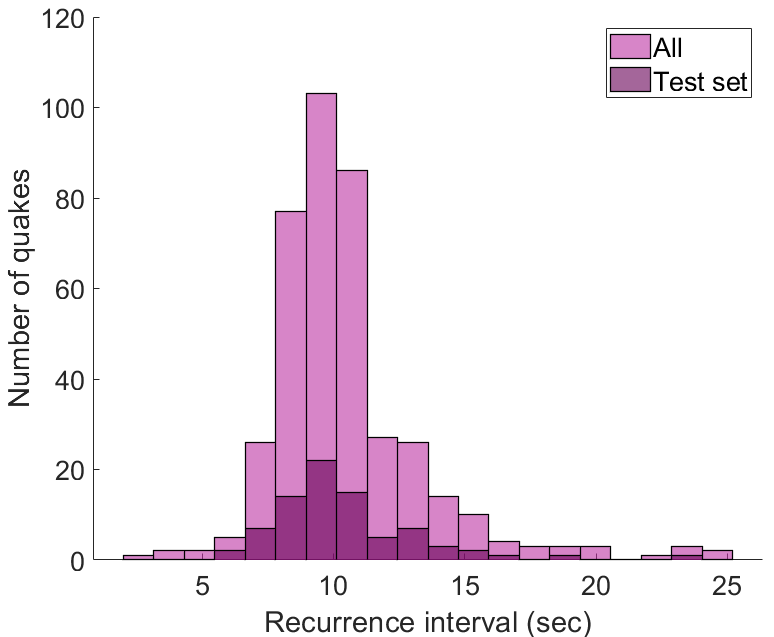}
    \caption{Recurrence intervals for train and test set}
    \label{lengths}
\end{figure*}

We divided the labquakes into train and test sets, with 80\% of the quakes for training and the remaining 20\% for testing. Unlike other studies, we divided the data based on entire cycles rather than on time windows. We did not use a validation set because of the limited data availability.

The test set was randomly chosen to have roughly the same recurrence interval distribution as the full dataset (Figure~\ref{lengths}). As with the Broom model data, all recurrence intervals and stress levels were included in a single dataset and trained upon simultaneously. The vast majority of recurrence intervals are around 10 seconds, which left very long/short recurrence intervals underrepresented in the training set. To address this, we experimented with techniques meant to alleviate issues with unbalanced datasets. We began by upsampling long and short cycles so that these quakes were seen by the network more than once per training epoch. This had little impact on the total MAE and substantially increased the network training time (Table S10 
in SI Section 5). 
Next, we tried pre-training by decreasing the number of average cycles by only including a small random subset of them in each training epoch. These network weights were then used to initialize another network that trained on the full dataset. Both the pre-training networks and the fine-tuned networks generally performed worse than the default training method (Table S10 
in SI Section 5). 
Finally, we experimented with loss weighting. We weighted average cycles so that they contributed less to the total loss than short and long cycles. Again, this did not improve the total MAE (Table S10 
in SI Section 5). 

\subsection{Network variations}

We tested a variety of network architectures, including variations of LSTM, attention networks, and meta learning. To improve the results, we experimented with network modifications such as different optimizers, gradient clipping \cite{pascanu2013}, weight decay \cite{loshchilov2019}, etc. In addition, to avoid overfitting, we employed dropout \cite{Srivastava2014}, early stopping \cite{Prechelt1998}, and ensembling \cite{Dietterich2000}. Table S9 
(in SI Section 5) 
shows the variety of hyperparameters and network modifications that we tested.

We began by forecasting with a simple multilayer LSTM network (Figures S16 and S17 
in SI Section 4). 
We used this network to test the input vector modifications (i.e. the stress slope) as well as the methods of handling the unbalanced recurrence intervals in the dataset. We also experimented with time-gated LSTM (TGLSTM), which is an LSTM variation designed to better handle irregularly sampled data by modifying the network according to the time interval between data points \cite{sahin2019} (see Figure S18 
in SI Section 4). 

We also forecasted TTF and shear stress with the attention network described previously. In addition, we tested a variation of this network that attempts increase efficiency by adding momentum \cite{nguyen2020}.

Finally, we tested various meta learning techniques that focus on learning from multiple datasets. In our experiments we used a domain adaptive meta learning (DAML) network \cite{DAML}. Before training, data must be divided into 'domains' representing some joint characteristic. We assigned cycle domains based on the normal stress (see Figure S15 
in SI Section 3). 
The step-up-step-down laboratory data consists of seven load levels for a total of seven domains. For comparison, we also experimented with using three domains based on recurrence interval (short, average, long). The training process begins by initializing a 'meta network' that can be any network type of the user's choosing (e.g. LSTM, attention, etc.). Weights from the meta network are used to initialize a new network for each domain (Figure S21 
in SI Section 4). 
At this stage, each domain network is an exact copy of the meta network. The domain networks are trained separately for some small predetermined number of epochs. After this training period, weight changes are averaged across the domains and used to update the meta network \cite{iMAML}. The domain networks are discarded and the process begins again until some number of meta-epochs is reached. We tested this network using our multilayer LSTM and attention networks as the base.

All networks were trained using the smooth L1 loss function. Results from these networks frequently contained illogical undulations in TTF that we wished to avoid. To address this, we added a penalty term to the loss function that encourages the network to produce TTF values with a slope of -1. The -1 slope comes from the simple fact that TTF should decrease by 1 second for every second that has elapsed. We calculated slopes using least squares on the denormalized TTF values for each sequence. Next, we found the difference between the actual slope and desired slope of -1 and renormalized this loss. We weighted the slope loss term to be less important than the L1 loss for the network update. In some cases, the addition of this term produced smoother results and was most beneficial early in the training.

\subsection{Shear stress results}

We trained the networks to simultaneously forecast shear stress and time-to-failure, but for simplicity we will discuss them one at a time. We report our results using the mean absolute error (MAE) calculated over each failure cycle. MAE is presented separately for shear stress and TTF, though they were combined during network training. It is important to note that we do not expect the trained network to have perfect accuracy. Due to the nature of the challenge, forecasting results rarely achieve very high accuracy. Instead, our goal is to present a technique that makes reasonably accurate forecasts and can be improved upon in the future.

\begin{figure*}[!htb]
	\includegraphics[width=\textwidth]{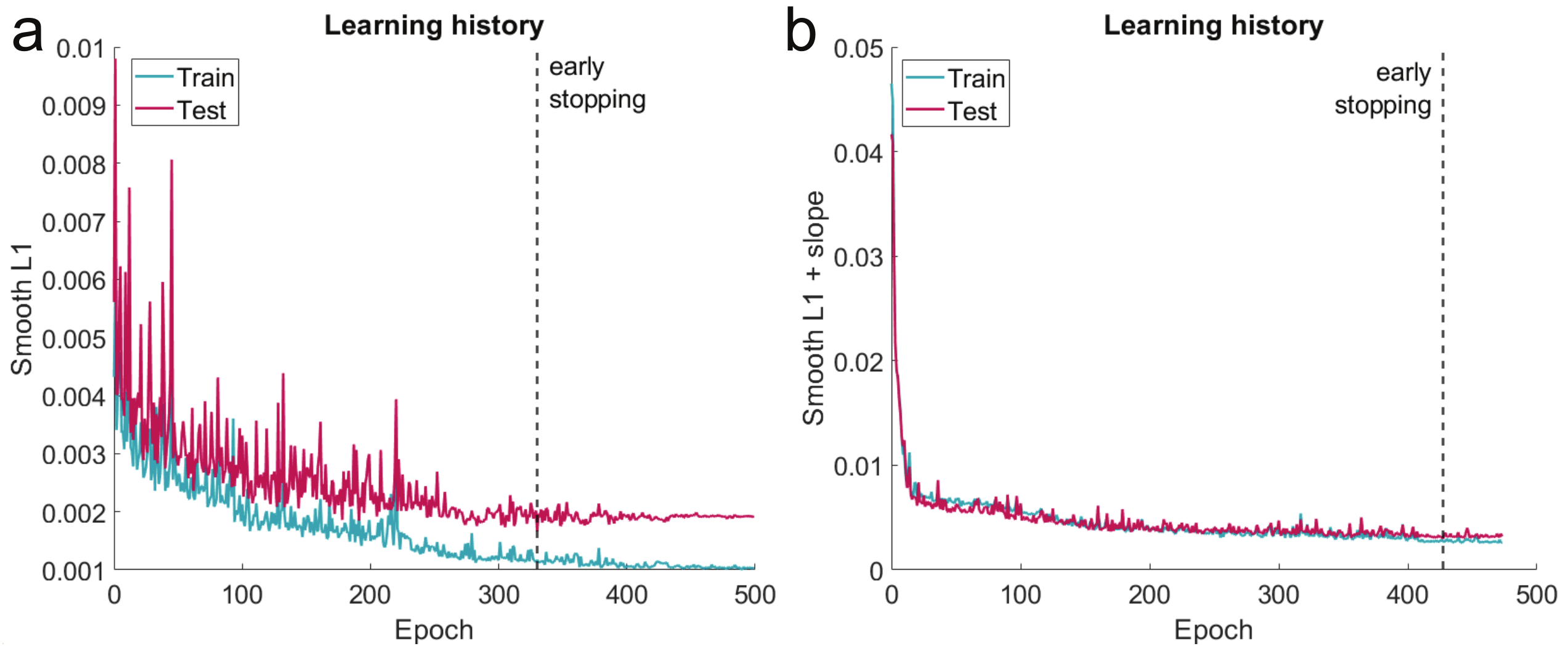}
	\caption{Shear stress training histories for the (A) LSTM and (B) attention networks showing loss over time for the train and test sets. The final network weights were taken from the point marked early stopping.}
	\label{shearHistories}
\end{figure*}

We began by training the LSTM network. The learning history for our best run is shown in Figure~\ref{shearHistories}a and the hyperparameters are listed in Table S11 
in SI Section 5. 
Though we ran the network for an additional ~150 epochs, we used the weights from the point marked ``early stopping'' at which the test loss stagnates. The large spikes in error, particularly near the beginning of training, are typical of recurrent networks and are exacerbated by the unbalanced dataset. For this network, we found no benefit from incorporating TTF slope into the loss function.

Next, we trained the attention network using the same training and testing sets as with the LSTM. The learning history for our best run is shown in Figure~\ref{shearHistories}b and the hyperparameters are listed in Table S11 
in SI Section 5. 
Once again, we use the network weights from the ``early stopping'' mark to make the final forecasts. This network did benefit from incorporating TTF slope into the loss function. Compared to the LSTM network, the attention network has fewer error spikes, but takes longer to converge.

\begin{figure*}[!htb]
	\includegraphics[width=\textwidth]{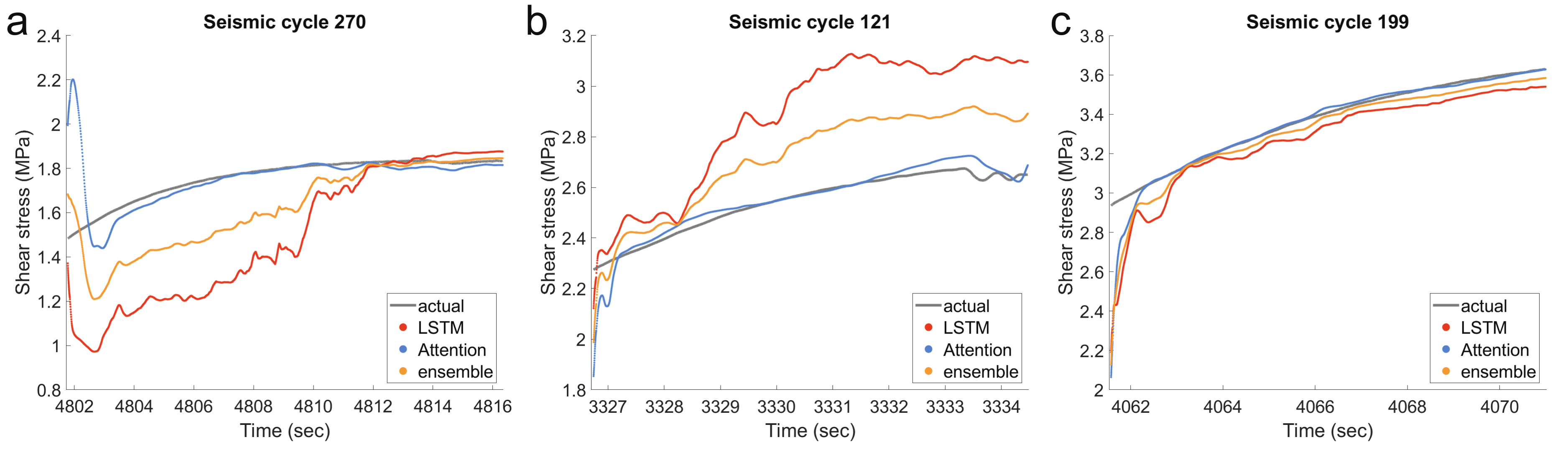}
	\caption{Actual vs. forecasted shear stress for an event with (A) low stress, (B) medium stress, and (C) high stress for each of the tested networks.}
	\label{stressResults}
\end{figure*}

Example cycles with the forecasted and measured shear stress are shown in Figure~\ref{stressResults}, with additional results in Figure S25 
in SI Section 5. 
As with the Broom data and the cumulative feature plots, these figures each portray a single, representative quake from the onset of stress to failure. The MAE for each run is listed in Table~\ref{stressTable}.

\npdecimalsign{.}
\nprounddigits{4}
\begin{table*}[!htb]
	\caption{Shear stress forecasting MAE by network for the test set$^{a}$}
	\centering
	\begin{tabular}{|c|n{1}{4}||n{1}{4}|n{1}{4}|n{1}{4}|n{1}{4}|n{1}{4}|n{1}{4}|n{1}{4}|}
		\hline
		&& \multicolumn{7}{c|}{\textbf{Stress MAE by domain}} \\ \hline
		\textbf{Name} & \textbf{Total} & \textbf{1} & \textbf{2} & \textbf{3} & \textbf{4} & \textbf{5} & \textbf{6} & \textbf{7} \\ \hline \hline
		LSTM & 0.1087013923 & 0.0532063863 & 0.0816841026 & 0.1617180854 & 0.1325188631 & 0.1353777302 & 0.1034869828 & 0.1033623795 \\ \hline  
		Attention & \textbf{0.0658} & 0.0394513217 & \textbf{0.0579} & 0.0944813662 & \textbf{0.0709} & \textbf{0.0632} & \textbf{0.0564} & 0.0787625789 \\ \hline   
		\rowcolor[HTML]{C0C0C0}
		Ensemble & 0.0699196080 & \textbf{0.0378} & 0.0617647564 & \textbf{0.0801} & 0.0793943859 & 0.0875601532 & 0.0644380491 & \textbf{0.0780} \\ \hline    
	\multicolumn{9}{c|}{\multirow{3}{*}{\parbox{\textwidth}{$^{a}$MAE is broken down by the 7 stress domains in the step-up-step-down laboratory experiment. Normal and shear stress increase from domain 1 to domain 7. The best result in each domain (as well as total MAE) is in bold.}}} \\
	\multicolumn{9}{c|}{}
	\end{tabular}
	\label{stressTable}
\end{table*}
\npnoround

The LSTM network produces a mix of acceptable and poor forecasts. For a number of cycles, the error is quite low, but for others the LSTM forecast is far from the true shear stress. Often, the forecasted stress is quite jagged, even after the initial network adjustment period. In some cases, the network suddenly becomes much more accurate in the last few seconds of the recurrence interval, whereas in others it drifts away from the true values. The network often doesn't even capture the general shape of the stress and instead meanders until failure is reached.

We can further examine the network forecasts by dividing the quake cycles by stress level. The nature of the data is such that low stress has the widest variety of recurrence intervals whereas high stress consists of mostly average recurrence intervals (see Figure S15b 
in SI Section 3). 
We divide the stress according to the steps in the step-up-step-down lab experiment. This produces seven domains where domain 1 has shear stress \textless 1 MPa and domain 7 has shear stress around 3-3.5 MPa. 

The LSTM network has the highest MAE in the middle domains (3-5) and a lower MAE at very low and very high stress. Visually, the high stress forecasts are the most accurate and least erratic. The network generally gets the shear stress shape correct and the forecasted values are more or less around the true stress. At medium stress, when the network is wrong, it is very wrong (see the y-scale in Figure~\ref{stressResults}). A similar phenomenon happens at low stress, but this is balanced against cycles that have very accurate forecasts. Clearly the network is not able to fully overcome the differences in recurrence interval, leading to large differences in quality in the domains with wide recurrence interval ranges.

It is important to acknowledge that we only obtain this interesting result because we trained the LSTM network on all stress conditions simultaneously. Had we only trained on a single stress level, high stress quakes, for example, the network would appear to perform very well and thus inspire false confidence. In the lab, stress conditions are highly controlled and straightforward, but in the real earth they are likely unknown and complex. For our network to perform well on the real earth, it must be able to generalize across multiple background stress conditions. 

We also note that the network performs particularly poorly on cycles from the second lab experiment, p4583. This behavior is not unexpected because this experiment was included to diversify the stress conditions and recurrence intervals. This experiment has fewer quakes overall, which makes learning them more challenging. However, it is quite apparent that LSTM struggles to adapt to new conditions. Perhaps given more data the network would perform better, but we have to be cognizant of the fact that in the real Earth data is limited and stress conditions are more complex.

After this disappointing performance, we turn to the forecasting results from the attention network. The attention network performs significantly better than the LSTM network, with a total MAE that is nearly 50\% smaller. In general, the attention forecasts are very smooth, without the jagged edges that defined the LSTM results. The attention forecasts for individual cycles are very accurate after the initial adjustment period, which tends to be fairly short. Even the less-accurate forecasts generally capture the shape of the shear stress curve. Unlike the LSTM network, once the attention network has settled on a pattern for a specific quake, it consistently follows that curve and doesn't drift away at the end of the cycle.

Dividing the cycles by stress, we see that the attention network performs well at all stress levels. Like the LSTM, it performs the worst on domain 3, but even there the attention network is a clear improvement. Visually, when the LSTM network performs poorly, the attention network is more likely to be slightly off in its forecasts. However, this is not always the case and the attention inaccuracies are much smaller than for the LSTM. This indicates that the attention network has a greater capacity to learn from a variety of stress environments than the LSTM network.

On the topic of the two lab experiments, the attention network performs much better on p4583 than the LSTM network does. In general, the p4583 quakes have higher attention error than the p4581 quakes, but the network performs better on p4583 than LSTM does on p4581.

There's an argument to be made that even when the attention network is wrong, its forecasts are more useful than those of the LSTM. A watch that is always off by 5 minutes is more useable than a watch that is erratic. In the same way that the first watch only needs a small modification to be accurate, the attention network likely also only needs small tweaks to correct inaccurate forecasts. This could come from the addition of another input feature or a minor architecture modification. This is comparatively simple to the work required to improve the LSTM network, which calls for significantly more data and even then may not produce accurate results.

Finally, we prepared an ensemble by averaging the results of the two networks. Even though the LSTM performs much worse than the attention network, this is a good practice because different networks often learn different aspects of the data. The combination gives a wider view of the data and can lead to better forecasts. Indeed, the ensemble performs better than its constituents in stress domains 1, 3, and 7. However, the attention network still reigns supreme overall.

Overall, we find that the attention network is easily able to forecast shear stress. The LSTM network performs well on some quakes, but the different stress conditions are simply too complex for it to learn with this small amount of data. The attention network, however, performs well under all conditions and thus is a good candidate for use in the real earth.

\subsection{Time-to-failure results}

As before, we report our results using the MAE calculated over each failure cycle. In our calculations, we omit the first 0.5s of each cycle since the network has not yet received enough data to make reasonable forecasts. Including this adjustment period in the error disproportionately impacts the short/long cycles and portrays their forecasts as worse than they are. With TTF forecasting, we determine that a cycle has been successfully forecasted if the MAE is less than 10\% of the recurrence interval. For comparison, the results for a control network that always outputs the average recurrence interval of the dataset are shown in Table~\ref{maeTable}.

\begin{figure*}[!htb]
	\includegraphics[width=\textwidth]{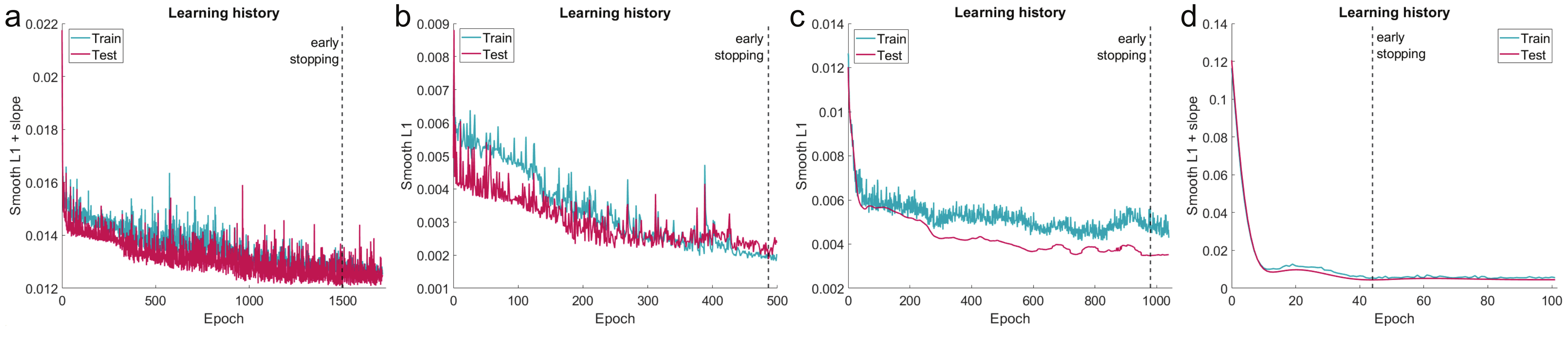}
	\caption{TTF training histories for the (A) LSTM (B) attention (C) DAML with LSTM and (D) DAML with attention networks showing loss over time for the train and test sets. The final network weights were taken from the point marked early stopping.}
	\label{ttfHistories}
\end{figure*}

\npdecimalsign{.}
\nprounddigits{4}
\begin{table*}[!htb]
	\caption{TTF forecasting MAE by network for the test set$^{a}$}
	\centering
	\begin{tabular}{|c|n{1}{4}|n{1}{4}|n{1}{4}|n{1}{4}|}
		\hline
		& \multicolumn{4}{c|}{\textbf{TTF MAE}} \\ \hline
		\textbf{Network} & \textbf{Total} & \textbf{\textless{}8.5s} & \textbf{8.5-11s} & \textbf{\textgreater{}11s} \\ \hline \hline
		Control & 2.0726462055 & 3.2164878020 & 1.6655161457 & 4.1710571120 \\ \hline
		LSTM & 1.1582365829 & 1.6203479529 & 0.8330546318 & 1.4429574932 \\ \hline
		LSTM ensemble & 1.1000370299 & 1.5295859671 & 0.8038503745 & 1.3530850541 \\ \hline
		Attention & 1.0504 & 1.5125 & \textbf{0.6861} & 1.4097761035 \\ \hline
		DAML (LSTM) & 1.3969 & 1.6649632802 & 0.9055746346 & 2.1400511921 \\ \hline
		DAML (attention) & 1.5065861117 & 1.9528957692 & 0.9223045264 & 2.2974439327 \\ \hline
		\rowcolor[HTML]{C0C0C0}
		Ensemble & \textbf{1.0400} & \textbf{1.4181} & 0.7640813848 & \textbf{1.2919} \\ \hline
	\multicolumn{5}{c|}{\multirow{3}{*}{\parbox{0.7\textwidth}{$^{a}$Test events were divided into recurrence interval bins for further comparison. Bold entries indicate the best MAE for each recurrence interval bin.}}} \\
	\multicolumn{5}{c|}{}
	\end{tabular}
	\label{maeTable}
\end{table*}
\npnoround

The learning histories are shown in Figure~\ref{ttfHistories} and the hyperparameters are listed in Table S11 
in SI Section 5. 
We again employ early stopping to select the appropriate network weights. We note that in the DAML runs the test data nearly always outperforms the train data. This was the case in every network configuration that we tested. The loss calculations for the test and train sets are identical, so this appears to be caused by the network itself.

\begin{figure*}[!htb]
	\includegraphics[width=\textwidth]{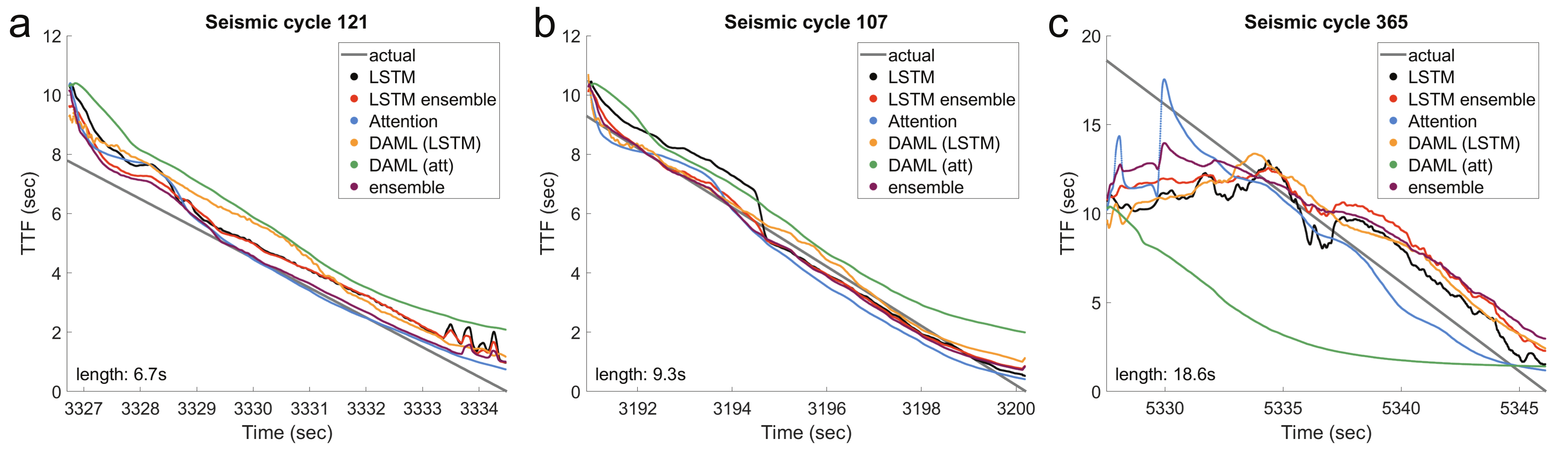}
	\caption{Actual vs. forecasted TTF for an event with a (A) short cycle, (B) average cycle, and (C) long cycle for each of the tested networks. Recurrence interval is noted in the lower left corner.}
	\label{ttfResults}
\end{figure*}

MAE for all runs is listed in Table~\ref{maeTable} and success percentages are listed in Table S12 
in SI Section 5. 
Forecasted vs. actual TTF for all networks on three example cycles are shown in Figure~\ref{ttfResults}, with additional examples in Figure S26 
in SI Section 5. 
The recurrence interval for each quake is noted in the bottom left corner.

Compared to the shear stress forecasts, the LSTM network performs better than expected. In general, the forecasts are relatively close to the true values and the overall shape is roughly a line with a slope of -1. We continue to see the jaggedness we saw in the shear stress forecasts, though to a lesser extent. The forecasts often drift away from the actual TTF (and the -1 slope line) in the last second or so of the cycle.

As discussed previously, the dataset consists of a wide range of recurrence intervals. We broke these down into short, average, and long cycles for further analysis. The LSTM network performs very well for average quakes, which have a recurrence interval of around 8.5-11 seconds. However, quakes with very short or very long cycles have much higher error. The network generally produces an output with the correct shape for these cycles, but fails to adjust the positioning on the y-axis. This may occur because the network prefers to decrease loss for the average case rather than risk larger loss by producing a generalized model. However, when compared to the control results in~\ref{maeTable}, it is clear that the network is an improvement over simple averaging.

Next, we combined the best performing LSTM runs into an ensemble. We did this because we observed that the LSTM struggled to learn both the short and the long quakes at once. In general, any given run would perform well on either the short or long quakes, but not both. For the LSTM ensemble, we specifically selected the best runs for each recurrence interval bin. The resulting ensemble performs better than the best single LSTM run at every recurrence interval.

The attention network clearly performs better than the LSTM network: MAE is lower and visually the forecasts are closer to the true values. Some of this improvement is likely due to the smoothing effects of the attention layer. Like with the shear stress forecasts, the attention network results are very smooth without any sudden directional changes. Even when the attention network misses the mark, it is generally closer to the true TTF than the LSTM network is.

The TTF forecasted by the attention network for the short and average cycles is more accurate than from the LSTM network and its ensemble. Surprisingly, the networks perform nearly the same on the long quakes, perhaps signalling that there are simply not enough cycles for complete learning. Though the attention network still struggles with some of the very long/short cycles, it produces successful forecasts for the majority of the test set.

Unfortunately, none of our DAML experiments were particularly successful (see Figures S23 and S24 
in SI Section 5). 
The total MAE for both network configurations is worse than the LSTM and attention networks. This is largely because DAML performs astoundingly poorly on the long cycles. On the short and average cycles, both DAML variations do about the same as LSTM. DAML has been used successfully on other datasets, but we appear to lack the data volume and computing power required to train it well. We believe this network's performance could improve given additional training data and more powerful computing resources. 

Finally, we calculated a mixed ensemble consisting of the best runs on short and long cycles across all the networks tested. Indeed, the ensemble improves the forecasting accuracy on the short and long cycles. The ensemble does unfortunately have some jagged edges due to its LSTM components, but on the whole it performs nicely.

These results pose a critical question: why do some quakes always have poor forecasts? For some quakes, this may be due to foreshocks before failure that muddy the cumulative features. Some of these foreshocks were large enough to be considered co-seismic stress drops and separated out, but this was not always the case. The attention network clearly combats this muddying by examining the feature trend over a window of previous time steps as well as the data at the current time step. For simple networks such as LSTM, a trend-based view may be beneficial.

For the long quake cycles in particular, poor performance is likely due to lack of data. The spread of long cycles is much greater than the spread of short cycles. As we saw with the Broom tests where only some normal stress values were seen during training (SI Section 2), 
the further the test data is from the training data, the less accurate the forecasts become. The extremely long quakes are seen very rarely, so we should expect higher error.

\section{Prior work} \label{comparison}

Recently, \citeA{rouetLeduc2017} used the random forest algorithm to forecast TTF for lab data using the continues AE signal for feature extraction (in contrast to event catalogs applied here). The input data consisted of 100 statistical features calculated from moving windows of raw data from 10 sequential quake cycles. The subsequent 10 cycles were used as the test set. The quake cycles had variable recurrence interval, but similar shear stress. The authors reported an R\textsuperscript{2} of 0.883 for the test data, a metric which we do not calculate here. The authors performed an additional test where the random forest was trained on one load level (5 MPa) and tested on another (8 MPa), which resulted in a test set R\textsuperscript{2} of 0.741. 

Los Alamos National Lab and the Department of Energy Office of Science (Geoscience program) sponsored a Kaggle competition in 2018-2019 with the goal of improving lab forecasting results \cite{johnson2021}. The winning solution used a combination of gradient boosted trees and a convolutional neural network to forecast TTF from raw seismic data. The team found that the two networks learned different components of the signal so averaging the two results gave a more complete view of the data. This setup achieved a MAE of 2.26s. Some teams published papers detailing their methods, including \citeA{holtz-unpub}, who used WaveNet for feature extraction and LSTM for forecasting with an MAE of 1.8, \citeA{zaidi2020}, who used a CNN and LSTM for an MAE of 1.51, and \citeA{brykov2020}, who used XGBoost for an MAE of 1.91. It is unclear if the reported MAE from these follow-up papers comes from the small ``public'' test set available to all during the competition, or the larger ``private'' test set, which was used to determine the competition winners.

Decision tree forecasting methods have been used in several subsequent papers, again employing the continuous AE signal. \citeA{hulbert2019} forecasted TTF, among other features, for fast and slow slip in the lab with an R\textsuperscript{2} of 0.88. \citeA{vanklaveren2020} used random forest to forecast TTF from rotary shear experiments on salt samples. After correcting for machine resonance, the team achieved an R\textsuperscript{2} of 0.85. \citeA{corbi2019} used gradient boosted regression trees on geodetic data from a laboratory subduction experiment. They report an overall R value of 0.3 for their TTF forecasting model, though most individual cycles are in the 0.7-0.8 range. They also report RRMSE values for each cycle, which range from mid-80s to low-20s. The average RRMSE for each of our LSTM and attention networks are in the range 28-33. \citeA{tanyuk2019} also forecasted TTF from biaxial lab data using statistical features. The authors experimented with several decision tree algorithms, and achieved the best MAE of 1.65s from random forest. Our MAE values are reported in Table~\ref{maeTable}, where the lowest MAE is 1.04s.

Other studies have used machine learning to forecast earthquakes without directly outputting time-to-failure. \citeA{corbi2020} used the same laboratory subduction setup as \citeA{corbi2019} with RUSBoost to forecast the alarm state of the fault (i.e. a binary classification of if the slip rate of the fault exceeds some threshold). The AETA Competition \cite{aeta} challenged participants to make binary forecasts each week given a combination of acoustic and electromagnetic data from the Sichuan and Yunnan provinces from the previous week. The top team for the 2020 competition achieved 75\% correctness.

\section{Discussion}

Overall, the forecasting scheme was successful. The attention network in particular was able to forecast TTF and shear stress within a reasonable amount of error for the majority of the test set. These results indicate that AEs from the red and blue clusters contain enough information to forecast time-to-failure without using the rest of the seismic trace. 

In general, the LSTM network performs better on the average and long cycles. The attention network (and the ensemble) performs roughly equally well on the short and long cycles, though the average cycles had substantially lower MAE. This difference is largely due to the fact that average cycles are overrepresented in the lab data. With the short and long cycles, the network simply does not have enough examples to adequately learn. With the Broom model, we found that the network was able to learn all recurrence intervals given a equally distributed training set. Given more short and long lab cycles, it is likely that forecasting performance would improve.

In their work with forecasting via random forest, \citeA{rouetLeduc2017} found that variance of the signal was the most important input feature they tested. Determining the most useful feature from an LSTM model is much more difficult. From the temporal AE results in section~\ref{temporal}, it appears that energy, which is related to variance, likely has great importance. We tested this by forecasting using cumulative energy as the only input feature. The forecasts of this network are clearly inferior to our previous results (Table S10 
in SI Section 5). 
This suggests that accurate forecasting is more complicated than simply monitoring AE energy. As we move to Earth and attempt to apply ML approaches to discern fault physics from seismic signals, in depth exercises analyzing laboratory and simulation data are key to informing how to proceed. This study represents a key element in this exercise.

\section{Conclusions}
In this study, we forecasted shear stress and TTF for laboratory earthquakes using machine learning. Through clustering, we found that a subset of acoustic emissions contain sufficient information for these tasks. Thus, network training does not require continuous seismic data. We obtained the most accurate forecasts using an attention network, though further improvements could be made. Our procedure produces more accurate forecasts than existing work and does so over a wider variety of normal stress conditions.

\section{Acknowledgments}

HJ and MdH were supported by U.S. Department of Energy, Office of Science, Office of Basic Energy Sciences, Chemical Sciences, Geosciences, and Biosciences Division under grant number DE-SC0020345 and PAJ and RAG by grant 89233218CNA000001, respectively. Special thanks to Erzsebet Merényi for the use of the NeuroScope software tools for CSOM training and to Michael Puthawala for his contributions to SI Section 1.

The laboratory data used in this study are publicly available at \url{https://scholarsphere.psu.edu/resources/a8e93370-2151-40e7-932e-4116d2f643bd}.

\nocite{benioff1951}
\nocite{varnes1989}
\nocite{bufe1993}
\nocite{bufe1994}
\nocite{hardebeck2008}
\nocite{gutenberg1944}
\nocite{tiampo2012}
\nocite{sangiorgio2020}
\nocite{gigante2019}
\nocite{raissi2019}
\nocite{wu2019}
\nocite{santeramo2018}
\nocite{zhu2017}
\nocite{trugman2013}
\nocite{brodsky2019}
\nocite{Reddi2018}
\nocite{keskar2017}

\clearpage
\bibliography{AEreferences}

\clearpage

\renewcommand{\thefigure}{S\arabic{figure}}
\renewcommand{\thetable}{S\arabic{table}}
\setcounter{figure}{0}  
\setcounter{table}{0}

\title{Supporting information for ``attention network forecasts time-to-failure in laboratory shear experiments''}

\authors{Hope Jasperson\affil{1}, David C. Bolton\affil{2}, Paul Johnson\affil{3}, Robert Guyer\affil{3,4}, Chris Marone\affil{2,5}, Maarten V. de Hoop\affil{6}}

\affiliation{1}{Rice University, EEPS}
\affiliation{2}{Pennsylvania State University, Department of Geosciences}
\affiliation{3}{Los Alamos National Laboratory, Geophysics Group}
\affiliation{4}{University of Nevada, Reno, Department of Physics}
\affiliation{5}{Sapienza Universitá di Roma, Dipartimento Scienze della Terra}
\affiliation{6}{Rice University, CAAM}

\date{}


\noindent\textbf{Contents of this file}
\begin{enumerate}
\item (Page~\pageref{appxRSF}-\pageref{appxRSF_end}) Rate-and-state friction and recurrent networks
\item (Page~\pageref{appxBroom}-\pageref{appxBroom_end}) Broom model forecasting benchmark
\item (Page~\pageref{appxLab}-\pageref{appxLab_end}) Laboratory data
\item (Page~\pageref{appxNetwork}-\pageref{appxNetwork_end}) Detailed network architecture
\item (Page~\pageref{appxStat}-\pageref{appxStat_end}) Statistical methods
\end{enumerate}

\newpage

\section{Rate-and-state friction and recurrent networks} \label{appxRSF}

Let us start with the ODEs for rate-and-state friction

\begin{align}
    \label{eqn:ode-form:1}
    \dot X &= V\\
    \dot V &= \frac{k}{M}(Z - X) + f(X,V,\theta)\\
    \dot \theta &= g(X,V,\theta)\\
    \label{eqn:ode-form:4}
    \dot Z &= V_0
\end{align}

Where $X$ is displacement, $V$ is the slip velocity, $k$ is the stiffness, $M$ is the slider mass, $\theta$ is the state variable, and $Z$ is the driver of the system. Function $f(X,V,\theta)$ determines the friction coefficient and $g(X,V,\theta)$ is the state evolution law.

Given that our problem is fully linear in $Z$, and non-linear in the other three quantities, we adopt a mixed implicit approach, implicit in $Z$, and explicit in $X, V$ and $\theta$. This is a popular approach for problems that contain both linear terms and nonlinear terms.

Let $S =\begin{bmatrix} X\\V\\\theta \end{bmatrix}$ and $S(t) = \begin{bmatrix} X\\V\\\theta \end{bmatrix}(t)$. Clearly, $Z$ can be computed without the need of a numerical solver. We then numerically approximate $S$ and apply an explicit Euler update in $X,V$ and $\theta$ as well as an implicit update of $Z$, 
with step size $\delta t$ so that $S_{t}$ evolves according to 

\begin{align}
    S_{t+\delta t} &= S_t + \delta t \dot S = \begin{bmatrix}X_t\\V_t\\\theta_t\end{bmatrix} + \delta t\begin{bmatrix}V_t\\\frac{k}{M}(Z_{t+\delta t} - X_t) + f(X_t,V_t,\theta_t)\\g(X_t,V_t,\theta_t)\end{bmatrix}\\
    &= \begin{bmatrix}X_t\\V_t\\\theta_t\end{bmatrix} + \delta t\begin{pmatrix}\begin{bmatrix}V_t\\\frac{k}{M}(Z_{t+\delta t} - X_t)\\0\end{bmatrix} + \begin{bmatrix}0\\f(X_t,V_t,\theta_t)\\g(X_t,V_t,\theta_t)\end{bmatrix}\end{pmatrix}\\
    &= \begin{bmatrix}X_t\\V_t\\\theta_t\end{bmatrix} + \delta t\begin{pmatrix}\begin{bmatrix}V_t\\-\frac{k}{M} X_t\\0\end{bmatrix} + \begin{bmatrix}0\\f(X_t,V_t,\theta_t)\\g(X_t,V_t,\theta_t)\end{bmatrix} + \begin{bmatrix}0\\\frac{k}{M}\\0\end{bmatrix}Z_{t+\delta t}\end{pmatrix}
    \label{eqn:final:ode}
\end{align}

If we consider recurrent networks of the form
\begin{align}
    \label{eqn:general-resnet}
    \begin{bmatrix}h_{t+1}\\\theta_{t+1}\end{bmatrix} = \begin{bmatrix}h_t\\\theta_t\end{bmatrix} + \mathbf{W}\begin{bmatrix}h_t\\\theta_t\end{bmatrix} + F(h_t,\theta_t) + \mathbf{U} x_{t + 1}
\end{align}
where $\mathbf{W}$ \& $\mathbf{U}$ are a fixed matrices and $x_t$, which comes from $Z$ evaluated at the time $t$, is an input that drives the system, then we can write Eqn. \ref{eqn:final:ode} in the form of Eqn. \ref{eqn:general-resnet} where 
\begin{align}
    \label{eqn:term-def}
    h_t = \begin{bmatrix}X_t\\V_t\end{bmatrix},\quad \mathbf{W} = \delta t\begin{bmatrix}0&1&0\\-\frac{k}{M}&0&0\\0&0&0\end{bmatrix},\quad F = \begin{bmatrix}0\\f(X_t,V_t,\theta_t)\\g(X_t,V_t,\theta_t)\end{bmatrix},\quad \mathbf{U} = \begin{bmatrix}0\\\frac{k}{M}\\0\end{bmatrix}.
\end{align}

In equation Eqn. \ref{eqn:term-def}, we use shallow neural networks to learn the functions $f$ and $g$ that best fit the experimental data. Everything else is fixed.
So we can see that by doing this mixed explicit (in $X,V$ \& $\theta$) and implicit (in $D$) Euler scheme as applied to the ODE specified in Eqn.s \ref{eqn:ode-form:1} - \ref{eqn:ode-form:4} can written as a version of the recurrent architecture given in \ref{eqn:general-resnet}. 

\label{appxRSF_end}
\clearpage

\section{Forecasting benchmark} \label{appxBroom}

We begin by testing our proposed forecasting method on a benchmark dataset. Our goals here are twofold: to test the capability of recurrent networks to learn patterns from metastable dynamical systems and to assess the robustness of the forecasting method. We perform these tests using a LSTM network. We do this because the LSTM is simpler and faster to train than the attention network. However, the LSTM network is also less powerful than the attention network \cite{bahdanau2015}. If the LSTM successfully learns the benchmark data, then we assume that the attention network would work as well.

A fair amount of prior work has been done on the topic of using machine learning to forecast outputs of dynamical systems \cite<e.g.>{sangiorgio2020, gigante2019, raissi2019, wu2019}. In general, these studies focus on forecasting the results of stable (stationary) systems. However, earthquake forecasting presents a different challenge because the dynamical system governing fault failure is metastable. In other words, small perturbations to the fault can produce catastrophic failure. In order to use LSTM networks to forecast earthquakes, we must first establish that they are capable of learning from data produced by metastable systems. In addition, prior studies use datasets that are regularly sampled. This approach is difficult to apply to earthquake forecasting. One could train the network on the entire continuous seismic trace, but that would require the network to also learn to recognize and disregard noise and other spurious signals in addition to the forecasting task. Even with efforts to denoise seismic data, environmental and travel-time effects remain. We simply cannot produce a clean, regular sampling of the fault dynamical system as is used in prior work. An alternative approach is to create an event-based dataset that only includes precursors and other important signals similar to the work described in \citeA{lubbers2018}. The downside of this approach is that the time steps between events are irregular. A handful of LSTM variations have been proposed to handle event-based data \cite<e.g.>{sahin2019, santeramo2018, zhu2017}, but the combination of irregularly sampled data and dynamical system forecasting has yet to be studied. Thus, this benchmark dataset will also serve as a test of LSTM's capabilities in this area.

As mentioned previously, our second goal with this benchmark is to assess the robustness of our forecasting method. Tectonic earthquakes nucleate over a wide spectrum of conditions, and thus, earthquake forecasting requires a robust method that is invariant to different conditions. Hence, an ideal method should be able to provide accurate forecasts regardless of setting. A benchmark dataset that includes a variety of conditions allows us to test the adaptive ability of LSTM. Should LSTM prove incapable of adapting, more advanced networks, such as meta learning, can be utilized.  

\subsection{Broom model}
\begin{figure*}[!htb]
	\includegraphics[width=\textwidth]{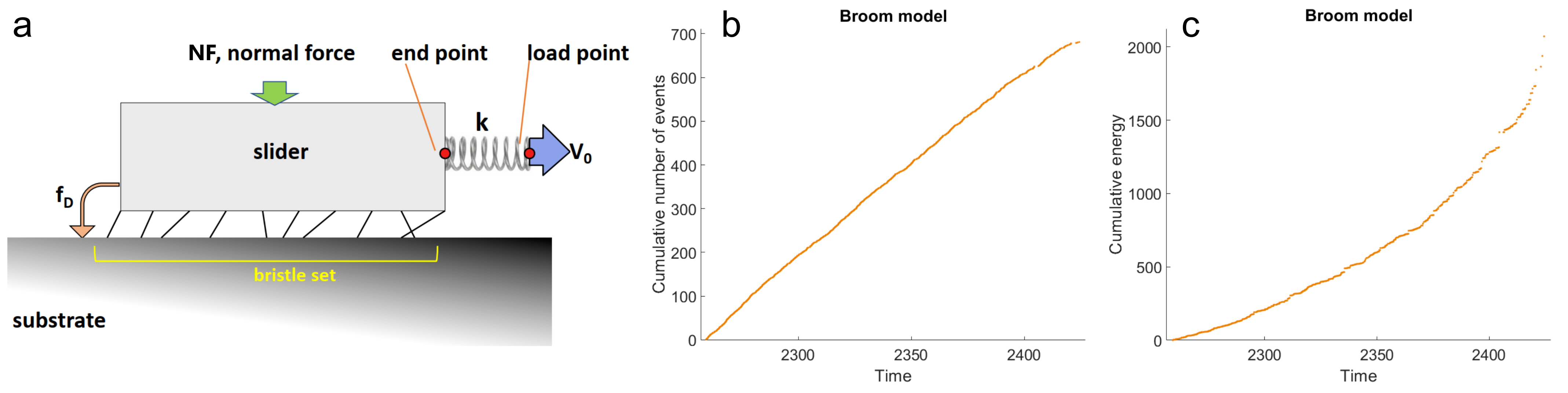}
	\caption{A) Schematic of the Broom model. B) Cumulative number of events and C) cumulative energy for a typical broomquake cycle.}
	\label{broom1}
\end{figure*}

We generate our benchmark dataset using the Broom model \cite{daub2011}. The Broom model is a variation of a  Knopoff-Burridge slider block model, that employs bristles acting fault interface asperities. It exhibits behavior of the type we want to examine, i.e., a slowly evolving dynamics that is punctuated by ``catastrophic'' slip events, almost periodic stick-slip behavior.  Among the variables describing the dynamics are a small set that can sensibly be tracked and used to test the efficacy of a LSTM network in catastrophe prediction. A schematic of the Broom model is shown along with results in Figure~\ref{broom1}a. A slider with mass $M$ experiences a variety of forces
\begin{enumerate}
\item
The slider, a rigid piece of material at $X$, is driven rightward by the force in the spring $k$. The right end of the spring, the load point, moves at a fixed velocity $V_0$. The left end of the spring, the end point, moves with the displacement of the slider, $X_{EP}=X$. The shear stress,
\begin{equation}
\sigma=\frac{k(V_0-X)}{A},
\end{equation}
where $A$ is the area of the slider-substrate interface, is one of the basic outputs in the description of the system.
\item The slider is coupled to the substrate through a system of brittle contacts called bristles. These represent the force structures in a fault gouge layer that carry the shear stress from the slider to the substrate.
The top end of each bristle is fixed at a permanent location on the slider, and thus moves with $X$. The bottom end of each bristle, at most moments of time, is at a fixed location on the substrate (which does not move), at $x_n$ for bristle $n$. Thus bristle $n$ exerts a force $\gamma(x_n-X)$ pulling the slider leftward ($\gamma$ is a spring constant). Each bristle is characterized by a critical length, e.g., $c_n$ for bristle $n$. 
After time $T_0$ the bristle $n$ re-sets with $x_n=X$ ($x_n$ re-sets at the zero force position) and a new value of $c_n$.   There are many bristles, i.e., a complex set of force structures in the gouge carry the shear stress. In this way the bristles, representing the force structures in the gouge, strengthen, fail, and are renewed. Failure of a force structure launches an acoustic broadcast, a contribution to the acoustic emission (AE). We take the energy in the acoustic  broadcast, $\Delta e_n$, to be proportional to the strength of the associated force structure at failure, i.e., $\Delta e_n\propto \gamma c_n$.
\item The slider is pushed against the substrate by the normal force, $FN$, which controls the strength of the ``brittle contacts'' through $\gamma$.  
\item There is in addition to brittle friction the possibility of an additive ductile friction, $f_D$, acting on the slider \cite{daub2011}.
\end{enumerate}

The phase space controlling possible behavior of the slider is $(V_0,FN,f_D)$ \cite{daub2011, trugman2013}. We employ the slider in the limit $(V_0,FN,0^{+})$ where the dominant motion is a nearly periodic sequence of stick-slip cycles. The slider motion in this limit provides the first data set we employ to test the prediction scheme we are advocating.

\subsection{Training data}

The dataset consists of 2000 stick-slip events at $V_0=0.05$ created at 10 different normal levels ($FN=15$ to $60$ in increments of 5). This produces recurrence intervals (i.e. the amount of time between two consecutive broomquakes) ranging from 50 to 290 unitless time steps (Table~\ref{broomT1}). At low $FN$, the data is relatively periodic, whereas at high $FN$ there is a much larger recurrence interval range. The recurrence intervals are not continuous between $FN$ values - the maximum recurrence interval for any $FN$ is 10 to 20 time steps shorter than the minimum recurrence interval for the next $FN$. This is not a perfect analogue of the real earth, but it will give the LSTM network the challenge of learning from multiple settings at once. At high $FN$, the LSTM will also have to learn the differences between events within the same $FN$ because of the relatively large recurrence interval range.

\begin{table*}[!htb]
    \caption{Recurrence interval statistics for each of the Broom model $FN$ values$^{a}$}
    \centering
    \begin{tabular}{|c|c|c|c|c|c|c|c|c|c|c|}
        \hline
        & \multicolumn{10}{c|}{\textbf{$FN$ value}} \\ \hline
        & \textbf{15} & \textbf{20} & \textbf{25} & \textbf{30} & \textbf{35} & \textbf{40} & \textbf{45} & \textbf{50} & \textbf{55} & \textbf{60} \\ \hline \hline
        Min & 50.0 & 69.4 & 94.0 & 118.3 & 141.1 & 159.4 & 192.0 & 223.7 & 250.8 & 274.9 \\ \hline
        Avg & 53.5 & 72.8 & 98.9 & 122.1 & 145.1 & 167.3 & 200.1 & 231.0 & 257.2 & 280.0 \\ \hline
        Max & 56.6 & 76.2 & 102.8 & 124.6 & 149.9 & 178.2 & 205.0 & 235.1 & 262.5 & 289.4 \\ \hline
        Range & 6.6 & 6.8 & 8.8 & 6.3 & 8.8 & 18.8 & 13.1 & 11.4 & 11.7 & 14.5 \\ \hline
    \multicolumn{11}{c|}{\multirow{3}{*}{\parbox{0.9\textwidth}{$^{a}$The Broom dataset consists of roughly 200 cycles for each $FN$ value. Recurrence interval and range increase with $FN$.}}} \\
    \multicolumn{11}{c|}{}
    \end{tabular}
    \label{broomT1}
\end{table*}

\begin{figure*}[!htb]
    \noindent
	\includegraphics[width=\textwidth]{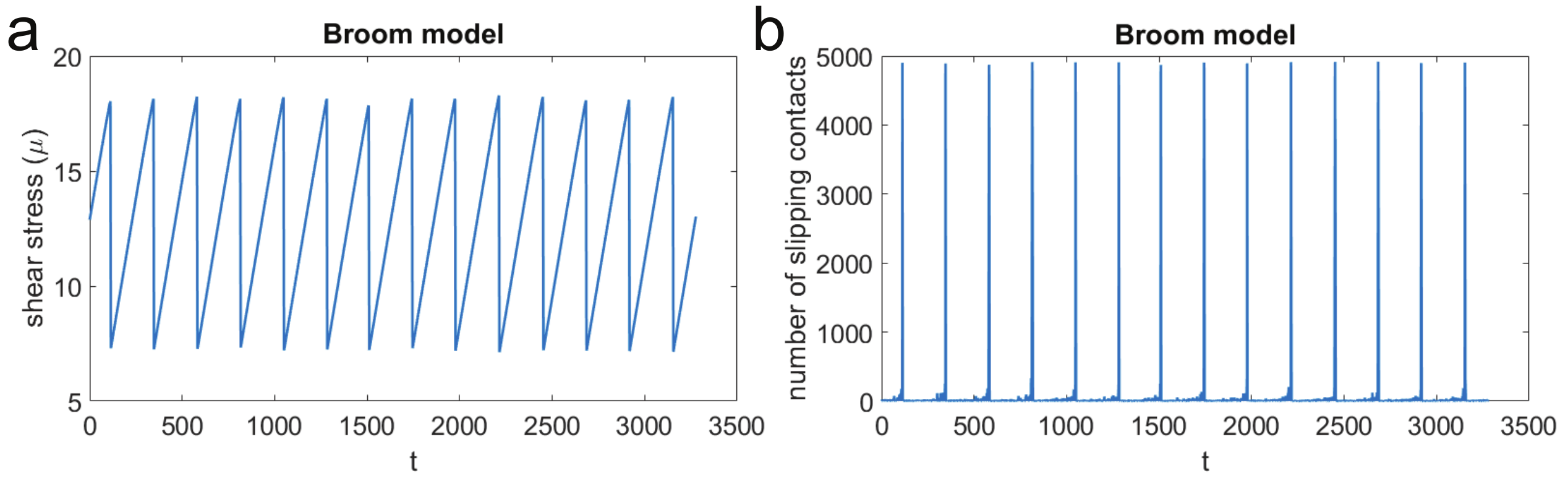}
	\caption{Snapshots from the Broom model of the A) shear stress and B) number of slipping contacts.}
	\label{broom2}
\end{figure*}

The other parameters of the Broom model are set to mimic the laboratory setup and are listed in Table~\ref{broomT2}. These parameters were carefully chosen in order to produce stick-slip events and avoid conditions that result in stable sliding. Figure~\ref{broom2} shows sample snapshots of the data produced by the Broom model. The broomquakes are quasi-periodic and exhibit shear stress typical of stick-slip events. At each time step, we calculate the number of slipping contacts and the amount of energy released, which will be the basis of our dataset. We focus on these features because the Broom model does not produce seismic data and we want to test the LSTM using features that are analogous to those calculated in the lab and field.

\begin{table*}[!htb]
    \caption{Broom model parameters}
    \centering
    \begin{tabular}{|c|c|c|c|c|c|c|}
        \hline
         \textbf{$Q$} & \textbf{Normal force ($FN$)} & \textbf{Velocity ($V_0$)} & \textbf{Num. bristles (NB)} & \textbf{$dt$} & \textbf{$T_0$} & \textbf{N1}  \\ \hline 
         0.8 & 15-60 & 0.05 & 5000 & 0.025 & 16 & 32768  \\ \hline
    \end{tabular}
    \caption{$Q$ is a parameter associated with ductile friction. $FN$ is the normal force. $V_0$ is the velocity of the slider. NB is the number of brittle contacts (bristles) on the slider. $dt$ is the sampling interval. Note that all times are unitless. $T_0$ is the amount of time between bristle resets to the zero force position. N1 is the number of timesteps in one sequence. Each run of the model has 5 sequences and only data from the last 4 are used.}
    \label{broomT2}
\end{table*}

\begin{figure*}[!htb]
    \noindent
	\includegraphics[width=\textwidth]{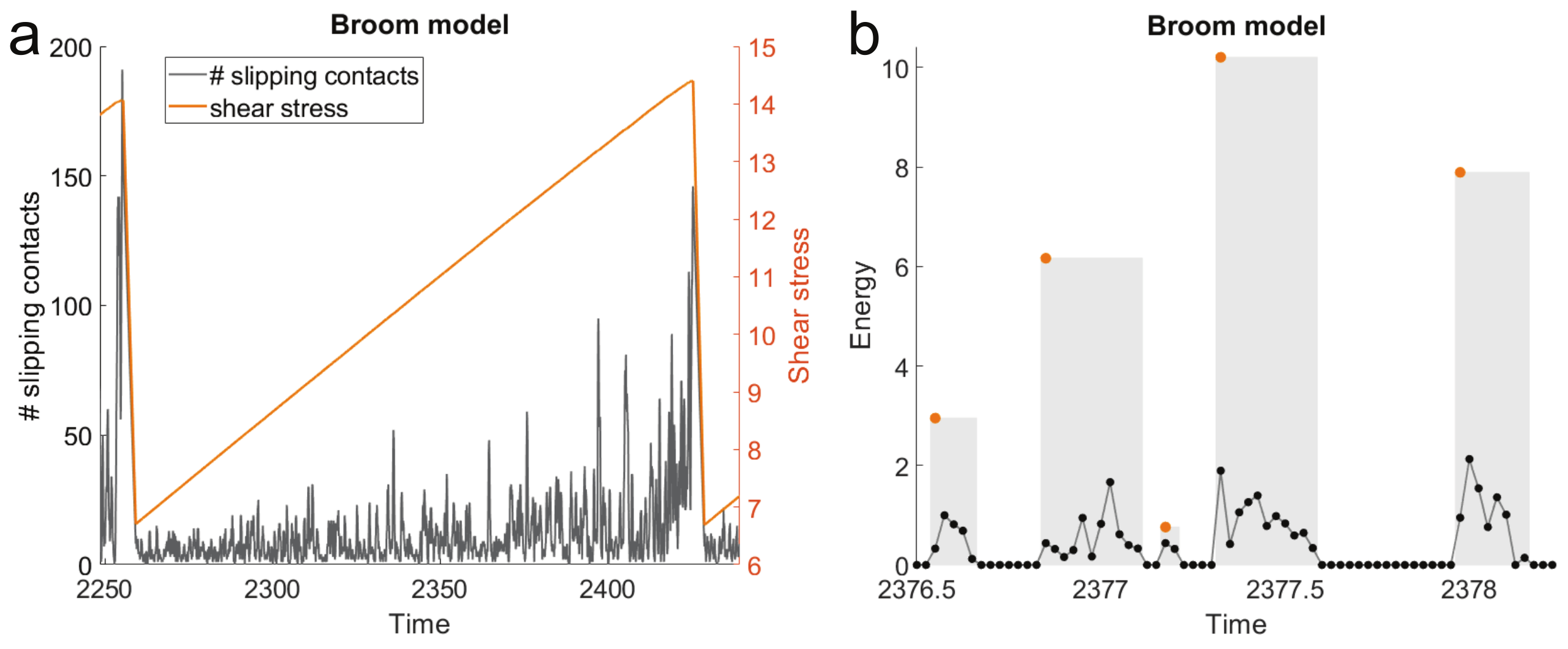}
	\caption{A) Zoomed in view of the Broom data showing a single stick-slip cycle. B) Example of the Broom model data grouping. Black circles show the energy released at each time step. Groups are indicated by they grey boxes. Energy for each group is summed and assigned to the first time step. This new event-based dataset is shown with orange circles.}
	\label{broom3}
\end{figure*}

In order to create an event-based dataset (i.e., catalog), we focus only on time steps where energy is released (i.e. at least one contact slips). In the data, contacts frequently fail in groups that span a few time steps. In seismic data, these slips would not be distinguishable (they are often closer together in time than the sampling frequency) and perhaps would even contribute to the same AE. Therefore, we preprocess the data by grouping time steps where multiple contacts fail (see Figure~\ref{broom3}). We determined the end points of each group by selecting time periods where at least two sequential timesteps have zero energy. The start point of each group is the first timestep where a nonzero energy is recorded. For each group, we summed the energy and assigned it to the group start time (red line). Each group represents one ``event'' in the cumulative number of events calculations. After this process, the Broom data resembles the event-based nature of the laboratory AE catalog. 

\begin{figure*}[!htb]
	\noindent
	\includegraphics[width=\textwidth]{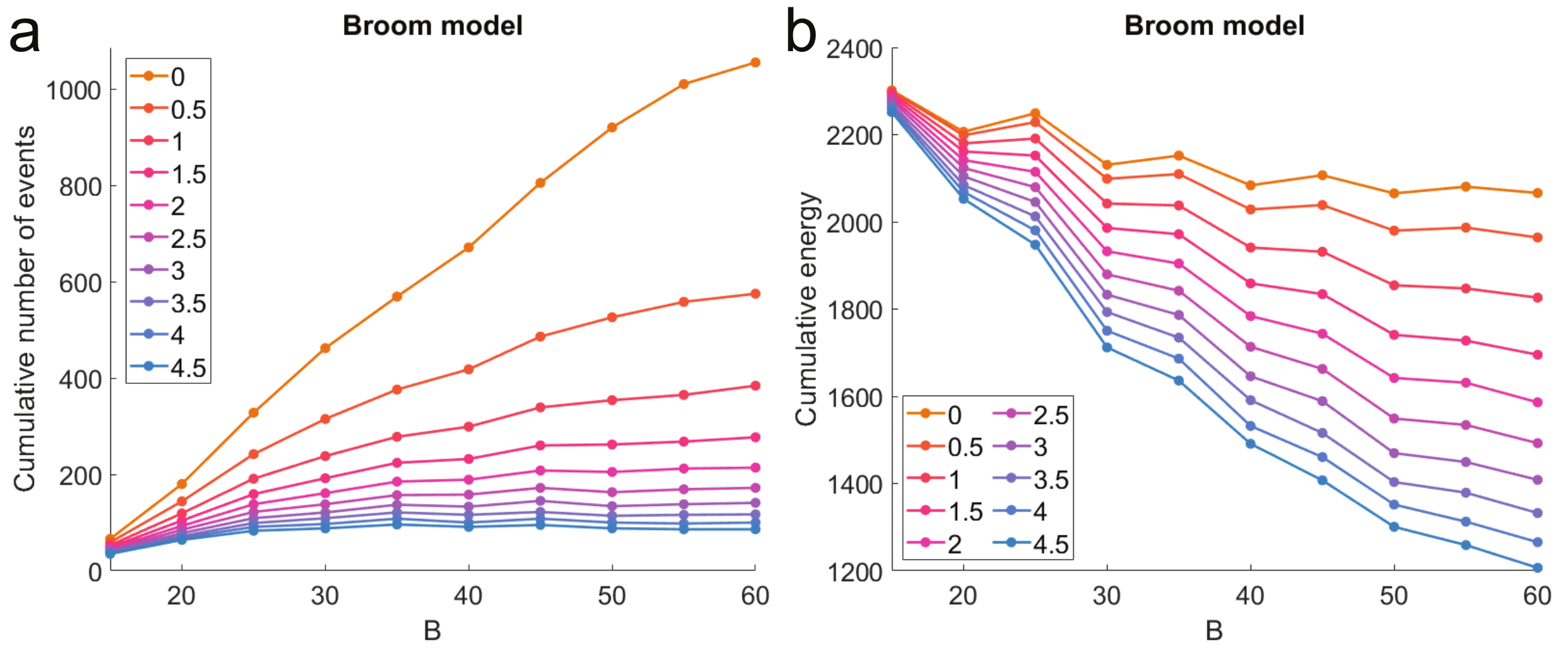}
	\caption{Total cumulative A) number of events and B) energy averaged across all events for each $FN$ value. The colors indicate the minimum energy threshold.}
	\label{broom4}
\end{figure*}

For the network training, we use two input features: cumulative energy and cumulative number of events. In this case, the events are the slip groups calculated previously. An example of these two features for a single broomquake is shown in Figure~\ref{broom1}. This figure, and many others in this paper, shows data over time for a single quake cycle. The left side of the x axis represents the onset of stress, and the the right side represents the failure point. We focus on details of only one slip cycle since the general features and characteristics of the slip cycles do not change as a function of time and/or slip displacement. For the broomquakes, the cumulative number of events is fairly linear whereas the cumulative energy has a power law shape and has a similar shape as the pre-seismic acoustic energy release in laboratory stick-slips. The linear number of events is unlikely to be useful for identifying trends associated with failure, although the raw number of events and/or slope of the line may give the LSTM a clue about the $FN$ value (see Figure~\ref{broom4}a). The cumulative energy curve will likely be the main driver of the actual forecasting due to its power law shape. As with the cumulative number of events, the individual energy values may also provide an indication of the $FN$ value (see Figure~\ref{broom4}b).

The cumulative values are calculated separately for each seismic cycle, excluding the failure itself. We normalize each feature to the range [0,1] using the maximum and minimum values across the whole dataset. Cycles are not normalized individually because that would diminish the impact of the differing recurrence intervals. For testing, we set aside 40 randomly chosen quakes for each $FN$ value for a total of 400 quakes. The remaining 1600 quakes are used for training. Note that we handle each cycle separately, so the testing and training set are not contiguous in time.

\begin{figure*}[!htb]
	\includegraphics[width=\textwidth]{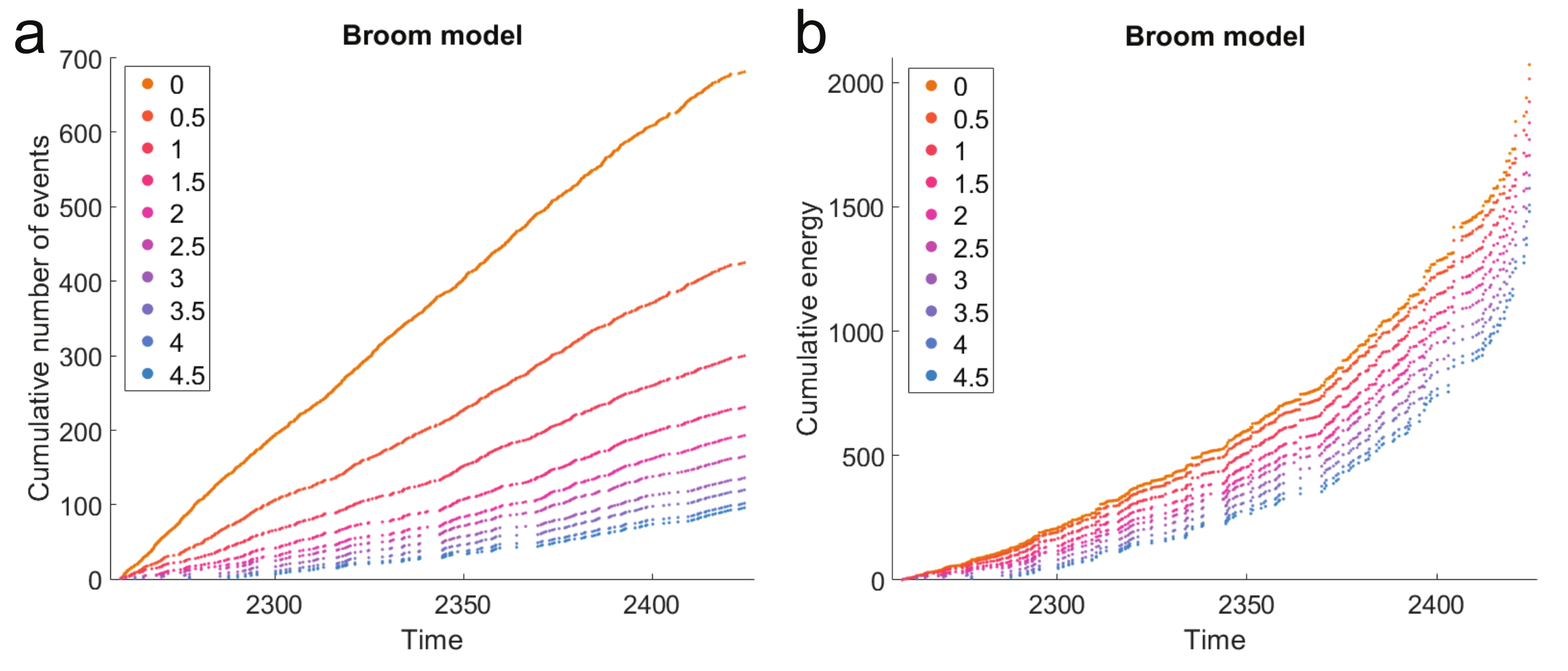}
	\caption{An example cycle showing the A) cumulative events and B) cumulative energy for the various energy thresholding datasets. For each dataset, only groups with energy above the value in the legend were included.}
	\label{broom5}
\end{figure*}

Detection of small earthquakes is a persistent problem in seismology \cite<e.g.>{ross2019, brodsky2019}. In many scenarios, the signal-to-noise ratio is low, which prevents detection of lower energy signals. In addition, smaller signals are more likely to be discarded during processing due to problems with determining the start or end time of the signal. As a result, seismic catalogs are generally incomplete, particularly with regard to small earthquakes. With the lab data and any future real-earth data, we assume that the LSTM network will only have some subset of the full dataset to work with. Our forecasting method will only be successful if it is able to make accurate forecasts despite this hurdle. To test the robustness of the forecasting scheme, we create several datasets with progressively increasing energy thresholds that determine if a group is included (Figure~\ref{broom5}). Groups with energy below the threshold are not used in the dataset. For example, if the threshold is 1, only groups with energy \textgreater 1 are included. Progressively eliminating more and more of the small Broom signals, the process of oblation, will simulate situations where LSTM only has access to a subset of the data \cite<e.g.>{lubbers2018}.

Figure~\ref{broom4} shows the average total cumulative number of events and energy by $FN$ for each of the minimum energy thresholds. These values were calculated by selecting the maximum cumulative energy (or number of events) for each broomquake, sorting by $FN$ value, and averaging. Looking at the 0 energy threshold (i.e. all timesteps are included), we see that the number of events increases roughly linearly with $FN$. Energy slightly decreases with $FN$ and the largest change occurs when$FN < 40$.

We included these same calculations for each of the energy thresholds, in which only events above a specified energy (see the legends) are included in the dataset. As the threshold is raised, larger $FN$ values lose a much greater proportion of events and energy than lower $FN$ values do. At the highest threshold (the blue lines), all $FN$ values contain approximately the same number of events. At the same threshold, the energy of the largest $FN$ values is greatly diminished resulting in a strong inverse relationship between energy and $FN$ value.

This figure illustrates why the combination of cumulative number of events and cumulative energy is a more powerful indicator of the $FN$ value than either feature is alone. At the 0 threshold, energy is roughly the same across $FN$ values, but the number of events greatly depends on $FN$. At the highest threshold, the opposite is true: energy depends on $FN$ whereas the number of events is uniform. Using both of these features together gives the LSTM network more information about the $FN$ value at every threshold than using just one feature would.

Finally, we test several situations in which the LSTM is only allowed to train on a subset of $FN$ values. The Broom model is not a perfect analog for real seismic data, so we have some concern that the LSTM will be successful because the model is too simple. If this is the case, the network should still make accurate forecasts for $FN$ values it has not seen. Conversely, if the LSTM performs poorly on these values, the Broom model can serve as a reasonable benchmark.

\subsection{Performance}

\begin{figure*}[!htb]
	\noindent
	\includegraphics[width=\textwidth]{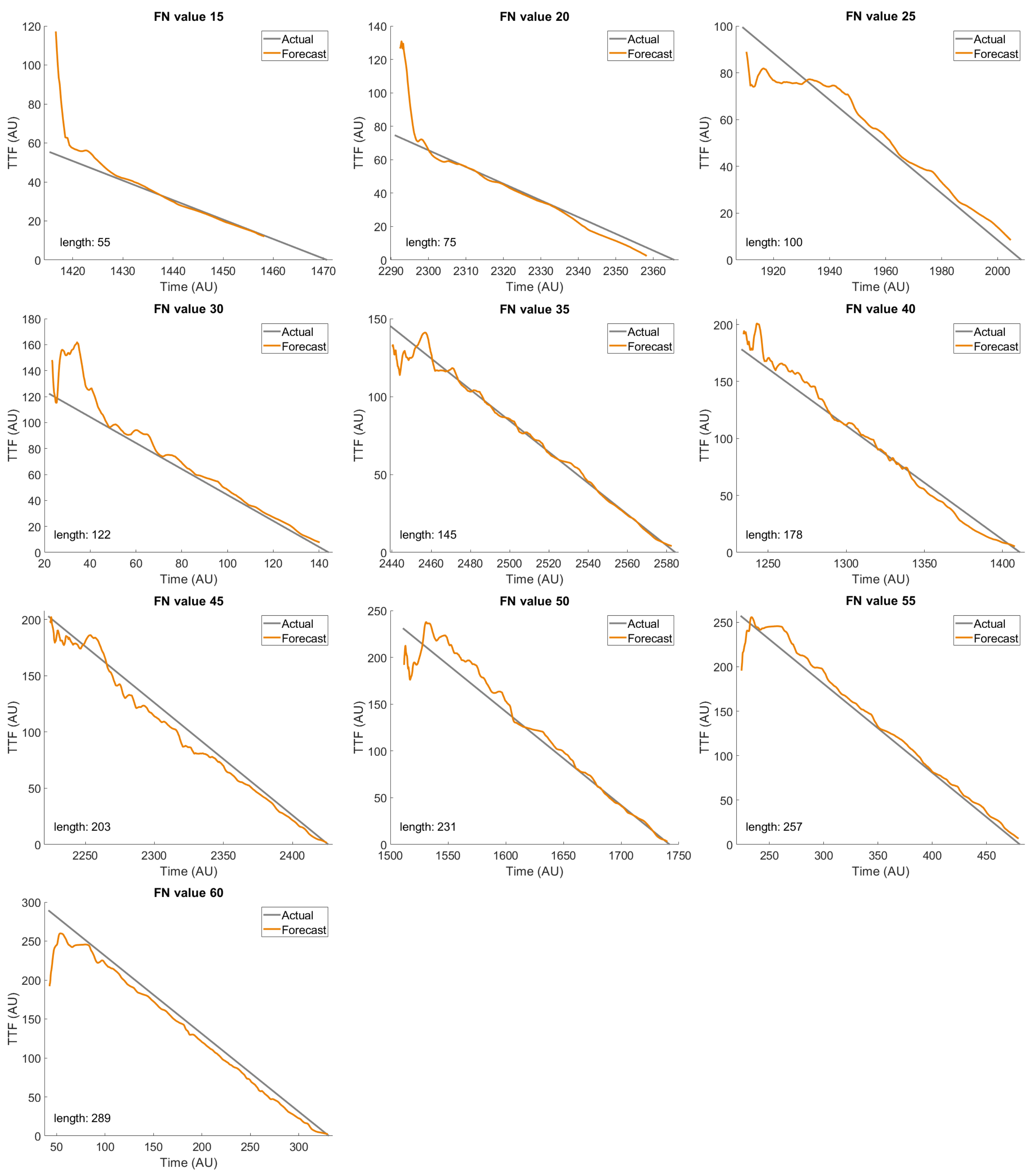}
	\caption{Typical cycles for each $FN$ value with forecasts from the LSTM network trained on the full Broom dataset. The recurrence interval is listed in the bottom left corner.}
	\label{broom6}
\end{figure*}

\begin{figure*}[!htb]
	\includegraphics[width=\textwidth]{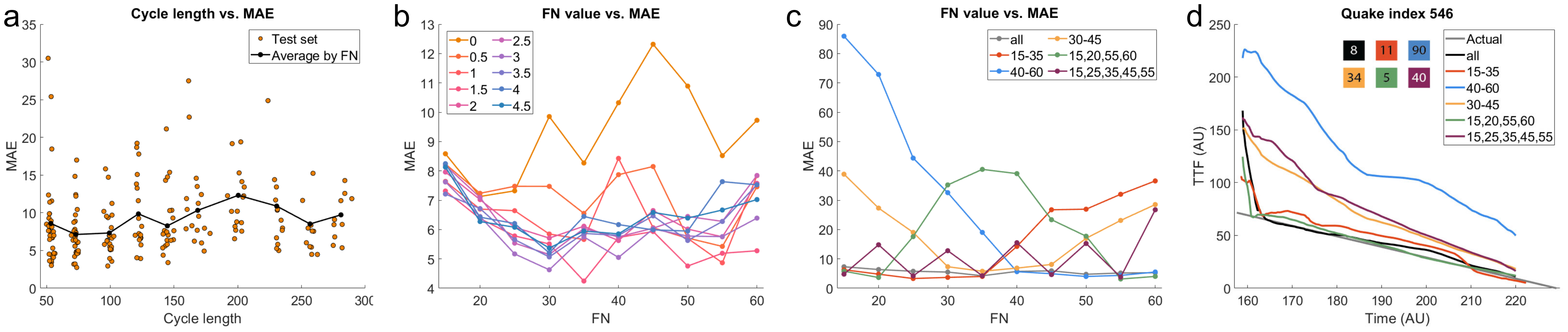}
	\caption{A) Scatter plot of MAE vs. recurrence interval for the Broom test set. The solid line shows average MAE for each $FN$ value. B) MAE for each $FN$ value for the energy thresholding datasets. The legend lists the threshold for each dataset. The LSTM performs roughly the same regardless of how many data points have been removed. C) MAE for each $FN$ value for the tests where the LSTM was trained on a subset of $FN$ values. The legend lists the $FN$ values that were included in the train set for each test. In general, the network performs well on the seen $FN$ values and poorly on the unseen ones. D) An example cycle from the tests in (C) that illustrates the error scale. The colored boxes show the error for each dataset.}
	\label{broom7}
\end{figure*}

The forecasting results are shown in Figure~\ref{broom6} and Table~\ref{broomT3}. The network hyperparameters are listed in Table~\ref{broomT3}. The LSTM forecasts TTF with a low MAE at all $FN$ values. (For a visualization of the error scale, see Figure~\ref{broom7}d.) The MAE is slightly higher for $FN$ values that have a larger recurrence interval range (i.e. $FN>=40$), indicating the LSTM is adept at discerning between $FN$ values, but has more difficulty assigning recurrence intervals within a given $FN$ value. However, overall no $FN$ value stands out as substantially better or worse than the others, showing that the LSTM was capable of adapting to and differentiating between the different stress conditions. 

\begin{table*}[!htb]
    \caption{Broom LSTM hyperparameters for energy threshold tests}
    \centering
    \begin{tabular}{ |c|c|c|c|c|c|c|c|c|c|c| } 
			\hline
			& \multicolumn{10}{c|}{\textbf{Broom LSTM run}} \\ \hline
			\textbf{Hyperparameter} & \textbf{1} & \textbf{2} & \textbf{3} & \textbf{4} & \textbf{5} & \textbf{6} & \textbf{7} & \textbf{8} & \textbf{9} & \textbf{10}\\ \hline \hline
			Number of train events & \multicolumn{10}{c|}{1674} \\ \hline
			Number of test events & \multicolumn{10}{c|}{400} \\ \hline
			Number of layers & \multicolumn{10}{c|}{3} \\ \hline
			Hidden size & \multicolumn{10}{c|}{6} \\ \hline
			Training epochs & 2963 & 3439 & 3946 & 3989 & 3975 & 3953 & 4953 & 3956 & 3000 & 3631 \\ \hline
			Starting learning rate & \multicolumn{10}{c|}{0.003} \\ \hline
			Sequence length & \multicolumn{10}{c|}{20} \\ \hline
			Learning rate factor & \multicolumn{10}{c|}{0.5} \\ \hline
			Mini-batch size & \multicolumn{10}{c|}{100} \\ \hline
			Learning rate patience & 50 & 50 & 100 & 100 & 50 & 100 & 100 & 100 & 100 & 100 \\ \hline
			Gradient clipping & \multicolumn{10}{c|}{0.5} \\ \hline
			Energy threshold & 0 & 0.5 & 1 & 1.5 & 2 & 2.5 & 3 & 3.5 & 4 & 4.5 \\ \hline \hline
			\textbf{MAE} & 9.293 & 7.158 & 6.538 & \textbf{5.611} & 6.450 & 6.437 & 5.861 & 6.222 & 6.579 & 6.432 \\ \hline
	\multicolumn{11}{c|}{\multirow{3}{*}{\parbox{\textwidth}{$^{a}$The threshold is the minimum energy required for a slip event to be included in the dataset. The best MAE is in bold. For an idea of the error scale, see Figure~\ref{broom7}d. For MAE broken down by $FN$ value, see Table~\ref{broomT5}.}}} \\
    \multicolumn{11}{c|}{}
    \end{tabular}
    \label{broomT3}
\end{table*}


\begin{sidewaystable*}[!htb]
    \caption{MAE for the Broom LSTM test sets$^{a}$}
    \centering
    \begin{tabular}{|c|c||c|c|c|c|c|c|c|c|c|c|}
        \hline
         && \multicolumn{10}{c|}{\textbf{FN value}} \\ \hline
         \textbf{Threshold} & \textbf{Total} & \textbf{15} & \textbf{20} & \textbf{25} & \textbf{30} & \textbf{35} & \textbf{40} & \textbf{45} & \textbf{50} & \textbf{55} & \textbf{60} \\ \hline \hline
         0 & 9.293 & 8.583 & 7.127 & 7.320 & 9.852 & 8.269 & 10.323 & 12.317 & 10.890 & 8.520 & 9.726 \\ \hline
         0.5 & 7.158 & 8.235 & 7.236 & 7.475 & 7.472 & 6.551 & 7.871 & 8.148 & 5.710 & 5.426 & 7.458 \\ \hline 
         1 & 6.538 & 7.643 & 6.697 & 6.643 & 5.847 & 5.661 & 8.426 & 6.063 & 5.700 & \textbf{4.867} & 7.837 \\ \hline 
         1.5 & \textbf{5.611} & 7.314 & 6.374 & 5.782 & 5.510 & \textbf{4.245} & 5.727 & \textbf{5.933} & \textbf{4.756} & 5.191 & \textbf{5.277} \\ \hline 
         2 & 6.450 & 7.961 & 7.020 & 6.080 & 5.709 & 6.114 & 5.625 & 6.648 & 6.015 & 5.753 & 7.575 \\ \hline 
         2.5 & 6.437 & 8.188 & 7.094 & 5.537 & 5.148 & 6.035 & 5.750 & 6.045 & 6.447 & 6.279 & 7.844 \\ \hline 
         3 & 5.861 & 7.622 & 6.444 & \textbf{5.169} & \textbf{4.628} & 5.749 & \textbf{5.050} & 6.020 & 5.778 & 5.759 & 6.390\\ \hline 
         3.5 & 6.222 & \textbf{7.211} & 6.715 & 5.657 & 5.067 & 5.868 & 5.802 & 6.466 & 5.627 & 6.274 & 7.533 \\ \hline 
         4 & 6.579 & 8.246 & 6.422 & 6.207 & 5.203 & 6.448 & 6.169 & 5.990 & 5.949 & 7.633 & 7.525 \\ \hline 
         4.5 & 6.432 & 8.129 & \textbf{6.273} & 6.078 & 5.370 & 5.949 & 5.852 & 6.585 & 6.392 & 6.667 & 7.023 \\ \hline 
    \multicolumn{12}{c|}{\multirow{4}{*}{\parbox{0.8\textwidth}{$^{a}$The threshold is the minimum energy required for a slip event to be included in the dataset. The table shows the total MAE as well as MAE broken down by $FN$ value. The best result for each $FN$ (as well as total MAE) is in bold. For an idea of the error scale, see Figure~\ref{broom7}d. We find that the LSTM network performs well for all thresholds and $FN$ values.}}} \\
    \multicolumn{12}{c|}{}
    \end{tabular}
    \label{broomT5}
\end{sidewaystable*}

Figure~\ref{broom6} shows forecasted vs. actual TTF for representative broomquakes for all $FN$ values. For all $FN$ values, the prediction error can be large very early in the cycle when the LSTM has only seen a few time steps. This behavior is expected because the LSTM begins by assuming the recurrence intervals will be of average length. In the absence of any other information, this choice will give the lowest error. For the broomquakes, the first forecast is generally between 150 and 200. As training progresses, the network adjusts as additional data comes in. As a result, cycles that are particularly short or long will have higher error because the network requires larger adjustments. After this short adjustment period, the broomquake forecasts quickly stabilize and follow very closely to the true TTF for the remainder of the cycle. This holds true regardless of $FN$ value or recurrence interval. Once the LSTM has settled on a forecasted duration for a particular cycle, it does not stray in its forecasted values. This illustrates the power of recurrence for time series forecasting; the LSTM uses knowledge from past time steps to continue to make stable and consistent forecasts throughout the remainder of the quake cycle. Without this memory component, the forecasts would likely be more jagged and inaccurate because the network would be forced to go through the adjustment phase whenever it received additional data.

Figure~\ref{broom7}a is a scatter plot of the MAE for each broomquake in the test set. The solid black line is the average MAE for each $FN$ value. Once again, no $FN$ value stands out as particularly better or worse than the others. At low $FN$, where the recurrence intervals are shortest and the range is lowest, the MAE spread is larger than at high $FN$. This is likely because the network must make a large initial recurrence interval adjustment and these short cycles have relatively fewer data points. This means that the network has a shorter amount of time to adjust its forecasts and that the adjustment period is proportionally longer and thus boosts the error.

\begin{figure*}[!htb]
	\noindent
	\includegraphics[width=\textwidth]{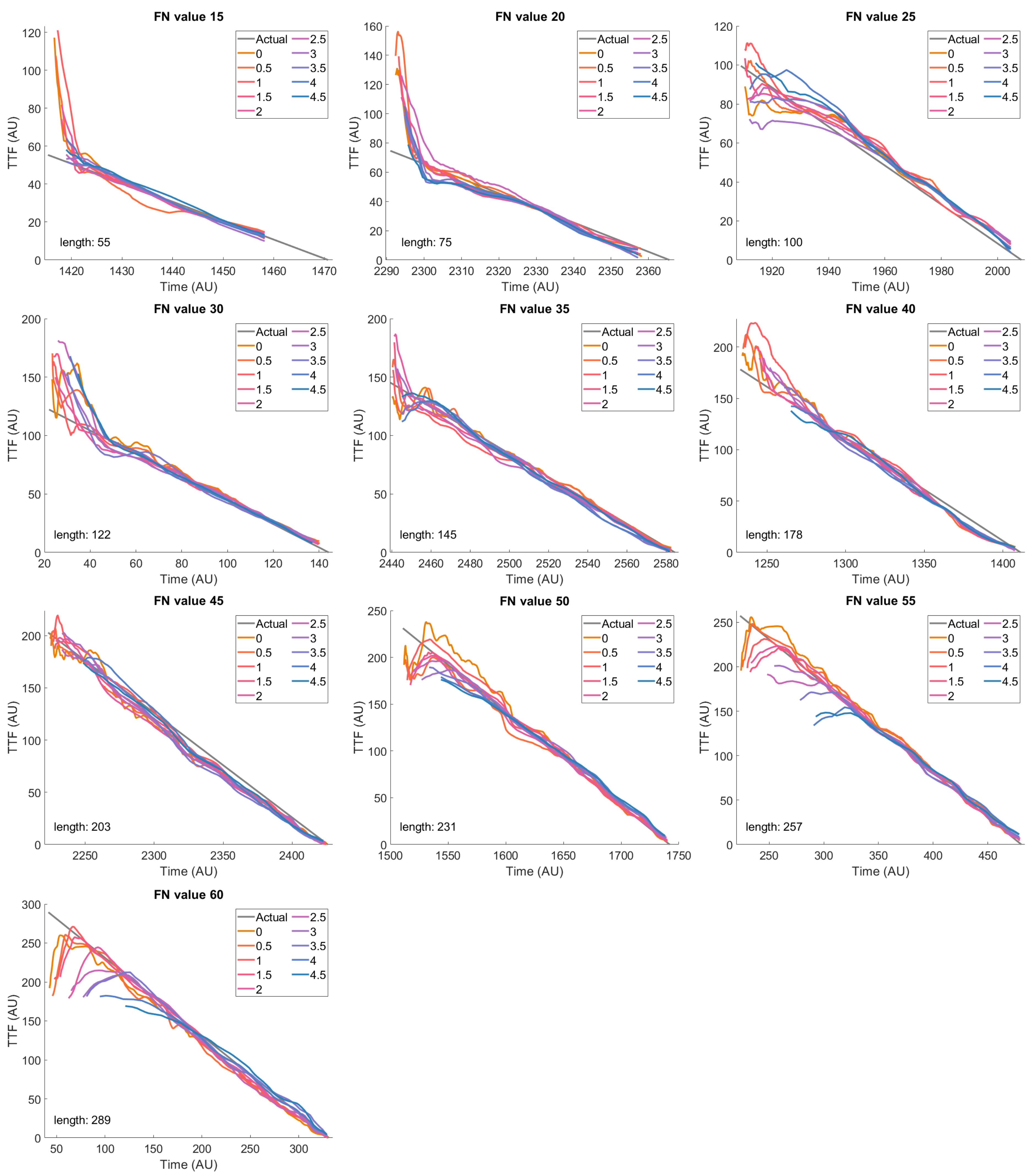}
	\caption{Typical cycles for each $FN$ value with forecasts from the LSTM networks trained on energy thresholded data. The recurrence interval is listed in the bottom left corner.}
	\label{broom8}
\end{figure*}

We also find that the network performs well across all tested energy thresholds, with no observed dropoff in accuracy (Table~\ref{broomT3} and Figure~\ref{broom7}b). The difference in total MAE between the threshold datasets is negligible and the MAEs by $FN$ value are also remarkably close. The network actually performs the worst on the full dataset (0 threshold), possibly because the memory component struggles with the volume of data. This is supported by the fact that the increase of MAE with $FN$ that we observed at 0 threshold is not present in the other datasets. The data reduction disproportionately impacts the longer broomquake cycles, bringing their number of data points closer to the average. Despite the good performance on all thresholds, there is some indication that there is an optimal number of data points for any given recurrence interval. At low $FN$, the best MAE is achieved at higher thresholds whereas high $FN$ has the best MAE at lower thresholds. These MAE differences are relatively small, but they suggest that longer recurrence intervals may be more difficult to forecast because they have a lower loss threshold before accuracy degrades. This may be partly because lower energy events tend to happen near the beginning of the cycle. If too many are removed, the effective start time of the cycle is pushed forward because there simply are no data points to learn from (see the blue curves for $FN=55$ in Figure~\ref{broom8}). As a result, the adjustment period takes up a proportionally longer amount of the recurrence interval. Overall, this test shows that the forecasting scheme should remain accurate even if a substantial portion of the data is lost.

\begin{figure*}[!htb]
	\noindent
	\includegraphics[width=\textwidth]{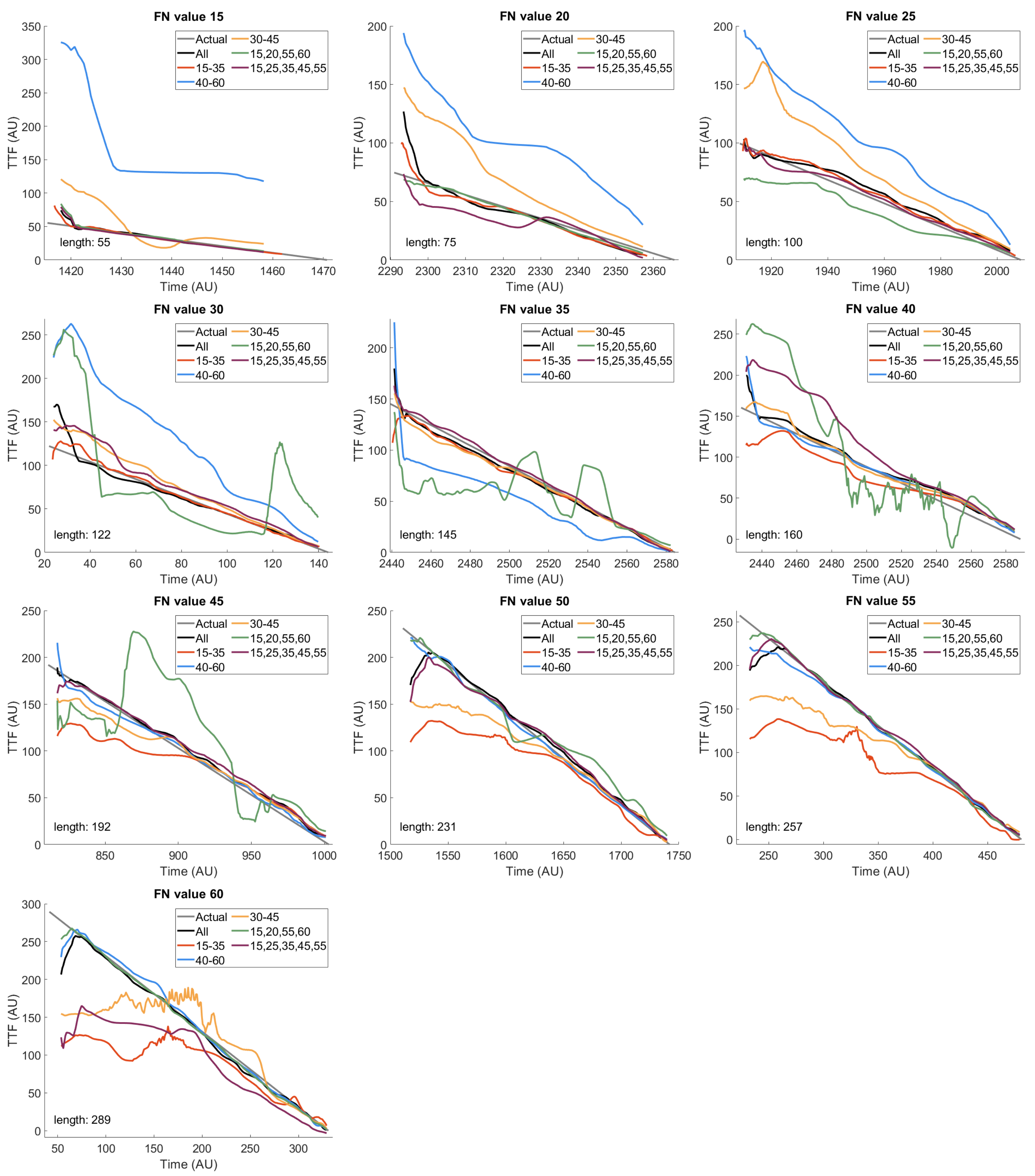}
	\caption{Typical cycles for each $FN$ value with forecasts from the LSTM networks trained on a subset of $FN$ values. The recurrence interval is listed in the bottom left corner.}
	\label{broom9}
\end{figure*}

The results of the tests where the network was trained on a subset of $FN$ values are shown in Figure~\ref{broom7}c, as well as in Tables~\ref{broomT6} and \ref{broomT7} and Figure~\ref{broom9}. We ran several tests in which different patterns of $FN$ values were used. We did this to get a generalized view of the network performance that is not unduly influenced by a specific $FN$ value pattern. The training set patterns we examined include the first/second contiguous half of $FN$ values, end values only, middle values only, and odd values. For each test, the forecast accuracy is markedly lower for the unseen $FN$ values. The magnitude of the MAE increases with distance from the training $FN$ values (i.e. for the $FN=15-35$ training set, $FN=60$ performs the worst of the unseen values). Tests where this distance is minimized (the odd value test) do produce lower MAE, but the MAE for the unseen $FN$ values is still much higher than for the seen values. These tests show that the broomquakes at each $FN$ value are sufficiently different that the LSTM cannot easily extrapolate to unseen $FN$ values. Thus, the Broom model is not problematically simple and can serve as a reasonable benchmark. This also shows that the LSTM has a remarkable ability to learn from multiple conditions at once. The great performance on the previous tests could only have occurred if the LSTM had learned each of the individual stress settings. If the LSTM was unable to handle multiple settings at once, we would have seen degraded performance like we see with these subset tests.

\begin{table*}[!htb]
    \caption{Broom LSTM hyperparameters for $FN$ subset tests}
    \centering
    \begin{tabular}{ |c|c|c|c|c|c|c| } 
			\hline
			& \multicolumn{6}{c|}{\textbf{Broom LSTM run}} \\ \hline
			\textbf{Hyperparameter} & \textbf{1} & \textbf{2} & \textbf{3} & \textbf{4} & \textbf{5} & \textbf{6} \\ \hline \hline
			Number of train events & 1674 & 869 & 805 & 659 & 670 & 841 \\ \hline
			Number of test events & 400 & 200 & 200 & 160 & 160 & 200\\ \hline
			Number of layers & \multicolumn{6}{c|}{3} \\ \hline
			Hidden size & \multicolumn{6}{c|}{6} \\ \hline
			Training epochs & 3989 & 2000 & 2000 & 2000 & 2000 & 2000 \\ \hline
			Starting learning rate & \multicolumn{6}{c|}{0.003} \\ \hline
			Sequence length & \multicolumn{6}{c|}{20} \\ \hline
			Learning rate factor & \multicolumn{6}{c|}{0.5} \\ \hline
			Mini-batch size & 100 & 10 & 10 & 10 & 10 & 10 \\ \hline
			Learning rate patience & \multicolumn{6}{c|}{100} \\ \hline
			Gradient clipping & \multicolumn{6}{c|}{0.5} \\ \hline
			Train set & all & 15-35 & 40-60 & 30-45 & 15,20,55,60 & 15,25,35,45,55 \\ \hline
    \end{tabular}
    \label{broomT6}
\end{table*}

\begin{sidewaystable*}[!htb]
    \caption{MAE for Broom $FN$ subset tests$^{a}$}
    \centering
    \begin{tabular}{|c|c||c|c|c|c|c|c|c|c|c|c|}
        \hline
         && \multicolumn{10}{c|}{\textbf{$FN$ value}} \\ \hline
          \textbf{Train set} & \textbf{Total} & \textbf{15} & \textbf{20} & \textbf{25} & \textbf{30} & \textbf{35} & \textbf{40} & \textbf{45} & \textbf{50} & \textbf{55} & \textbf{60} \\ \hline \hline
         all & 5.611 & 7.314 & 6.374 & 5.782 & 5.510 & 4.245 & 5.727 & 5.933 & 4.756 & 5.191 & 5.277 \\ \hline 
         15-35 & 15.863 & 6.302 & 4.790 & 3.304 & 3.689 & 4.004 & \cellcolor[HTML]{C0C0C0}{14.246} & \cellcolor[HTML]{C0C0C0}{26.749} & \cellcolor[HTML]{C0C0C0}{26.937} & \cellcolor[HTML]{C0C0C0}{32.025} & \cellcolor[HTML]{C0C0C0}{36.579} \\ \hline
         40-60 & 27.925 & \cellcolor[HTML]{C0C0C0}{85.923} & \cellcolor[HTML]{C0C0C0}{72.880} & \cellcolor[HTML]{C0C0C0}{44.358} & \cellcolor[HTML]{C0C0C0}{32.578} & \cellcolor[HTML]{C0C0C0}{19.028} & 5.642 & 4.945 & 3.999 & 4.369 & 5.527 \\ \hline
         30-45 & 18.194 & \cellcolor[HTML]{C0C0C0}{38.874} & \cellcolor[HTML]{C0C0C0}{27.325} & \cellcolor[HTML]{C0C0C0}{19.030} & 7.330 & 5.784 & 6.865 & 8.103 & \cellcolor[HTML]{C0C0C0}{16.999} & \cellcolor[HTML]{C0C0C0}{23.109} & \cellcolor[HTML]{C0C0C0}{28.526} \\ \hline
         15,20,55,60 & 19.009 & 5.799 & 3.652 & \cellcolor[HTML]{C0C0C0}{17.592} & \cellcolor[HTML]{C0C0C0}{35.211} & \cellcolor[HTML]{C0C0C0}{40.515} & \cellcolor[HTML]{C0C0C0}{39.071} & \cellcolor[HTML]{C0C0C0}23.352 & \cellcolor[HTML]{C0C0C0}17.741 & 3.142 & 4.015 \\ \hline
         15,25,35,45,55 & 10.679 & 4.792 & \cellcolor[HTML]{C0C0C0}14.807 & 4.190 & \cellcolor[HTML]{C0C0C0}12.788 & 4.246 & \cellcolor[HTML]{C0C0C0}15.523 & 4.651 & \cellcolor[HTML]{C0C0C0}15.255 & 3.786 & \cellcolor[HTML]{C0C0C0}26.748 \\ \hline
    \multicolumn{12}{c|}{$^{a}$White cells indicate seen (training and testing) $FN$ values and grey cells indicate unseen (testing only) $FN$ values} \\
    \multicolumn{12}{c|}{}
    \end{tabular}
    \label{broomT7}
\end{sidewaystable*}

We find that the LSTM network is clearly capable of forecasting failure from the Broom model data. In general, the network forecasts time to failure very accurately and does so consistently across all stress levels and recurrence intervals. With these tests, we show that LSTM networks have the ability to learn from irregularly sampled metastable seismic systems. The Broom model is a simple example of such a system, but the excellent network performance suggests that a similar method could be used on more complex data.

These results also demonstrate the robustness of the forecasting method. The network performs with roughly the same accuracy across all stress levels and recurrence intervals. This leads us to conclude that the network could provide accurate forecasts even when trained on seismic data from different geological settings. Given enough data, the network could theoretically forecast failure for many different fault systems after a single training session. A caveat is that noise and fault complexity in Earth may degrade the model predictions.

\label{appxBroom_end}
\clearpage

\section{Laboratory data} \label{appxLab}

\begin{figure*}[!htb]
	\begin{center}
		\includegraphics[width=3in]{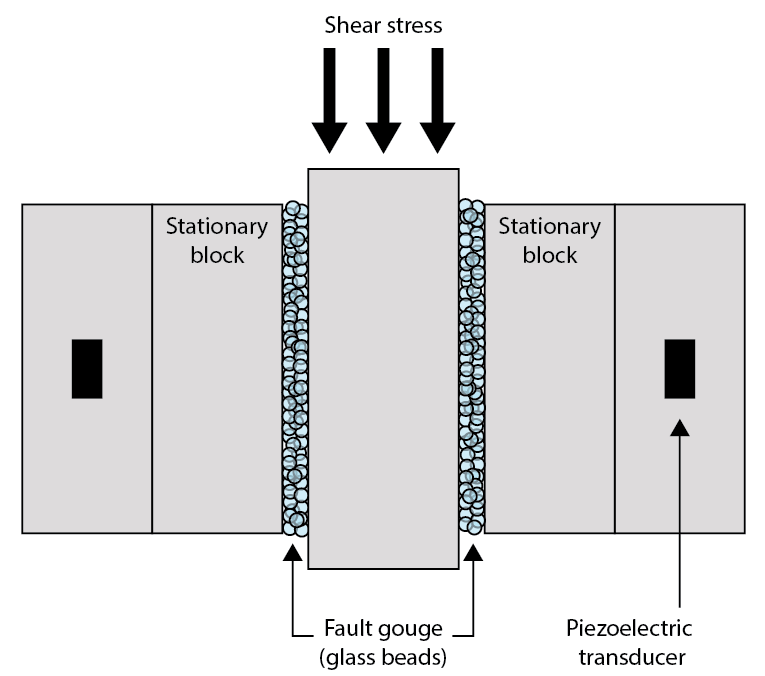}
	\end{center}
	\caption{Schematic of the biaxial shearing system used to create the laboratory dataset.}
	\label{lab1}	
\end{figure*}

We use data generated by a biaxial deformation apparatus in a double-direct shear (DDS) configuration (Figure~\ref{lab1}). In this configuration, two layers of simulated fault gouge (i.e., glass beads) are held together by three steel reinforcement blocks. Adjacent to the DDS are steel blocks containing piezoelectric transducers. The transducers are epoxied inside blind holes (black squares in schematic) and placed 22 mm from the edge of the fault zone. Prior to shearing, the DDS is placed inside a load frame and a constant normal stress is applied to the sample. After allowing the sample to compact, the center block is sheared downward at a constant velocity producing periodic stick-slip events (i.e., laboratory earthquakes). 

\begin{figure*}[!htb]
	\noindent
	\includegraphics[width=\textwidth]{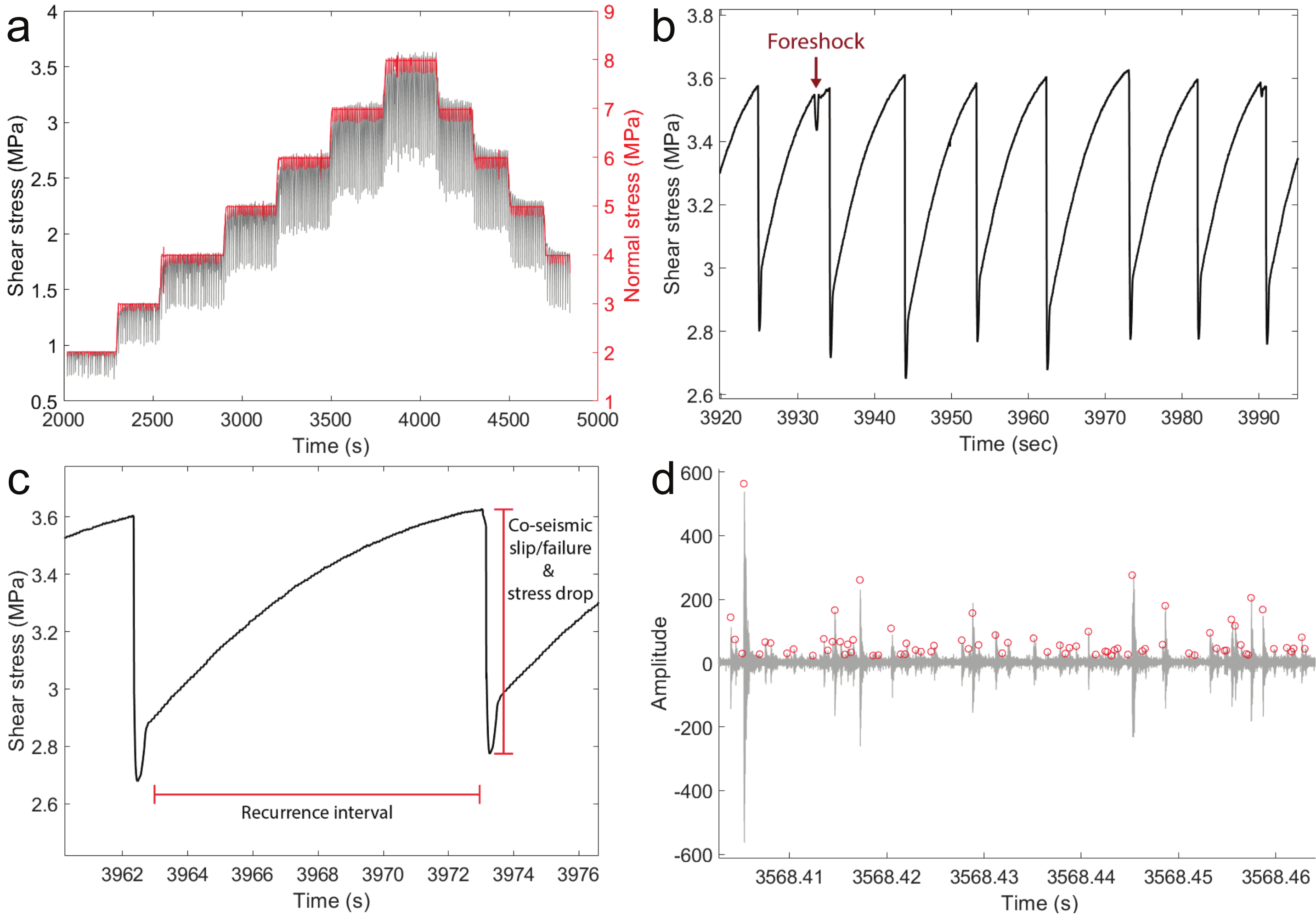}
	\caption{A) Shear (grey) and normal (red) stress for the step-up step-down experiment (p4581). B) Zoom  of several stick-slip cycles at a normal stress of 8 MPa. One cycle shows an example of a foreshock. C) Shear stress for a single seismic cycle. The recurrence interval is calculated as the time difference from the minimum shear stress of the previous slip cycle to the peak shear stress of the current cycle.  D) Snapshot of the continuous AE data for one channel. Red circles represent cataloged AEs.}
	\label{lab2}
\end{figure*}

We use data from Experiment p4581, where the fault zone normal stress was systematically increased from 2-8 MPa in steps of 1 MPa and subsequently decreased back to 2 MPa  (Figure~\ref{lab2}a). We plot zooms of several laboratory seismic cycles in (Figure~\ref{lab2}b). For each laboratory seismic cycle, we focus our analysis on data defined by the inter-seismic period (highlighted with red line in Figure \ref{lab2}c). In other words, we disregard data associated with the co-seismic slip phase. Note, the recurrence interval scales systematically with normal stress. In Figure~\ref{lab2}d we plot the continuous AE data for one channel along with the detected/cataloged AEs marked by red circles.

\begin{figure*}[!htb]
    \centering
    \includegraphics[width=5in]{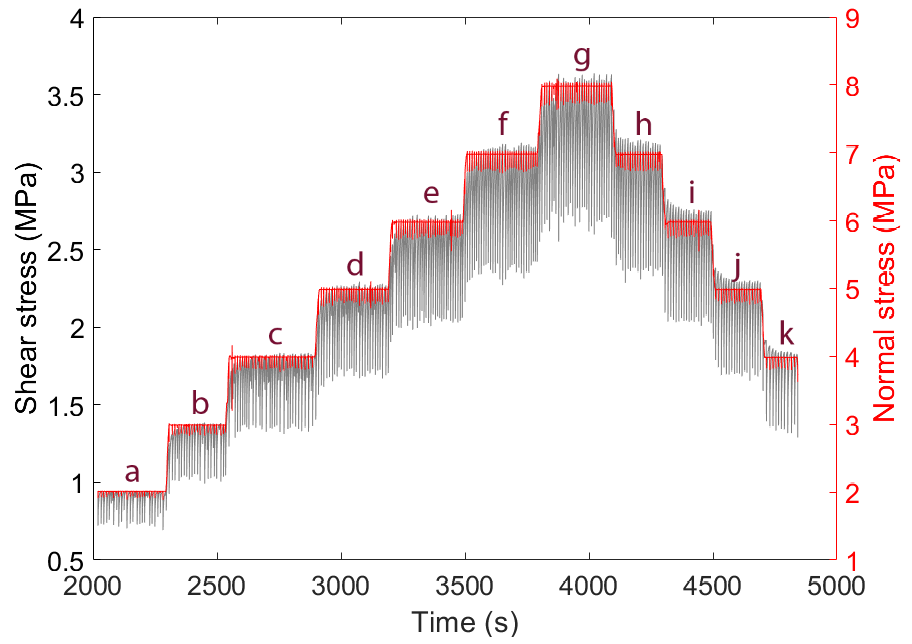}
	\caption{Reference figure for use with Figures~\ref{lab4} and \ref{lab5}. The letters indicate the step from which the example quakes were taken.}
	\label{lab3}
\end{figure*}

\begin{figure*}[!htb]
	\noindent
	\includegraphics[width=\textwidth]{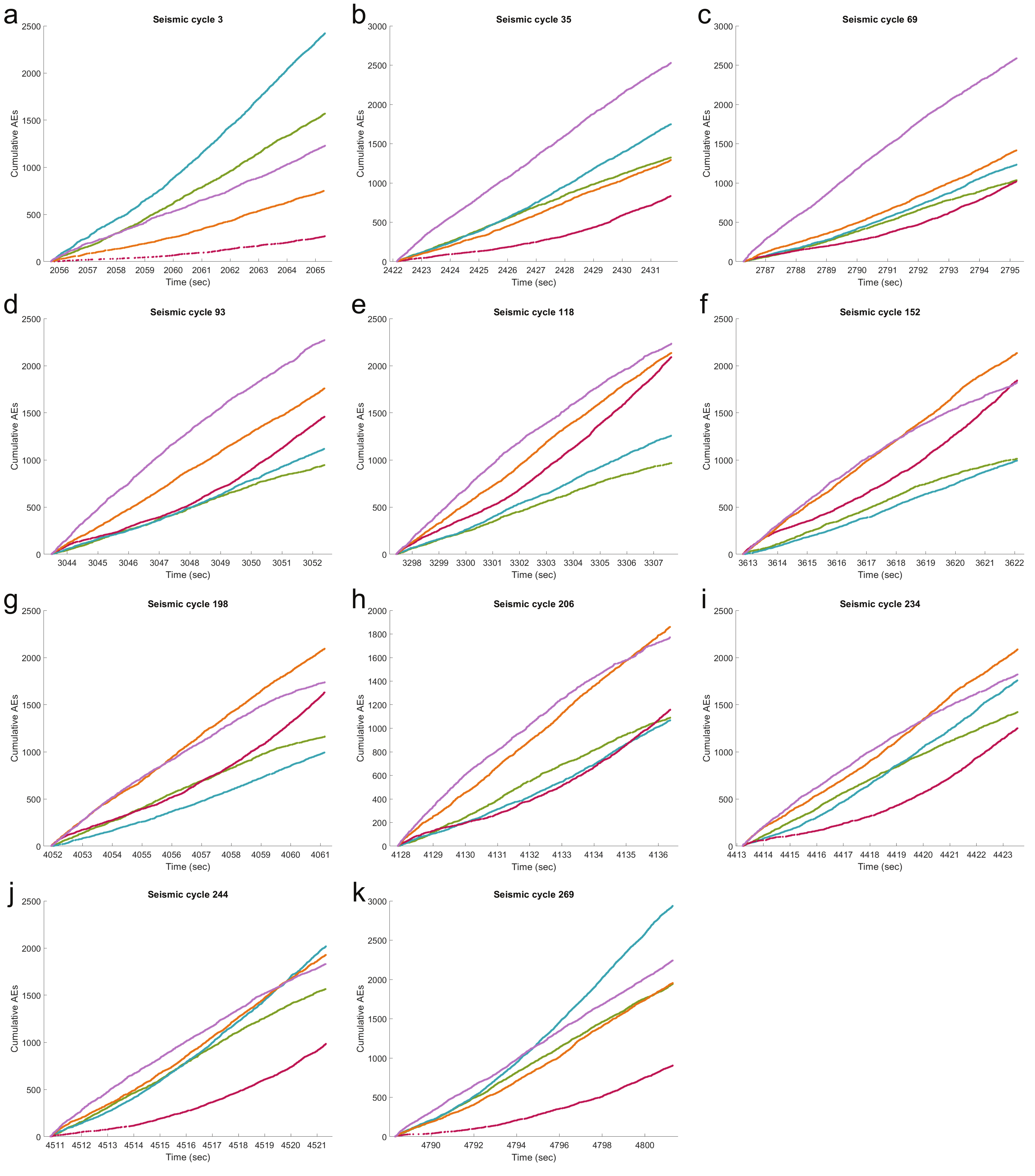}
    \caption{Cumulative number of AEs for example quakes from each shear stress step of the experiment. See Figure~\ref{lab3} for the corresponding steps.}
    \label{lab4}
\end{figure*}

\begin{figure*}[!htb]
    \noindent
	\includegraphics[width=\textwidth]{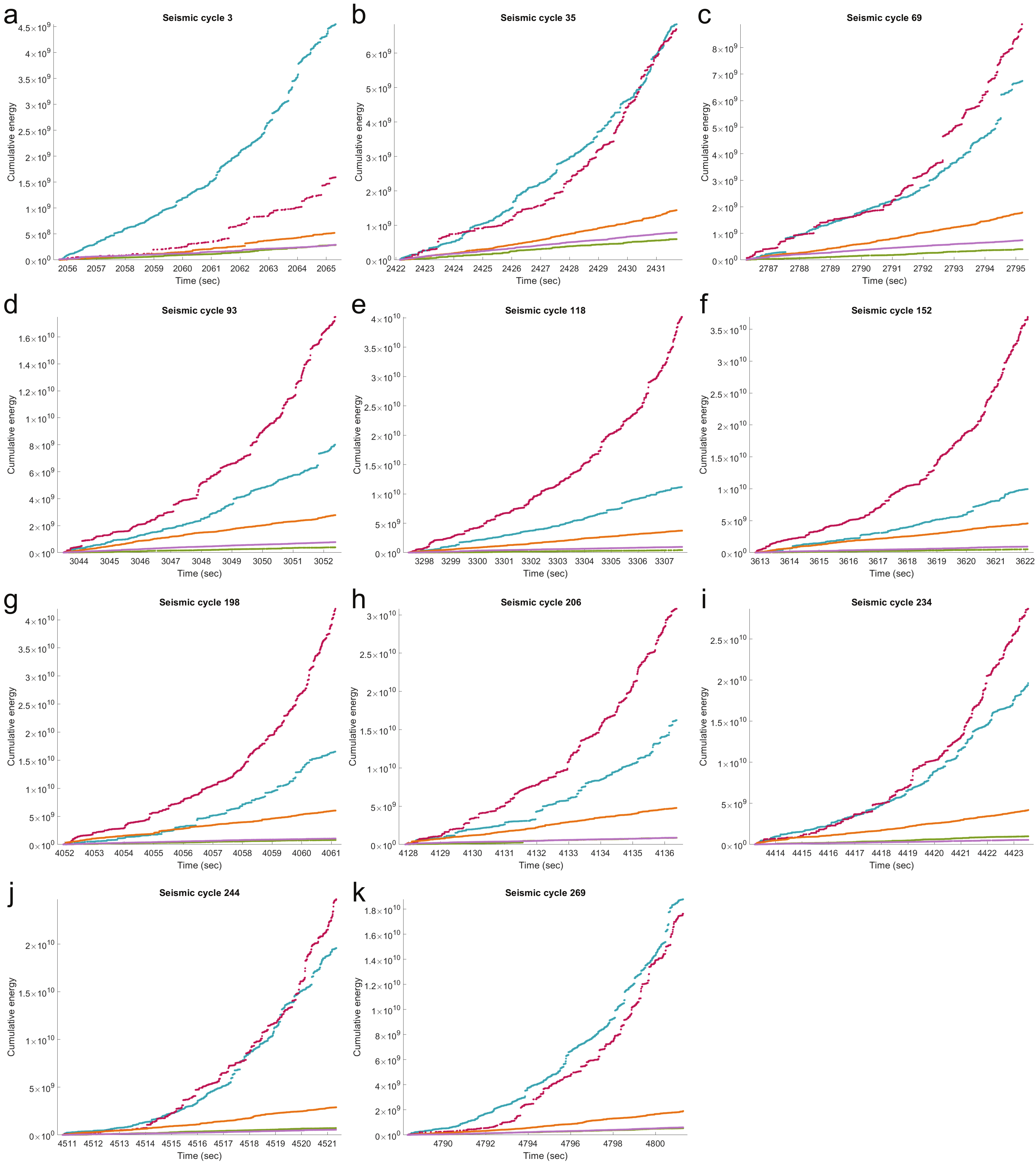}
    \caption{Cumulative energy for a representative slip cycle at a given normal stress. Colors represent different clusters (see main text for details). See Figure~\ref{lab3} for the corresponding steps.}
    \label{lab5}
\end{figure*}

Figures~\ref{lab4} and \ref{lab5} show the evolution of cumulative number of AEs and energy, respectively, over the course of the step-up-step-down experiment. We observe that each normal stress presents a different pattern of AE production. For example, as normal stress increases, so does the relative proportion of red cluster AEs. The patterns on the ``step-down'' side of the experiment do not exactly match those on the ``step-up'' side, which could be linked to a shear strain effect.

\begin{figure*}[!htb]
    \noindent
    \includegraphics[width=\textwidth]{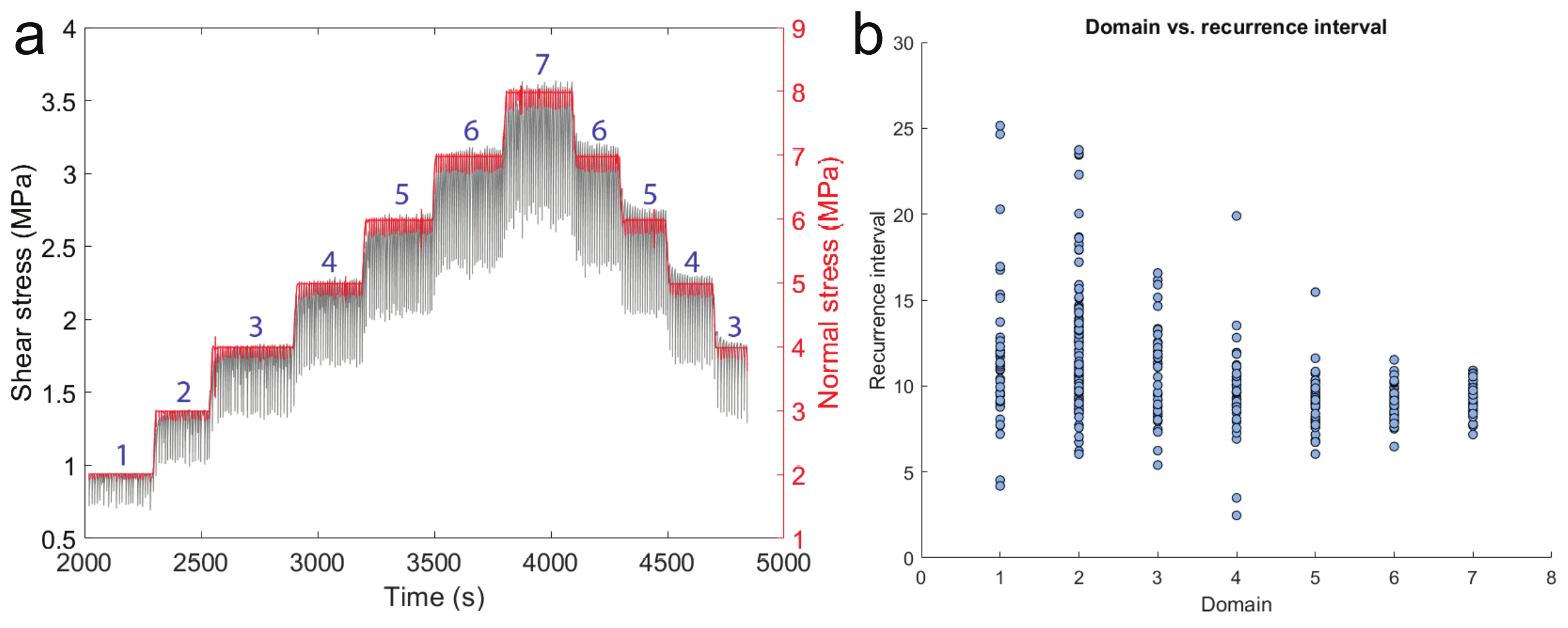}
	\caption{A) The seven DAML shear stress domains. B) Recurrence interval as a function of normal stress. Note, the spread in recurrence intervals at low normal stresses, relative to those at high normal stresses }
    \label{lab6}
\end{figure*}

In Figure~\ref{lab6}, we plot the recurrence interval for all slip cycles at a given normal stress. In general, as the normal stress increases the quakes become more periodic, with most cycles hovering around the average recurrence interval. At low normal stresses slip cycles are significantly more varied and display a wide range of recurrence intervals.

\label{appxLab_end}
\clearpage

\section{Detailed network architecture} \label{appxNetwork}

\begin{table*}[!htb]
    \caption{CSOM training parameters} 
	\centering
	\begin{tabular}{|c|c|c|c|c|c|}
		\hline
		Steps: & 0-1.5M & 1.5M-4M & 4M-8M & 8M-16M & 16M-20M \\ \hline \hline
		Alpha & 0.3 & 0.2 & 0.15 & 0.1 & 0.05 \\ \hline
		Beta & 0.8 & 0.5 & 0.3 & 0.1 & 0.05 \\ \hline
		Gamma & 5 & 4 & 3 & 2 & 0.5 \\ 
		\hline
	\end{tabular}
	\label{netT1}
\end{table*}

Table~\ref{netT1} lists the training parameters for the 15x15 CSOM. We trained in five stages, each of increasing length and decreasing parameter values. We experimented with a variety of lattice sizes, stage lengths, and parameter values.

\begin{figure*}[!htb]
	\noindent
	\includegraphics[width=\textwidth]{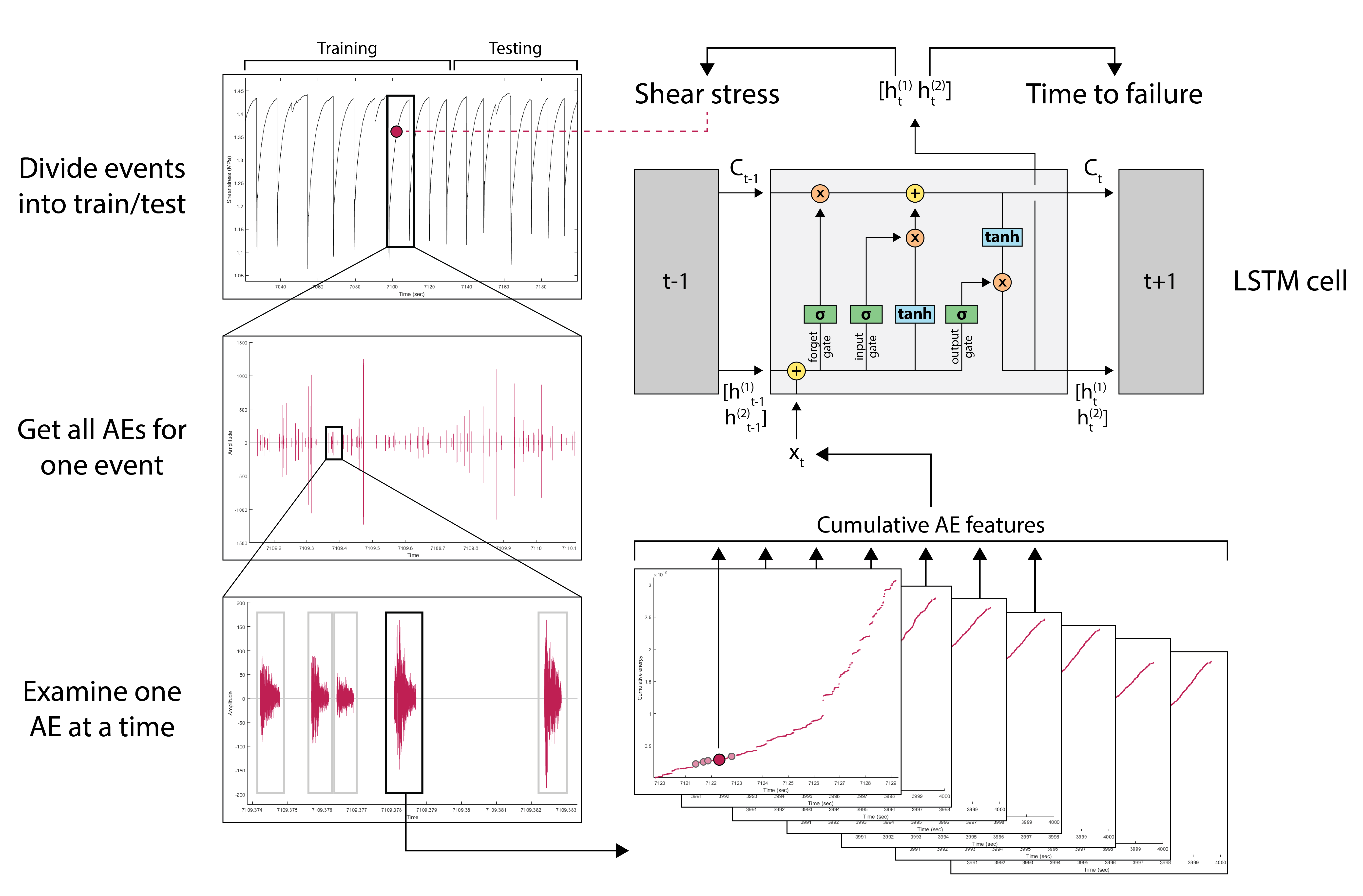}
	\caption{Simplified version of the LSTM training process.}
	\label{net1}
\end{figure*}

Figure~\ref{net1} is a schematic of the LSTM training process. Counterclockwise from the top left, the steps are as follows. First, we divide the quakes into the training and testing sets. Unlike other studies, we divided the data based on entire cycles rather than on time windows. We used 80\% of the quakes for training and the remaining 20\% for testing. We did not use a validation set because of the limited data availability. Note that though the training and testing sets in the figure are consecutive, we assigned quakes randomly. 

Second, we gathered all the AEs for a given quake. For simplicity, the figure shows only the red cluster AEs, but we also used the blue cluster in the training. During the training process, the LSTM examines one AE at a time in the order in which they were produced. The input vector consists of the values of the cumulative features over the recurrence interval for the given AE.

\begin{figure*}[!htb]
    \centering
	\includegraphics[height=7in]{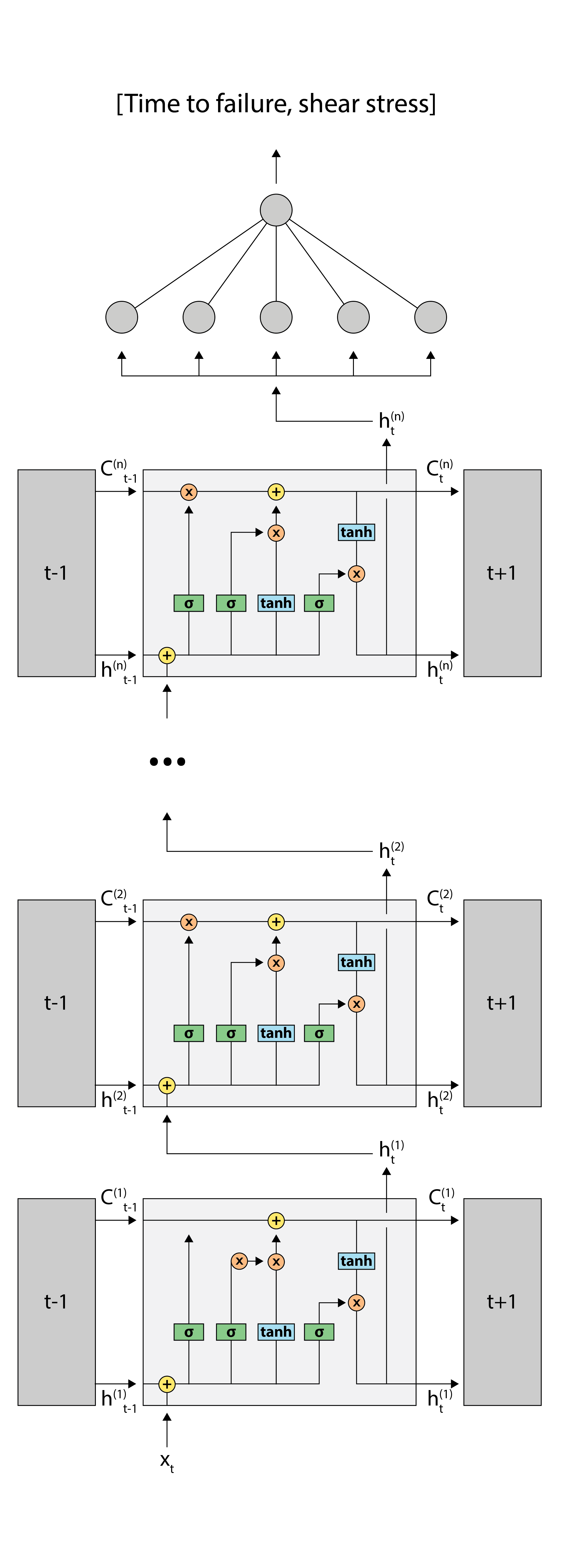}
	\caption{Multilayer LSTM network architecture}
	\label{net2}
\end{figure*}

The LSTM network consists of a series of LSTM cells followed by a single fully connected layer (Figure~\ref{net2}). The inputs (\(x_t\)) are the cumulative AE features and the outputs (\(h_t\)) are forecasts for TTF and shear stress. This, and all subsequent networks, were implemented using Pytorch.

\begin{figure*}[!htb]
	\noindent
	\includegraphics[width=\textwidth]{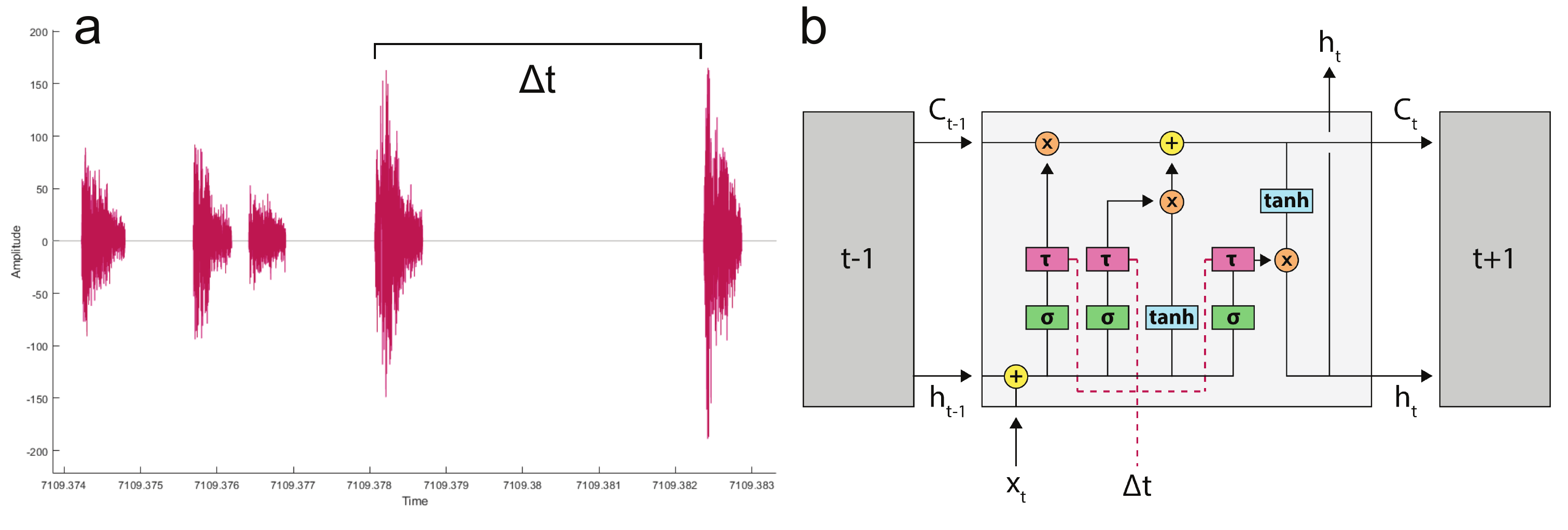}
	\caption{A) Example of data with variable amounts of time between AEs. B) TGLSTM network architecture}
	\label{net3}
\end{figure*}

We experimented with using Time-Gated LSTM (TGLSTM) in addition to the vanilla LSTM. The AE data is event-based, which means that the time steps between events are irregular (Figure~\ref{net3}). Vanilla LSTM assumes that the data has a consistent sampling frequency, which may degrade its performance on the lab dataset. TGLSTM attempts to solve this problem by modifying the LSTM cell to take the time between events into account. In particular, it adds an additional time gate (pink boxes) to the three existing gates that takes in the time distance between events. The time gates modulate the information within the cell based on how much time has passed. For example, if the time between two events is relatively large, they are likely less related than vanilla LSTM would assume. One way the TGLSTM time gates would respond would be to decrease the impact of the memory of the first event on the forecast for the second.

\begin{figure*}[!htb]
    \centering
	\includegraphics[width=0.7\textwidth]{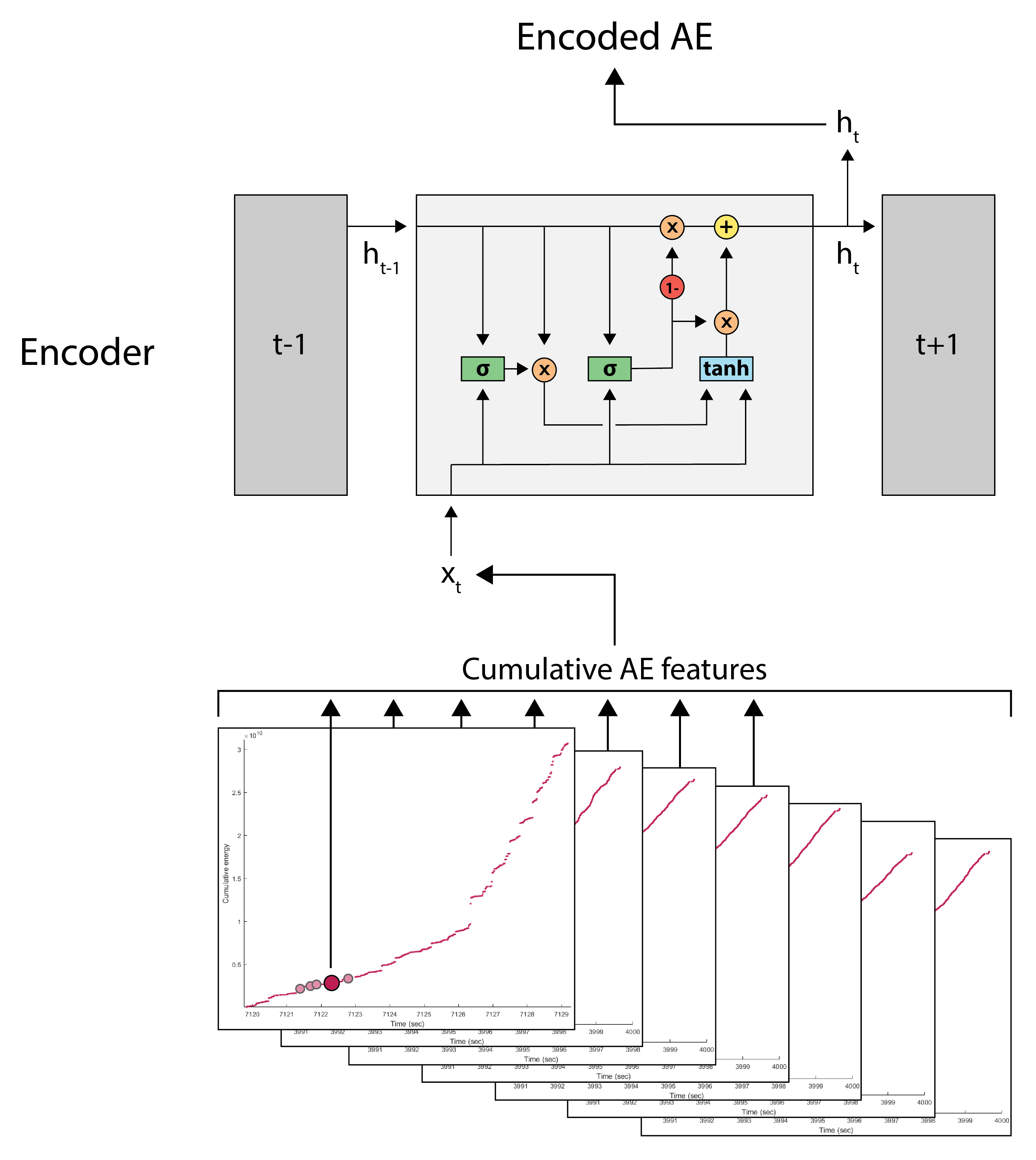}
	\caption{The encoder component of the attention network. The encoder consists of one or more GRUs.}
	\label{net4}
\end{figure*}

\begin{figure*}[!htb]
    \centering
	\includegraphics[height=7in]{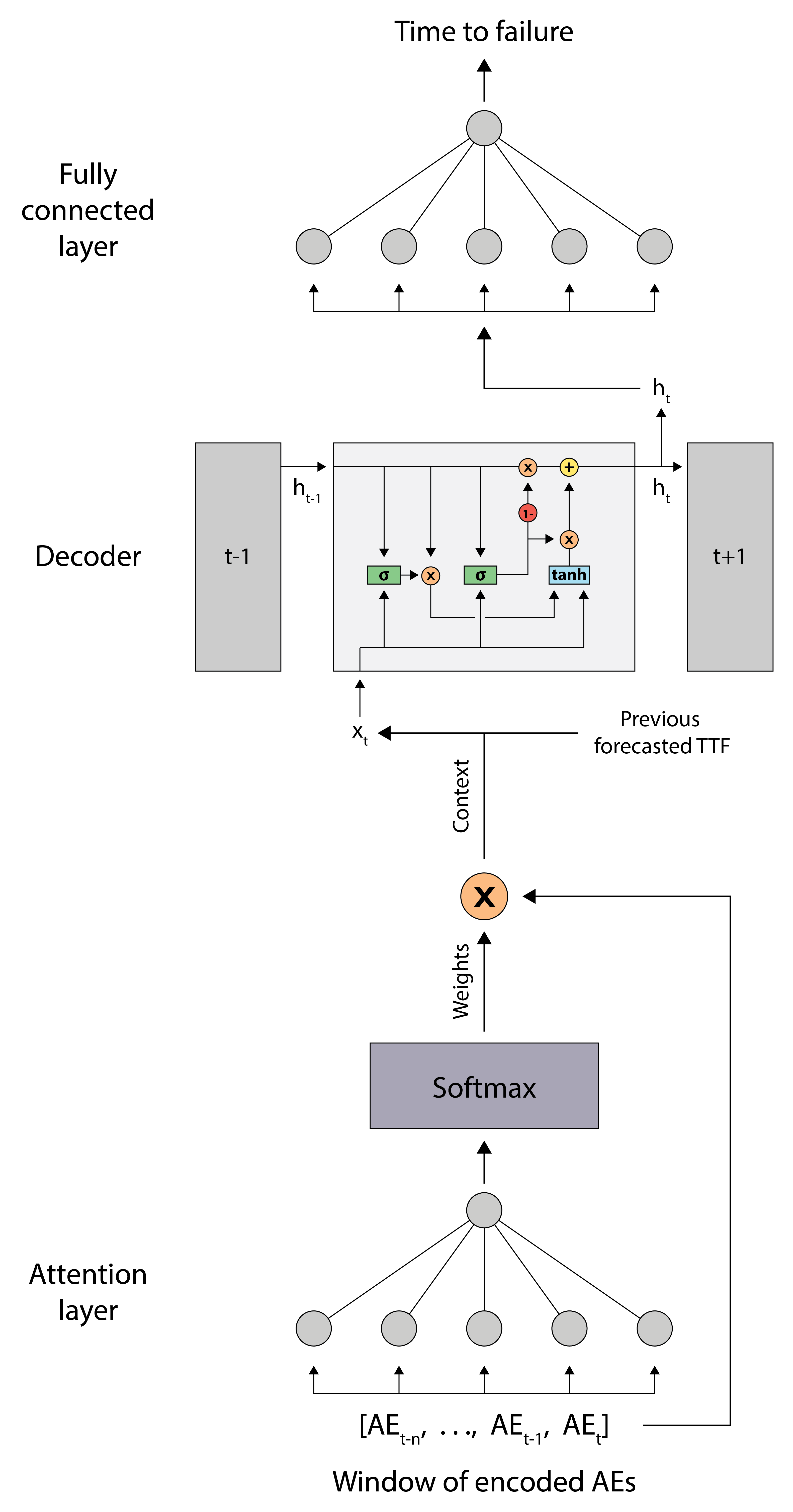}
	\caption{The attention and decoder components of the attention network. The attention layer is a fully connected layer followed by softmax and matrix multiplication. The decoder is similar to the encoder, with the addition of a final fully connected layer.}
	\label{net5}
\end{figure*}

Figures~\ref{net4} and \ref{net5} show the network architecture for the attention network. The network has three components: an encoder (Figure \ref{net4}) an attention layer, and a decoder (Figure \ref{net5}). The encoder works much the same way as the LSTM network, though it uses simpler GRU cells rather than LSTM cells. In the first step of the training process, each AE in the cycle is encoded into a latent space.

Next, encoded AEs are fed into the attention layer. For each input AE, the attention layer examines a window of previous AEs. The purpose of the attention layer is to assign relative importances (weights) to each AE in the window. Thus, the network learns which data points to ``pay attention'' to and which to ignore. The weights must be recalculated for each input AE. The weights are multiplied by their respective window AEs to produce a vector called the context.

Finally, the context and the previous hidden state (i.e. the previous output) are used as inputs for the decoder. Again, the decoder consists of one or more GRU cells. A final fully connected layer transforms the GRU output into the forecasted TTF.

\begin{figure*}[!htb]
    \centering
	\includegraphics[width=\textwidth]{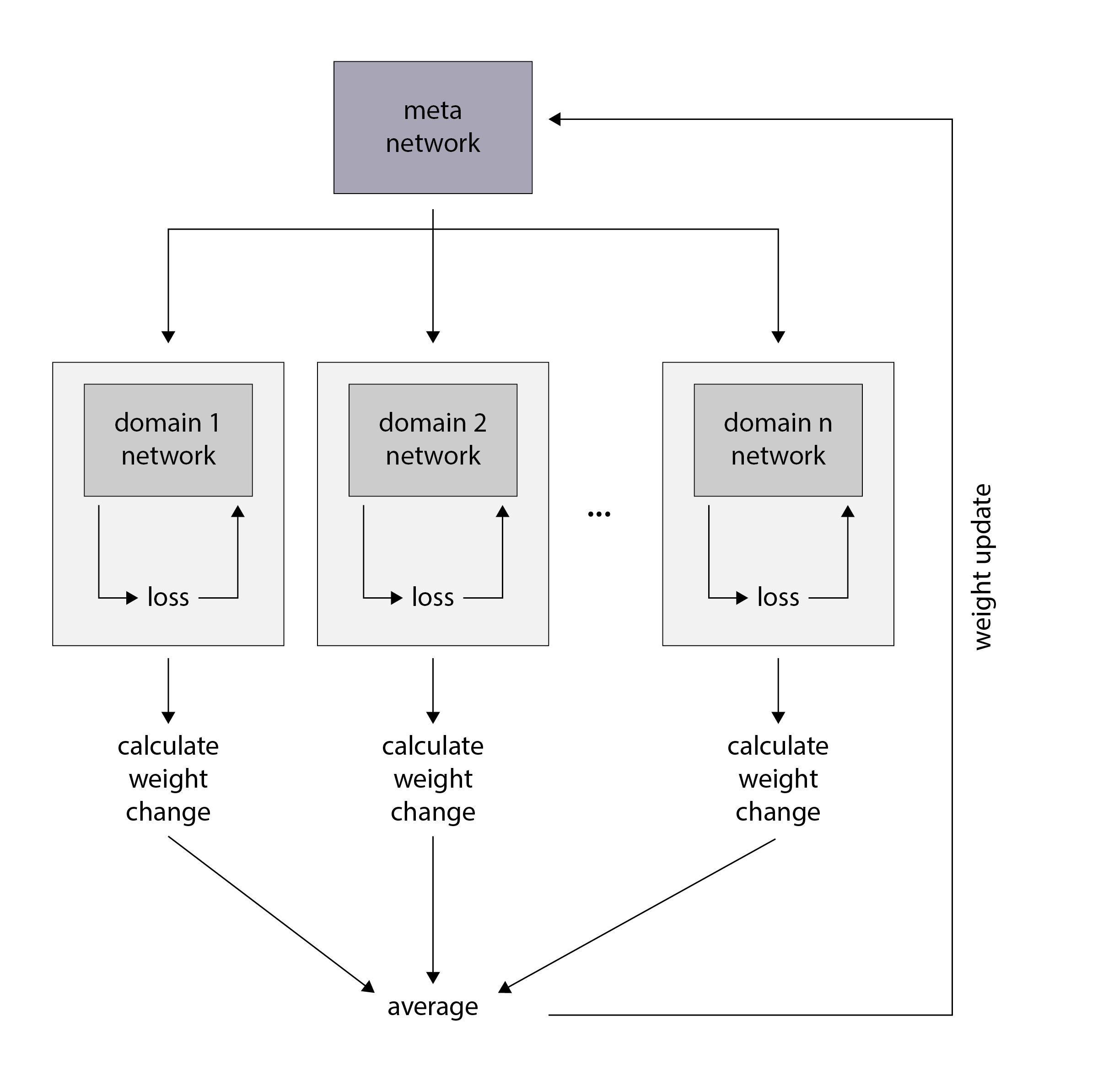}
	\caption{DAML cell}
	\label{net6}
\end{figure*}

Figure~\ref{net6} is a schematic for the domain adaptive meta learning network (DAML). First, the data is divided into domains representing differences in some feature. With the lab data, we assigned domains based off of quake shear stress. We also experimented with assigning domains based on recurrence interval. Next, we initialize the meta network. The meta network can be any network architecture with any number of layers. Here, we did two tests using the LSTM and attention networks.

Each domain is trained separately on a copy of the meta network. Thus, each domain network has the same starting weights. Each domain network is trained as normal for a predetermined number of epochs, which is usually relatively small. At the end of domain training, we calculate the weight change between the meta network and the trained domain networks. These changes are averaged and used to update the weights of the meta network. The process continues as the domain networks make fresh copies of the meta network for another round of training. Training proceeds until some number of meta epochs (i.e. number of meta network weight updates) is reached.

The purpose of DAML is to better learn multiple types of data at once. By having separate domain networks, the training difficulty is greatly reduced. Averaging the weight changes ensures that all data types play a role regardless of the size of the domain. Theoretically, the meta network weights will move towards a representation where forecasts for all domains are reasonably accurate.

\label{appxNetwork_end}
\clearpage

\section{Training with stress slope and forecasting} \label{appxForecast}

\begin{table*}[!htb]
	\caption{LSTM hyperparameters and architecture variations tested} 
	\centering
	\begin{tabular}{ |c|c| } 
		\hline
		\textbf{Component} & \textbf{Value} \\ \hline \hline
		Number of layers & 1-10, 20, 30, 40\\ 
		Hidden size & 1-6, 10, 20 \\ 
		Initial learning rate & 0.0001-0.01 \\
		Learning rate factor & 0.01, 0.05, 0.1, 0.25, 0.5, 0.75, 0.9 \\
		Learning rate patience & 5, 10, 30, 50, 70, 100, 200\\
		Training epochs & up to 3000 \\
		Sequence length & 10, 25, 50, 75, 100, 150, 200, 250, 300, 400, 500 \\
		Mini-batch size & 1, 2, 5, 10, 20 \\
		Gradient clipping \cite{pascanu2013} & 0.1, 0.25, 0.5, 0.75 \\
		Dropout probability \cite{Srivastava2014} & 0.25, 0.5, 0.75\\
		Weight decay \cite{loshchilov2019} & 0.01, 0.005 \\
		Adam with AMSGrad \cite{Reddi2018} & - \\
		Adam followed by SGD \cite{keskar2017} & - \\
		Upsampling rare cases & 1-5 times \\
		Loss weighting for average cases & 0.25, 0.5 \\
		Pre-training on rare cases & - \\
		\hline
	\end{tabular}
	\label{trainT1}
\end{table*}

\begin{figure*}[!htb]
	\noindent
	\includegraphics[width=\textwidth]{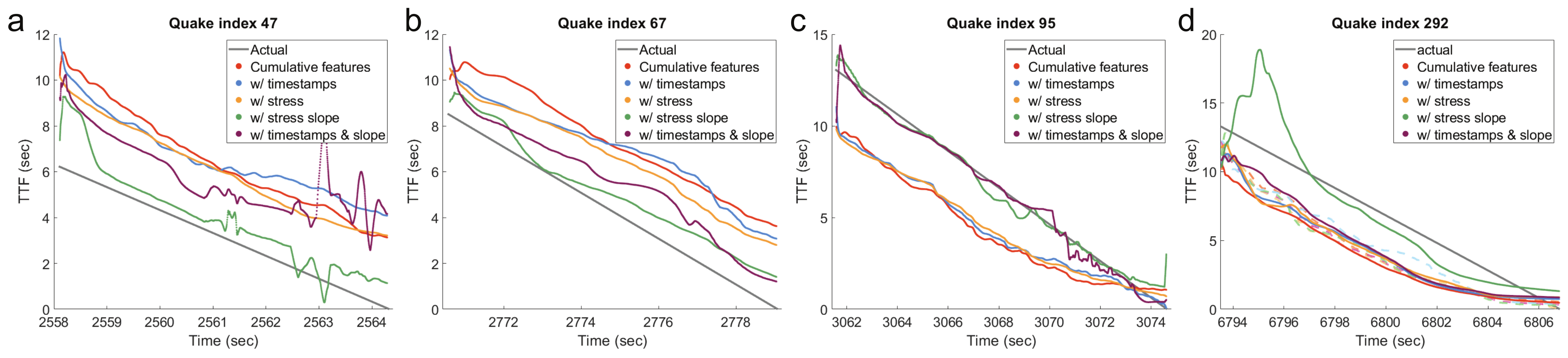}
	\caption{Comparison of actual vs. forecasted time-to-failure for several input datasets with a A) short cycle, B) average cycle, and C) long cycle. In A), LSTM forecasts are greatly improved by using stress slope. In B), all LSTM datasets are incorrect, but the stress slope results have the least error. D) A cycle where the attention network (solid lines) improved over the LSTM (dotted lines)results.}
	\label{train1}
\end{figure*}

Table~\ref{trainT1} lists the network modifications we tested on the lab data. All tests were performed using the LSTM network. We tested these modifications one at a time in order to assess performance. MAE for some of these tests is listed in Table~\ref{trainT2}. In general, network modifications did not improve forecasting results on short and long recurrence intervals. The exception is the inclusion of the slope of the shear stress as an input feature. Using the stress slope alongside the cumulative waveform features led to reduced MAE in all recurrence interval categories. Figure~\ref{train1} shows several example cycles with all tested input variations.

We found that, in many cases, short and average quake recurrence intervals performed best under different circumstances than long recurrence intervals. For example, upsampling generally improved forecasts on long recurrence intervals, but reduced performance on short and average recurrence intervals. This is an argument in favor of using an ensemble or a network like DAML that allows different recurrence intervals to have different training setups.

Table~\ref{trainT3} lists the hyperparameters used for all lab data forecasting networks.

\npdecimalsign{.}
\nprounddigits{4}
\begin{table*}[!htb]
    \caption{MAE from various LSTM architecture tests$^{a}$}
    \centering
    \begin{tabular}{|c|c|n{1}{4}|n{1}{4}|n{1}{4}|n{1}{4}|}
         \hline
		\textbf{Test Parameter} & \textbf{Value} &  \textbf{Total} & \textbf{\textless{}8.5s} & \textbf{8.5-11s} & \textbf{\textgreater{}11s} \\ \hline \hline
		
		\multirow{5}{*}{Input features} & Cumulative & 1.2499007581 & 1.7441183611 & 0.8739976491 & 1.6081029823 \\ \cline{2-6}
		& w/ timestamps & 1.2519341327 & 1.8865957616 & \textbf{0.8259} & 1.6036491163 \\ \cline{2-6}
		& w/ shear stress & 1.2458122708 & 1.6734404230 & 0.8660935786 & 1.6597274818 \\ \cline{2-6}
		& w/ stress slope & \textbf{1.1582} & \textbf{1.6203} & 0.8330546318 & 1.4429574932 \\ \cline{2-6}
		& w/ slope + timestamps & 1.1602220638 & 1.7255127860 & 0.8425267794 & \textbf{1.3556} \\ \cline{2-6} \hline \hline
		
		\multirow{6}{*}{Upsampling} & None & \textbf{1.3262} & \textbf{1.6169} & \textbf{0.9491} & 1.8346646983 \\ \cline{2-6}
		& x1 & 1.3546386134 & 1.8774243469 & 1.0228239150 & 1.6078952316 \\ \cline{2-6}
		& x2 & 1.4135133646 & 1.9245745143 & 1.0832241181 & 1.6723847263 \\ \cline{2-6}
		& x3 & 1.3704165090 & 1.8216235996 & 1.0427646405 & 1.6677831010 \\ \cline{2-6}
		& x4 & 1.4046359157 & 1.9080744487 & 1.0947452758 & 1.6301082041 \\ \cline{2-6} 
		& x5 & 1.4662148593 & 1.9581700612 & 1.2158024970 & \textbf{1.5865} \\ \cline{2-6} \hline \hline
		
		\multirow{5}{*}{Subset of avg events} & All avg & \textbf{1.3262} & \textbf{1.6169} & \textbf{0.9491} & 1.8346646983 \\ \cline{2-6}
		& 80 avg & 1.4157615440 & 1.7908581207 & 1.1673726451 & 1.6171610225 \\ \cline{2-6}
		& 40 avg & 1.4758845888 & 1.7340874760 & 1.1857095262 & 1.8420712448 \\ \cline{2-6}
		& 20 avg & 1.5475985967 & 1.7430618464 & 1.3681700950 & 1.7479888274 \\ \cline{2-6}
		& 10 avg & 1.5668520425 & 1.9046143437 & 1.4303211824 & \textbf{1.5819} \\ \cline{2-6} \hline \hline
		
		\multirow{3}{*}{Loss weighting} & Unweighted & \textbf{1.3262} & \textbf{1.6169} & \textbf{0.9491} & 1.8346646983 \\ \cline{2-6}
		& 0.5 avg & 1.3603261895 & 1.9632025634 & 1.0323458991 & \textbf{1.5480} \\ \cline{2-6}
		& 0.25 avg & 1.3705873488 & 1.8802783879 & 1.0701561336 & 1.5734534586 \\ \cline{2-6} \hline \hline
		
		\multirow{2}{*}{Energy only} & All features & \textbf{1.1582} & \textbf{1.6203} & \textbf{0.8331} & \textbf{1.4430} \\ \cline{2-6}
		& Energy only & 1.5777283783 & 1.9677905693 & 1.2999448719 & 1.8243607515 \\ \cline{2-6} \hline
    \multicolumn{6}{c|}{$^{a}$Bold entries are the best MAE for each test type in each TTF category.} \\
    \multicolumn{6}{c|}{}
    \end{tabular}
    \label{trainT2}
\end{table*}
\npnoround

\begin{table*}[!htb]
	\caption{Hyperparameters for the networks used to forecast shear stress and TTF} 
	\centering
	\begin{tabular}{|c|c|c||c|c|c|c|}
	\hline
	    & \multicolumn{2}{c||}{\textbf{Stress}} & \multicolumn{4}{c|}{\textbf{TTF}} \\ \hline
		\textbf{Hyperparameter} & \textbf{LSTM} & \textbf{Att.} & \textbf{LSTM} & \textbf{Att.} & \textbf{DAML (LSTM)} & \textbf{DAML (att)} \\ \hline \hline
		Number of train events & \multicolumn{2}{c||}{316} & \multicolumn{4}{c|}{316} \\ \hline
		Number of test events & \multicolumn{2}{c||}{80} & \multicolumn{4}{c|}{80} \\ \hline
		Number of layers & \multicolumn{2}{c||}{3} & \multicolumn{4}{c|}{3} \\ \hline
		Hidden size & \multicolumn{2}{c||}{6} & \multicolumn{4}{c|}{6} \\ \hline
		Training epochs & 500 & 474 & 1712 & 500 & 1040 & 101 \\ \hline
		Domain epochs & - & - & - & - & 10 & 10 \\ \hline
		Starting learning rate & \multicolumn{2}{c||}{0.003} & 0.0001 & 0.003 & 0.003 & 0.003 \\ \hline
		Sequence/window length & \multicolumn{2}{c||}{300} & \multicolumn{4}{c|}{300} \\ \hline
		Learning rate factor & \multicolumn{2}{c||}{0.5} & 0.75 & 0.5 & 0.5 & 0.5 \\ \hline
		Mini-batch size & \multicolumn{2}{c||}{10} & \multicolumn{4}{c|}{10} \\ \hline
		Learning rate patience & 30 & 50 & 100 & 50 & 100 & 15 \\ \hline
		Gradient clipping & - & 0.5 & - & 0.5 & 0.5 & 0.5 \\ \hline
		Number of upsamples & 3 & 0 & 0 & 0 & 0 & 0 \\ \hline
	\end{tabular}
	\label{trainT3}
\end{table*}

\begin{table*}[!htb]
	\caption{Percent of quakes correctly forecasted$^{a}$} 
	\centering
	\begin{tabular}{|c|c|c|c|c|}
		\hline
		\textbf{Network} & \textbf{Total} & \textbf{\textless{}8.5s} & \textbf{8.5-11s} & \textbf{\textgreater{}11s} \\ \hline \hline
		Control & 42.5 & 0 & \textbf{82.93} & 0 \\ \hline \hline
		LSTM & 60 & 23.81 & 75.61 & 66.67 \\ \hline
		LSTM ensemble & 65 & 23.81 & \textbf{82.93} & \textbf{72.22} \\ \hline
		Attention & 65 & 23.81 & \textbf{82.93} & \textbf{72.22} \\ \hline
		DAML (LSTM) & 50 & 19.05 & 78.05 & 22.22 \\ \hline
		DAML (attention) & 45 & 4.76 & 68.29 & 38.89 \\ \hline
		Ensemble & \textbf{66.25} & \textbf{28.57} & \textbf{82.93} & \textbf{72.22} \\ \hline
	\multicolumn{5}{c|}{$^{a}$Bold entries are the highest percent in each category} \\
	\multicolumn{5}{c|}{}
	\end{tabular}
	\label{trainT4}
\end{table*}

Table~\ref{trainT4} lists the percentage of quakes deemed to be correctly forecasted by each network. We determined that a quake is correctly forecasted if its MAE is less than 10\% of its recurrence interval. The control network always outputs the average recurrence interval for comparison. Interestingly, both ensembles are equally correct on the average and long quakes despite having somewhat different MAE. For earthquake forecasting, small changes in MAE may not be worth the additional training effort.

\begin{figure*}[!htb]
	\noindent
	\includegraphics[width=\textwidth]{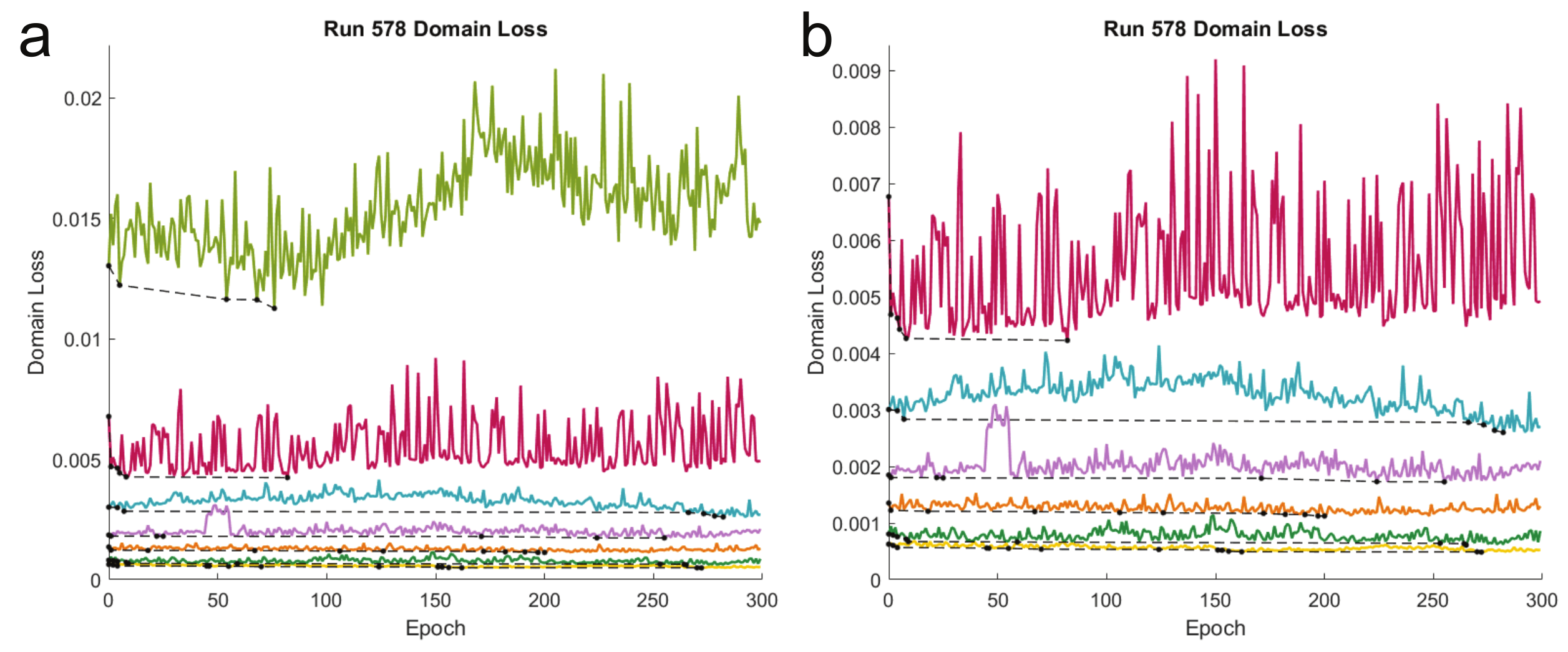}
	\caption{A) Test set loss for each domain over the DAML (LSTM) training process. B) Zoom. Black circles indicate progressive minima. The colors are the same as in Figure~\ref{train3} (In order of increasing stress: light green, teal, red, orange, purple, yellow, dark green.)}
	\label{train2}
\end{figure*}

\begin{figure*}[!htb]
	\noindent
	\includegraphics[width=\textwidth]{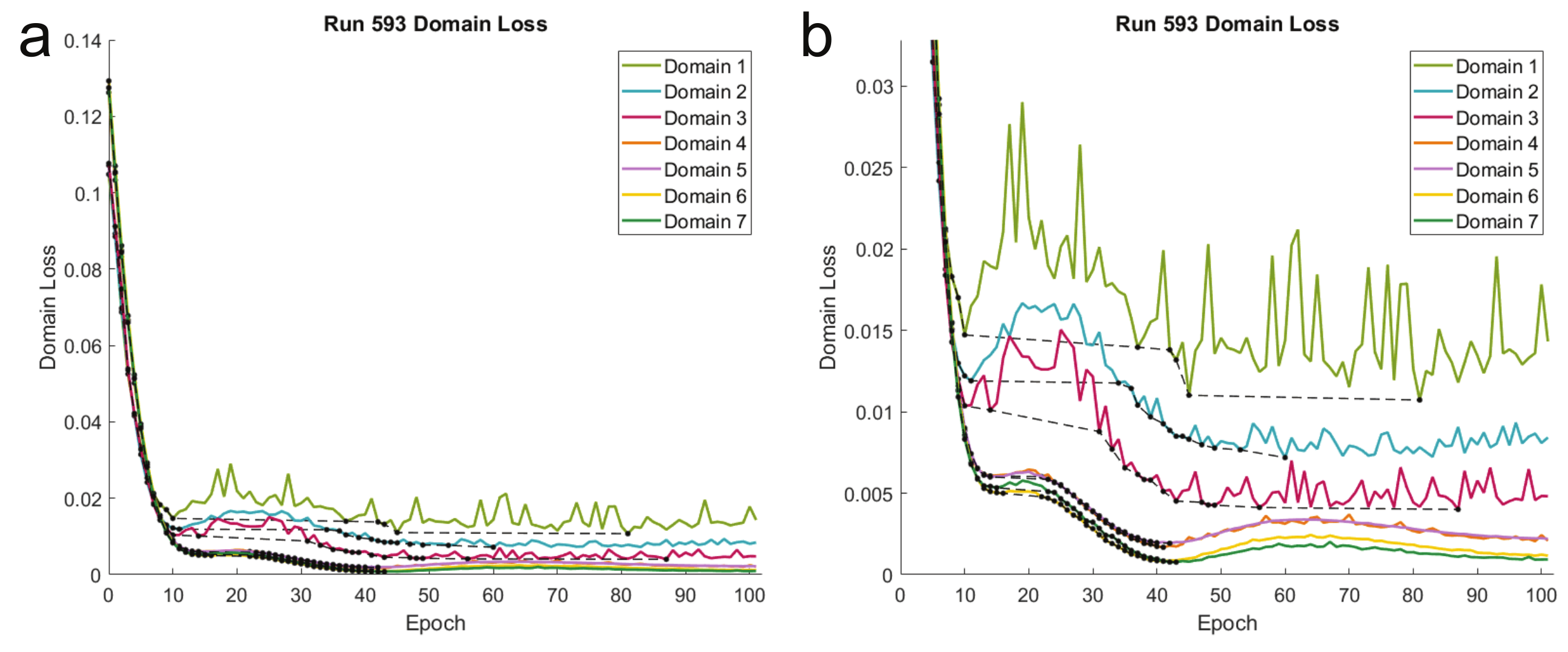}
	\caption{A) Test set loss for each domain over the DAML (attention) training process. B) Zoom. Black circles indicate progressive minima.}
	\label{train3}
\end{figure*}

Figures~\ref{train2} and \ref{train3} show test set loss by domain (Figure~\ref{lab6}) for the two DAML network architectures. The three lowest stress domains have the highest loss, likely due to the wide recurrence interval range. As this range decreases with increased stress, loss decreases as well. With more data, perhaps the low stress domains could be improved.

\begin{figure*}[!htb]
	\noindent
	\includegraphics[width=\textwidth]{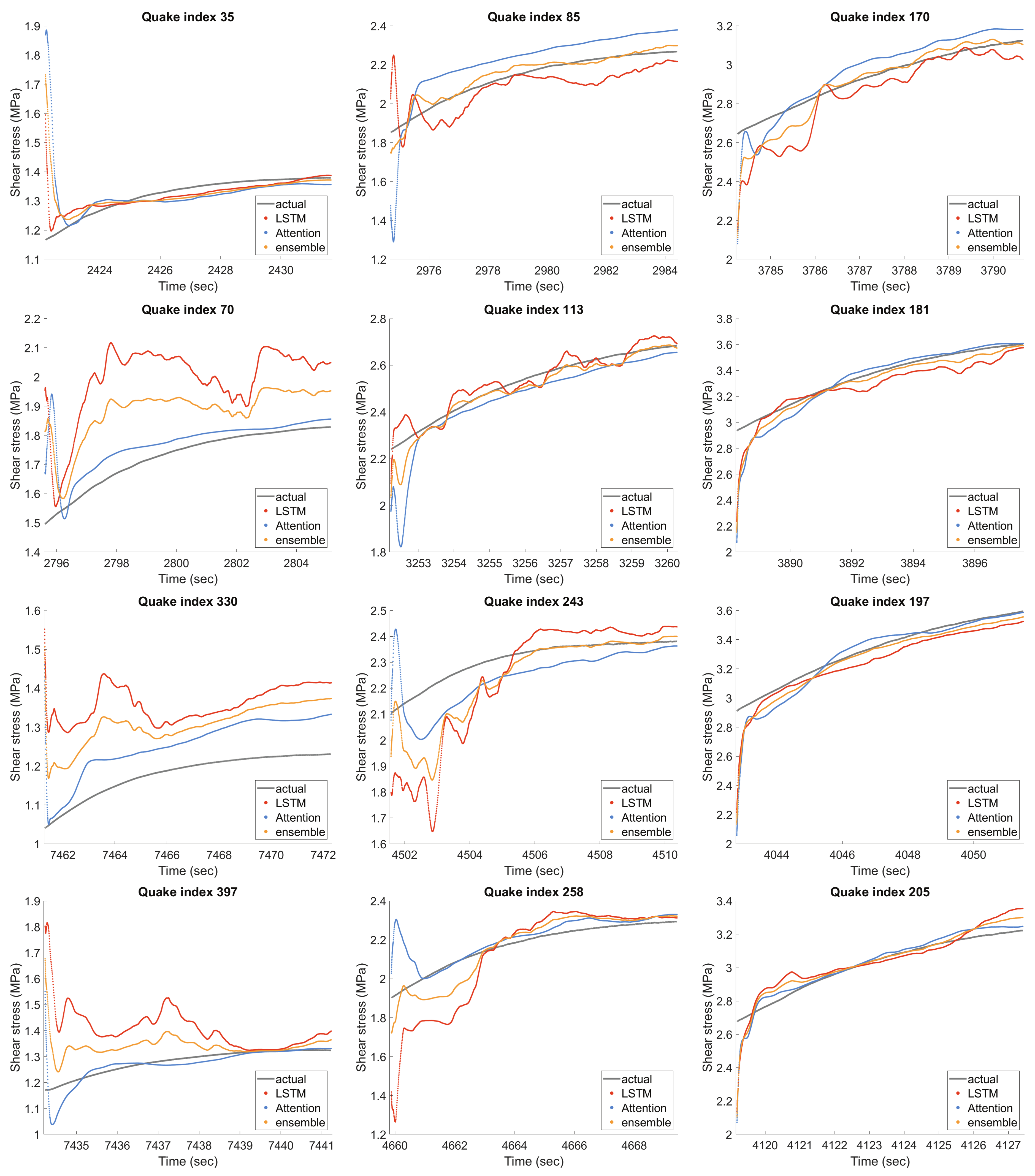}
	\caption{Actual vs. forecasted shear stress for events with (left column) low stress, (middle column) medium stress, and (right column) high stress.}
	\label{train4}
\end{figure*}

\begin{figure*}[!htb]
	\noindent
	\includegraphics[width=\textwidth]{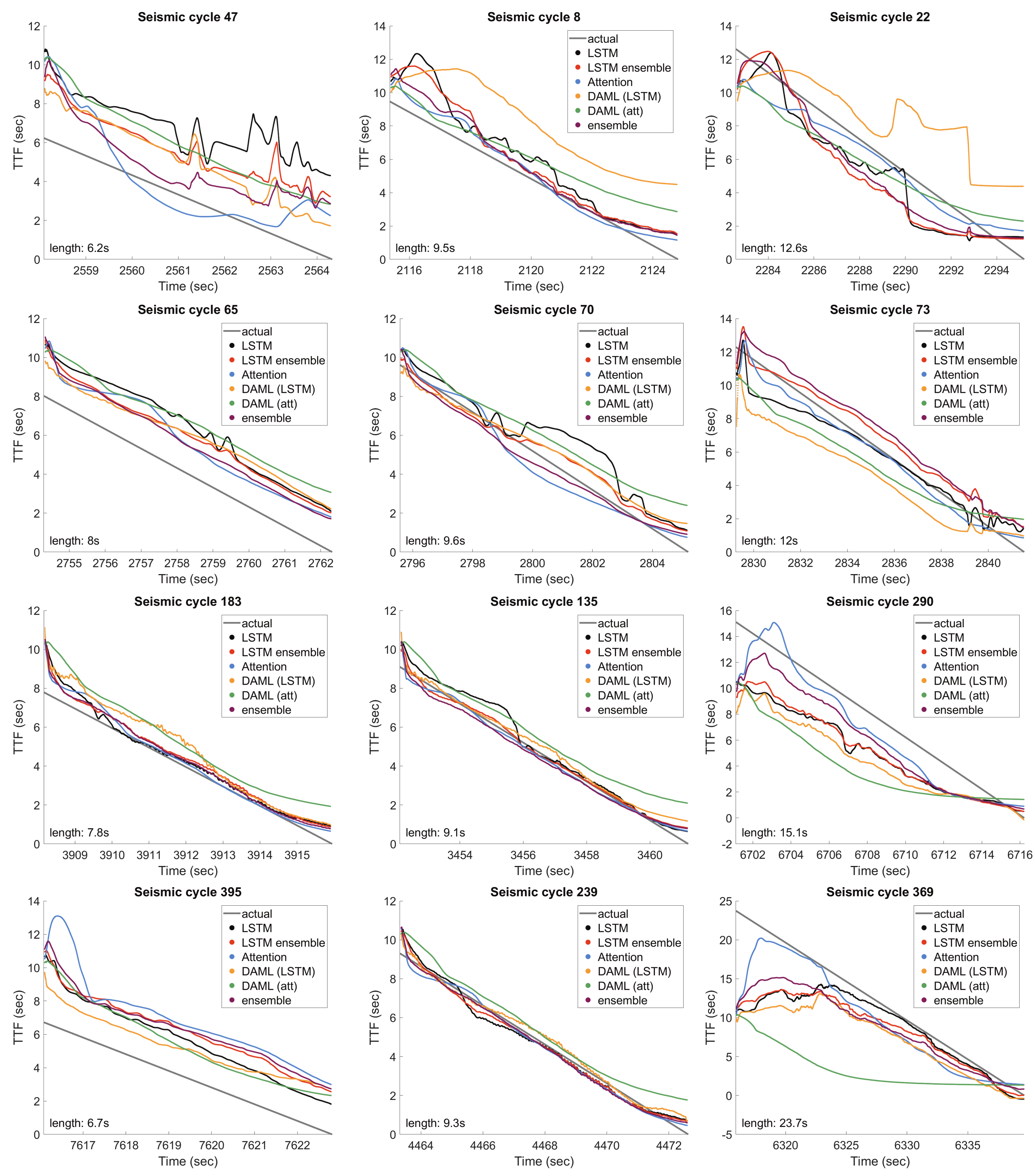}
	\caption{Actual vs. forecasted TTF for events with a (left column) short recurrence interval, (middle column) average recurrence interval, and (right column) long recurrence interval.}
	\label{train5}
\end{figure*}

Figures~\ref{train4} and \ref{train5} show forecasting results for additional test set quakes. Stress results are divided into low, medium, and high stress and TTF results are divided into short, average, and long recurrence interval.

\label{appxForecast_end}
\clearpage

\section{Statistical methods} \label{appxStat}

\begin{figure*}[!htb]
	\noindent
	\includegraphics[width=\textwidth]{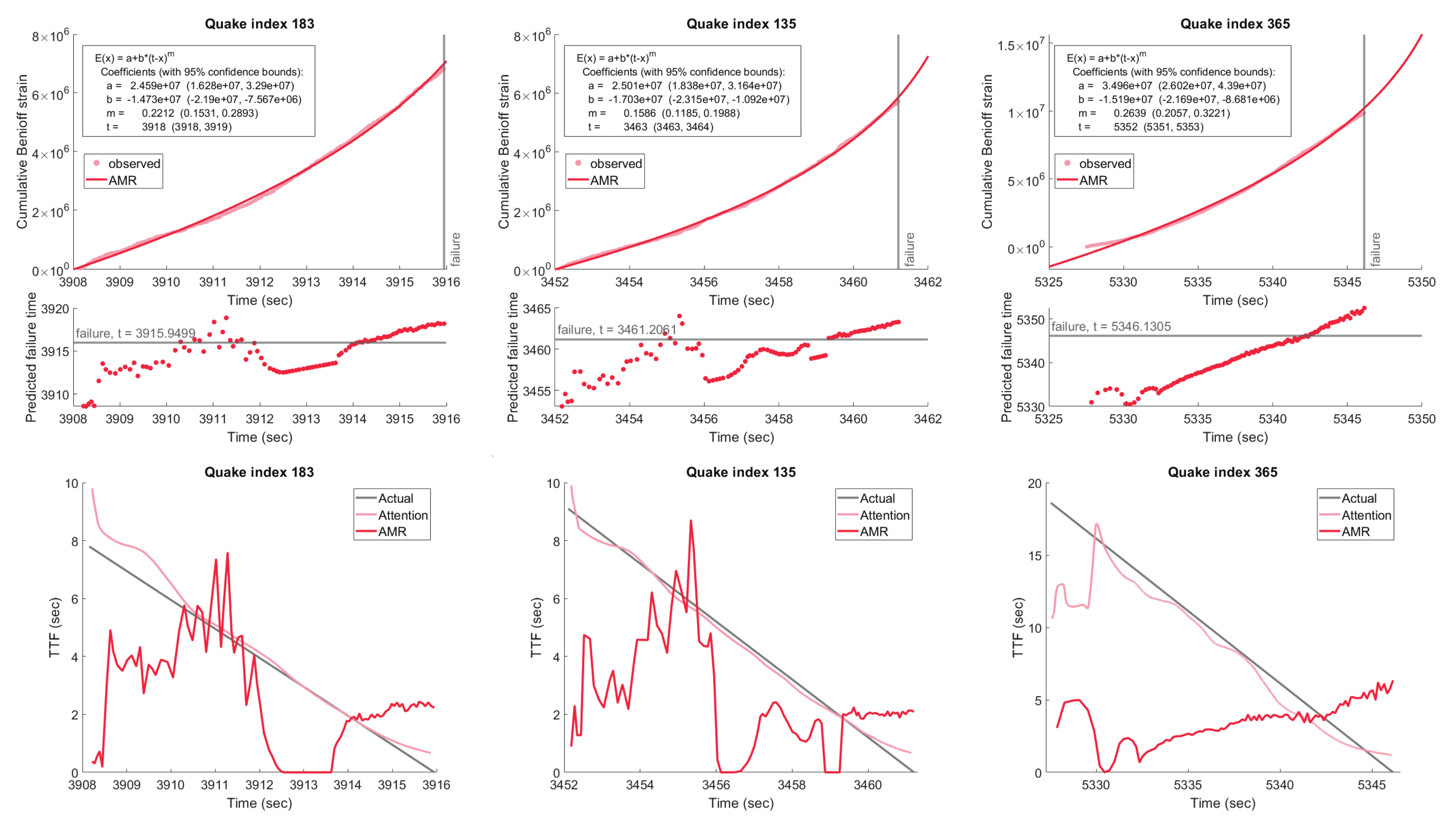}
	\caption{(top) Sample AMR results for three test set quakes. The top plots show the final AMR curve compared to the true cumulative Benioff strain. The lower scatter plots show TTF predictions over the quake cycle. (bottom) AMR TTF predictions compared to the true TTF and LSTM forecasts for a (L to R) short, average, and long recurrence interval.} 
	\label{stat1}
\end{figure*}

The Gutenberg-Richter law and accelerating moment release (AMR) are two statistical methods that some have suggested could be used for earthquake forecasting. AMR attempts to forecast TTF by tracking the energy released from seismic precursors \cite{benioff1951, varnes1989, bufe1993, bufe1994}. AMR was widely used for many years, but quickly fell out of fashion after its forecasts were proven to be statistically insignificant \cite{hardebeck2008}. We include this method here for its historical significance. We calculated TTF using AMR on the test set quakes used in our LSTM and attention networks. To compare to our network results, we calculated \(t_c\) and TTF at several timesteps using only previously recorded data. The results (Figure~\ref{stat1}) show that AMR performs quite badly. The AMR curve generally fits the observed strain values well, but the \(t_c\) prediction is off by many seconds and even fails to capture the linearly decreasing nature of the true TTF.

The Gutenberg-Richter law defines the relationship between earthquake magnitude and the number of earthquakes expected to occur as

\[log(N) = a-bM\]

where \(N\) is the number of earthquakes with magnitude \(\geq M\) and \(a\) and \(b\) are constants \cite{gutenberg1944}. Our focus is on the b-value, which has been shown to vary with the type of seismicity as well as with changing fault conditions \cite{tiampo2012}. Some have suggested that the b-value could be used to forecast earthquakes, although there currently is no method to directly tie it to the time of failure. Despite this, we calculated the b-value over time for our test set to show its general trends.

\citeA{riviere2018} did extensive work on this topic in the lab, showing that the law still holds in that setting. We followed the procedure laid out in that paper for determining b-values. AE magnitude is defined as log(A) where A is the maximum amplitude. The b-value is obtained by fitting a line to magnitude vs. log(number of quakes) and taking the slope. We used a magnitude range of [1.5-2.7] and a moving window of 2000 AEs shifted every 100 AEs. As with AMR, we calculated b-values for the LSTM test set quakes.

\begin{figure*}[!htb]
	\noindent
	\includegraphics[width=\textwidth]{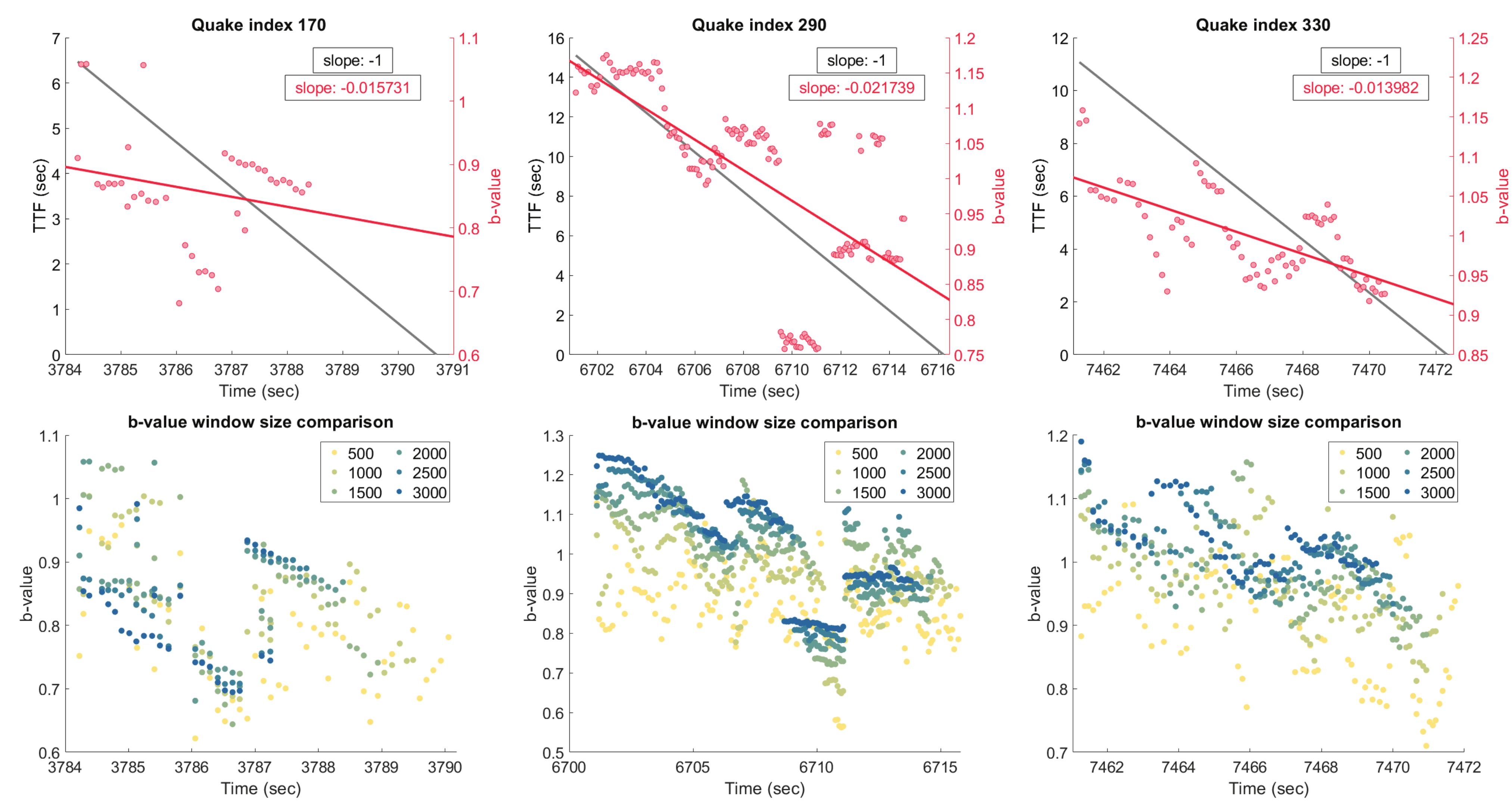}
	\caption{(top) Sample b-value results for three test set quakes. The grey line is actual TTF (for reference), the red circles are calculated b-values, and the red line is the b-value trendline. (bottom) B-values for the same three quakes calculated at different window lengths, ranging from 500 to 3000 AEs.} 
	\label{stat2}
\end{figure*}

Our results are shown in Figure~\ref{stat2}. Our results agree with \citeA{riviere2018} in that the b-value generally decreases over the quake cycle, though this is not always the case. Beyond this, there is no clear way to connect the b-values to either the TTF or time of failure. We also found that the b-values depend heavily on the AE window size. Larger windows often produce larger b-values. This dependency indicates that the b-value may not be a very good tool for TTF forecasting, even if such a method existed.

\label{appxStat_end}

\end{document}